\documentclass[astrosymb,trackchanges]{aastex7}

\usepackage{comment}
\usepackage{multirow}
\usepackage{arydshln}
\usepackage{ulem}

\usepackage{mathtools}

\newcommand\T{\rule{0pt}{2.6ex}}       
\newcommand\B{\rule[-1.2ex]{0pt}{0pt}}

\usepackage [english]{babel}
\usepackage [autostyle, english = american]{csquotes}
\MakeOuterQuote{"}

\def\simgt{\lower.5ex\hbox{$\; \buildrel > \over \sim \;$}}
\def\simlt{\lower.5ex\hbox{$\; \buildrel < \over \sim \;$}}

\newcommand\chandra{{\it Chandra}}
\newcommand\xmm{{\it XMM-Newton}}

\newcommand\swift{{\it Swift\/}}
\newcommand\nustar{{\it NuSTAR}}

\newcommand\nicer{{\it NICER}}
\newcommand\ixpe{{\it IXPE}}
\newcommand\xrism{{\it XRISM}}

\newcommand\VLA{{\it VLA}}
\newcommand\meerkat{{\it MeerKAT}}
\newcommand\EP{{\it Einstein Probe}}

\newcommand\src{{MAXI J1744-294}}
\newcommand\axj{{AX J1745.6-2901}}

\newcommand{\Fexxv}{Fe~{\sc xxv}}
\newcommand{\Fexi}{Fe~{\sc xi}}
\newcommand{\Fexii}{Fe~{\sc xii}}

\newcommand{\Fexxvi}{Fe~{\sc xxvi}}
\newcommand{\Caxx}{Ca~{\sc xx}}

\newcommand{\Nixxvii}{Ni~{\sc xxvii}}
\newcommand{\Nixxviii}{Ni~{\sc xxviii}}
\newcommand{\kmps}{km s$^{-1}$}
\newcommand{\Fei}{Fe~{\sc i}}

\newenvironment{rotatepage}%
    {\clearpage\pagebreak[4]\global\pdfpageattr\expandafter{\the\pdfpageattr/Rotate 90}}%
    {\clearpage\pagebreak[4]\global\pdfpageattr\expandafter{\the\pdfpageattr/Rotate 0}}%

\shorttitle{XRISM spectroscopy of the Galactic center II. MAXI J1744-294 aka T37}
\shortauthors{Parra et al.}
\setwatermarkfontsize{170pt}
\graphicspath{{./}{figures/}}

\begin{document}

\title{XRISM spectroscopy of a crowded Galactic center region - II.\\
Narrow emission lines in the Black Hole candidate MAXI J1744-294/Swift J174540.2-290037}


\correspondingauthor{Maxime Parra}
\email{maxime.parrastro@gmail.com}

\correspondingauthor{Shifra Mandel}
\email{ss5018@columbia.edu}

\author[0009-0003-8610-853X]{Maxime Parra}
\affiliation{Department of Physics, Ehime University, 2-5, Bunkyocho, Matsuyama, Ehime 790-8577, Japan}
\email{maxime.parrastro@gmail.com}

\author[0000-0002-6126-7409]{Shifra Mandel}
\affiliation{Columbia Astrophysics Laboratory, Columbia University, New York, NY 10027, USA}
\email{ss5018@columbia.edu}

\author[0009-0003-0653-2913]{Kai Matsunaga}
\affiliation{Department of Physics, Graduate School of Science, Kyoto University, Kitashirakawa Oiwake-cho, Sakyo-ku, Kyoto 606-8502, Japan}
\email{matsunaga.kai.i47@kyoto-u.jp}

\author[0000-0002-9709-5389]{Kaya Mori} 
\affiliation{Columbia Astrophysics Laboratory, Columbia University, New York, NY 10027, USA}
\email{km211@columbia.edu}

\author[0000-0002-6797-2539]{Ryota Tomaru}
\affiliation{Department of Earth and Space Science, Graduate School of Science, Osaka University, 1-1 Machikaneyama, Toyonaka, Osaka 560-0043, Japan}
\email{r.tomaru.sci@osaka-u.ac.jp}

\author[0000-0002-3252-9633]{Efrain Gatuzz}
\affiliation{Max-Planck-Institut f\"ur extraterrestrische Physik, Gie{\ss}enbachstra{\ss}e 1, 85748 Garching, Germany}
\email{egatuzz@mpe.mpg.de}

\author[0000-0002-2218-2306]{Paul A. Draghis}
\affiliation{MIT Kavli Institute for Astrophysics and Space Research, Massachusetts Institute of Technology, Cambridge, MA 02139, USA}
\email{pdraghis@mit.edu}

\author[0000-0001-8195-6546]{Megumi Shidatsu} 
\affiliation{Department of Physics, Ehime University, 2-5, Bunkyocho, Matsuyama, Ehime 790-8577, Japan}
\email{shidatsu.megumi.wr@ehime-u.ac.jp}

\author[0000-0003-4580-4021]{Hideki Uchiyama}
\affiliation{Faculty of Education, Shizuoka University, 836 Ohya, Suruga-ku, Shizuoka, Shizuoka 422-8529, Japan}
\email{uchiyama.hideki@shizuoka.ac.jp}

\author[0000-0003-1130-5363]{Masayoshi Nobukawa}
\affiliation{Faculty of Education, Nara University of Education, Nara, 630-8502, Japan}
\email{nobukawa@cc.nara-edu.ac.jp}

\author{Tahir Yaqoob}
\affiliation{NASA/Goddard Space Flight Center, Greenbelt, MD 20771, USA}
\affiliation{Center for Research and Exploration in Space Science and Technology, NASA/GSFC (CRESST II), Greenbelt, MD 20771, USA}
\affiliation{Center for Space Science and Technology, University of Maryland, Baltimore County (UMBC), 1000 Hilltop Circle, Baltimore, MD 21250, USA}
\email{tahir@umbc.edu}

\author{Charles~J.~Hailey}
\affiliation{Columbia Astrophysics Laboratory, Columbia University, New York, NY 10027, USA}
\email{chuckh@astro.columbia.edu}

\author[0000-0002-2006-1615]{Chichuan Jin} 
\affiliation{National Astronomical Observatories, Chinese Academy of Sciences, Beijing 100101, China}
\affiliation{School of Astronomy and Space Science, University of Chinese Academy of Sciences, Beijing 100049, China}
\affiliation{Institute for Frontier in Astronomy and Astrophysics, Beijing Normal University, Beijing 102206, China}
\email{ccjin@bao.ac.cn}

\author[0009-0008-1132-7494]{Benjamin Levin} 
\affiliation{Columbia Astrophysics Laboratory, Columbia University, New York, NY 10027, USA}
\email{bsl2134@columbia.edu}

\author[0000-0003-0293-3608]{Gabriele Ponti} 
\affiliation{INAF - Osservatorio Astronomico di Brera, Merate, Italy}
\affiliation{Max-Planck-Institut f\"ur extraterrestrische Physik, Gie{\ss}enbachstra{\ss}e 1, 85748 Garching, Germany}
\email{gabriele.ponti@inaf.it}

\author[0000-0003-1621-9392]{Mark Reynolds} 
\affiliation{Department of Astronomy, Ohio State University, 140 West 18th Ave., Columbus, OH 43210}
\affiliation{Department of Astronomy, University of Michigan, 1085 S. University Ave., Ann Arbor, MI 48109}
\email{reynolds.1362@osu.edu}

\begin{abstract} 
Narrow, highly ionized X-ray emission lines in black hole low-mass X-ray binaries (BH-LMXBs) are rare and have been observed in only a few sources, during unusual, heavily obscured accretion states. We report on a detailed high-resolution spectral analysis of emission line features from the first \xrism{} observation of a BH-LMXB candidate in a bright soft state, MAXI J1744–294/Swift J174540.2-290037, in the central parsec region of our galaxy. The source was observed as part of an extensive, coordinated multi-wavelength campaign on its recurring X-ray outburst in early 2025. By carefully modeling the contributions of multiple point sources and diffuse emission within the XRISM/Resolve field of view, and combining these data with broadband X-ray coverage from \xmm{} and \nustar{} (Paper I), we identified a narrow ($\sigma \sim 500$–$1000$ km s$^{-1}$), static emission component intrinsic to the system. This component likely arises from a highly ionized (log $\xi \gtrsim 5.5$) photoionized plasma in the inner disk atmosphere, and is accompanied by a weak, narrow neutral Fe I K$\alpha$ line at 6.4 keV. We also detected at least three narrow emission features at atypical energies between 6.7 and 7.1 keV. The lack of corresponding rest-frame atomic transitions points toward highly ionized blueshifted Fe lines with outflow velocities of $-1300$ to $-6000$ km s$^{-1}$, which we model with multiple layers of photoionized or collisional plasma. 
We explore scenarios in which these unprecedented features are produced by multiple phases in a jet and/or a disk wind, and discuss potential similarities between MAXI J1744-294 and the exotic microquasar SS 433.

\end{abstract}


\section{Introduction} \label{sec:intro}

Low-mass X-ray binaries (LMXBs, \citealt{Bahramian2023_LMXB_review}) are a phase of binary stellar evolution during which a stellar mass compact object accretes high amounts of matter from the Roche-Lobe overflow of a low-mass stellar companion ($\lesssim $ 1 M$_\odot$). This process leads to the formation of an accretion disk around the compact object, whose electromagnetic emission turns these objects into some of the brightest X-ray sources in the sky.
Among them, transient black hole LMXBs (BH-LMXBs) exhibit erratically repeating "outbursts" \citep{Hameury2020_DIM_review}, periods with high levels of mass transfer lasting months to years, interspersed with years or decades of "quiescence", with negligible accretion rates. 

During these outbursts, BH-LMXBs exhibit a rise of more than 5 orders of magnitude in X-ray luminosity, along with a range of spectral and timing evolutions in this energy band, linked to the reconfiguration of an accretion-ejection structure much more complex than a simple accretion disk. The most notable dichotomy is a sequential evolution through so-called "hard" and "soft" spectral-timing states (\citealt{Done2007_BHXRB_accretion}) at specific luminosities. The "hard" state, found during the source's initial rise and return to quiescence, shows an X-ray spectrum dominated by a comptonized component with $\Gamma\sim1.5$ and a cutoff at $\sim100$keV, interpreted as the emission of an optically thin plasma region in the direct vicinity ($\lesssim100$ Rg -- gravitational radii) of the BH, dubbed the corona. This X-ray emission is notably accompanied by optically thick synchrotron emission from compact jets \citep{Fender2004_BHXRB_jet}, which extends from the radio band to the infrared, and complex absorption and emission profiles tracing the presence of a cold equatorial wind in the Optical-Infrared (OIR), with a much higher mass load than jets but outflow velocities of "only" $\sim1000$ \kmps{}.
Meanwhile, the "soft" spectral state showcases an X-ray spectrum dominated by thermal emission from a geometrically thin, optically thick accretion disk, with an inner radius close to or at the innermost stable circular orbit (ISCO) of the BH, along a weak comptonized tail with no observable X-ray cutoff \citep{Cangemi2023_hardtail_multi}. It appears after a progressive spectral-timing transition from the "hard" state lasting a few weeks, during which the jet emission becomes dominated by the optically thin synchrotron emission of discrete ejections and eventually disappears. This state shows different cold wind OIR signatures, and additional absorption signatures from highly ionized lines become observable in the X-ray band, originating in a hot equatorial wind with speeds of a few $\sim100$ \kmps. 

While this phenomenological template matches most outbursting BH-LMXBs, the behavior of the accretion and ejection structure, along with the physical mechanisms at the origin of the state transitions and outburst "track", remains very poorly understood. On one hand, the radio emission becoming dominated by the discrete ejecta generally masks the disappearance of the jet core during the hard to soft state transition (see e.g. \citealt{Hughes2025_radio_1,Hughes2025_radio_2} for a high-quality radio coverage of bright outburst). On the other hand, the most direct (absorption) signatures of hot and cold winds remain mutually exclusive, and have only been detected in highly inclined systems \citep{Munoz-Darias2026_wind_review}. The jet emission could in theory also be seen in X-rays, via narrow, highly ionized emission lines with relativistic blueshifts, created by collisional ionization of the hot plasma in the jet. Yet for now, the only system to reliably exhibit these features is the atypical, highly inclined Super-Eddington source SS 433 \citep{Fabrika2004_SS433_review}.

In this context, the advent of high-resolution X-ray spectroscopy and microcalorimeters largely opens up the parameter space necessary to track and link the evolution of outflows. Such instruments allow precise characterization of much weaker absorption of emission lines from X-ray winds, which may then be detected much closer to the hard state or in much less equatorial systems. The first telescope with such capabilities, \xrism, has already uncovered the presence of cold iron emission lines in the hard states of a few disk-fed black hole X-ray binaries (BH-XRBs), such as Cygnus X-1 \citep{Yamada2025a_CygnusX-1_XRISM} and the obscured GRS 1915+105 \citep{Miller2025_GRS1915+105}, likely originating from distant reflection. Coverage of the soft state has however remained particularly elusive, with a single observation in that state before 2025, obtained for 4U 1630-47,  a (highly inclined) BH-LMXB candidate \citep{Miller2025_4U1630-47}. It led to the discovery of multiple distinct highly ionized absorption phases in a very low-Eddington soft state, which was particularly puzzling as one phase only appeared at the end of the observation, the velocities ranged between few 100 km s$^{-1}$ in outflow \textit{and} similar values in inflow, and the total column densities of the absorbers reached $N_H=2\times10^{23}\;\mathrm{cm}^{-2}$. All of these elements clashed with the expectations of a continuous highly ionized wind structure \citep{Woods1996_wind_thermal_init_2,Fukumura2021_magneticwind_BHLMXB} and its potential conversion into an atmosphere at low luminosities. Since then, additional observations of this accretion state have been sought after across a larger panel of inclinations and luminosities, in order to bring new answers to the presence of lines in the soft state, and complete the view of the accretion-ejection structure in BH-LMXBs. Meanwhile, the first \xrism{} observations of SS 433 have provided new insights into the structure of its atypical inner jet \citep{Shidatsu2025_SS433}, and its link with more distant, colder signatures in OIR and below \citep{Sakai2026_SS433_jet_X-ray_vs_OIR}. 

MAXI J1744-294, previously known as Swift J174540.2-290037 during its initial 2016 outburst \citep{Mandel2026_AtelT37}, is among the most recent additions to the rapidly growing list of XRB transients found in the direct vicinity of the Galactic center.  Its latest outburst, which was discovered on January 2 2025 by \textit{MAXI} \citep{Kudo2025} and shortly after confirmed by \swift{} \citep{Heinke2025}, became the subject of an intense multi-wavelength monitoring campaign, including \nustar{}, \xrism{}, \xmm{}, \chandra{} \citep{Mandel2025a,Mandel2025b,Mandel2025c,Mandel2025d}, \nicer{} \citep{Jaisawal2025_MAXIJ1744_NICER}, \EP{} \citep{Wang2025_MAXIJ1744_EP} and \ixpe{} \citep{Marra2025_MAXIJ1744-294_IXPE} in X-rays, as well as \meerkat{} \citep{Grollimund2025} and the \VLA{} in radio \citep{Michail2025_MAXIJ1744_VLA} and \textit{Keck} in near-infrared (\citealt{Mandel2026_ApJ}, hereafter M26). The comprehensive broadband study of the outburst presented in M26 and previous studies of the first outburst of this source \citep{Mori2019} all point towards a highly spinning BH-LMXB at moderate inclinations, and highlights a high level of contamination from diffuse emission and X-ray transients in the Galactic center, which requires advanced background modeling and source-region selection even for telescopes with high (relative to X-rays) angular resolution, such as \xmm{} and \nustar{}.

This paper is the second part of a series focusing on the \xrism\ observation of the Galactic center performed on March 3 2025 to study the properties of \src{}, and its complementary broadband X-ray coverage with \xmm{} and \nustar{}. Naturally, the field of view (FoV) of  \xrism{}'s micro-calorimeter Resolve \citep{Ishisaki22_Resolve,Porter2024_Resolve} and its soft X-ray imager CCD Xtend \citep{Mori2022_Xtend,Noda2025_Xtend,Uchida2025_Xtend} covered a range of point sources, including the eclipsing, wind-emitting Neutron Star (NS)-LMXB AX J1745.6-2901, and several sources of diffuse emission, with notable line contributions from the supernova remnant (SNR) Sgr A East and the Galactic center X-ray emission (GCXE). The limited angular resolution of \xrism, the differences in spectral energy distribution (SED) between the very soft \src{} and very hard \axj{}, along with the multiple sources of emission lines within the diffuse emission, all result in heavily contaminated spectra blending the features of all the sources in the entirety of the Resolve FoV. In order to uncover the real spectral features of each object, in Parra et al. (submitted to ApJ), hereafter Paper I, we analyzed and disentangled the respective contributions of each source. To achieve this, we optimized the region selection for the different instruments and notably the Resolve array, and compared different methodologies to model the spatial-spectral-mixing (SSM) and diffuse emission underlying \src{}. We used state-of-the-art tools and models and leveraged the previous \xrism\ observation of the Galactic center, analyzed in \cite{XrismCol2025_GC_obs_diffuse} and \cite{Tanaka2026_AXJ}, which provides a crucial view of the diffuse emission at the location of \src{} before its 2025 outburst. 

We thus refer to Paper I for a detailed description of the observations, region selection, data analysis, background modeling techniques, along with comprehensive estimates and discussions of the systematics of our methodologies. In parallel, we refer to Gatuzz et al. (submitted to A\&A, hereafter Paper III) for a detailed analysis of the interstellar medium features in our spectra. This work focuses on \src{}: after a short reminder of our methodology in Section \ref{sec:methodo}, in Section \ref{sec:empirical}, we present the results of the phenomenological modeling of its time-averaged March 03 spectra with \xmm{} + \nustar{}, Xtend, and Resolve, in increasing order of spectral resolution. In Section \ref{sec:anal_phys}, we compare several types of physical models to the main line features detected in Resolve's high-resolution \src{} spectrum. We discuss the physical interpretation of our results as outflow signatures, their robustness, their contextualization within the current high-resolution observational landscape, and implications for past and future studies of soft-state BH-XRBs in Section \ref{sec:discu}. We summarize our findings in Section \ref{sec:conclu}. 

Furthermore, we note that a recent paper on that same \xrism\ observation \citep{spreadingmisinformationontheinternet} presented a range of unusual results in the iron band, including a combination of broad and narrow emission lines, as well as absorption lines, which were all attributed to MAXI J1744-294. However, this paper analyzed the entire pixel array spectrum as originating from MAXI J1744-294, without considering the other sources in the FoV.  Paper I and the present analysis of \src{} unambiguously show that the line emission \textit{integrated over the entire array} is dominated by a blend of the diffuse emission and AX J1745.6-2901's absorption features, and none of the line features discussed in \cite{spreadingmisinformationontheinternet} are intrinsic to MAXI J1744-294.

\section{Methodology}\label{sec:methodo}
Here, we recall the complementarity between the different instruments and the challenges encountered in their respective analysis, detailed in Paper I and M26. Our main objective is the description of any potential line features in the \xrism{} observation (hereafter "DDT observation") of \src{} (hereafter M1744), thanks to the high spectral resolution of \xrism{}'s microcalorimeter Resolve. Due to Resolve's limited bandpass of $\sim2-10+$keV, Xtend provides the highest spectral resolution at energies below 2 keV, and \xmm{} and \nustar{} provide a more complete view of the full X-ray continuum, with higher effective area, coverage of hard X-rays, and more robust calibration. This continuum, which is studied in detail in M26, along with its evolution across the entire outburst, will be paramount to compare the line properties to photoionization models in Section ~\ref{sub:phys_photo}.
The first challenge in our analysis is common to all instruments: the position of the source in the Galactic center leads to a large degree of contamination by several sources of diffuse emission in the vicinity. These sources include continuum components, but most importantly strong, narrow emission lines from many elements, and most particularly highly ionized iron. These contributions must be removed to assess the intrinsic SED and line features of M1744, and are thus estimated from previous observations of the same field of view. For \nustar{}, direct background subtraction is sufficient, but in \xmm{}, Xtend and Resolve, we systematically modeled the background emission from archival observations, notably including a \xrism{} pointing of the Galactic center during the Performance Verification phase (hereafter "PV" observation), before the outburst of M1744 \citep{XrismCol2025_GC_obs_diffuse}, whose FoV does not fully coincide with our newer observation.

The second challenge, more specific to \xrism{}'s Resolve due to its low angular resolution and restricted region selection, is the high degree of spatial-spectral mixing between the point spread functions of M1744, \axj{} (hereafter AXJ), and the multiple sources of diffuse emission. To tackle this problem and its systematics in the best way possible, the entire \xrism{} Resolve analysis is repeated for two different regions, each with a different approach to background modeling. The first region (hereafter "big" M1744 region) aims to maximize the signal-to-noise ratio of M1744, and is thus taken from all the pixels in the FoV of the \xrism{} DDT observation, except for the few pixels dominated by AXJ, with calibration issues, or outside of the FoV of the PV observation. Its background is fitted empirically from the same region transposed in the archival \xrism{} observation, using a single source with a uniform spatial distribution over the field of view, whose model includes a simple absorbed powerlaw continuum and 35 individual line components.
The second region (hereafter "small" M1744/BH region) aims to minimize the contamination from other sources, and is thus taken from the four pixels closest to M1744. Its background is fitted physically from the same region in the PV observation, using 2 sources. The first is the GCXE and is thus spatially uniform over the FoV, and the second represents the SNR Sgr A East, using for angular distribution a \Fexxv{} K$\alpha$ flux map of the SNR obtained with Chandra. Both sources are modeled with two layers of overionized plasma and few empirical lines for the neutral Fe contributions, building on the methodology of \cite{XrismCol2025_GC_obs_diffuse}. Since AXJ was already in outburst in the PV phase observation, each Resolve diffuse emission "background" region is fitted simultaneously with an \axj{} dominated region, which is modeled empirically, to consider both the off-axis contribution of the diffuse sources in the AXJ region, and the off-axis contribution of AXJ in the background region. The same approach is used in the DDT observation to distinguish AXJ and M1744, as will be shown in Section \ref{sub:empirical_resolve}. In Resolve and \nustar{}, the contribution of AXJ is directly subtracted with a background region off-axis from AXJ.

Finally, the dust scattering haloes (DSH) surrounding M1744 (M26) and AXJ \citep{Jin2017_DSC} significantly affect the spectral shape of the continuum of each source, but among our instruments, we have only developed models deconvolving their contributions for \xmm{} and \nustar{}. Since the distribution of the dust in the FoV is highly complex, we choose not to include any "generic" DSH modeling, such as xscat \citep{Smith2016_xscat_dust_scattering_model}, in our \xrism{} analysis, and we will thus rely on the \xmm{}-\nustar{} results for any definitive estimates of the source SED. The \xrism{} analysis will only focus on the characterization of the line features, on which the DSH will have a negligible impact. 

In this entire work, we will use the C-statistic. Uncertainties are quoted at a 90\% significance level unless stated otherwise. We use the convention of negative velocities for blueshifts, and positive velocities for redshifts.

\section{Phenomenological modeling} \label{sec:empirical}

\begin{figure*}[t!]
\centering
    \includegraphics[clip,trim=0cm 0cm 0.cm 0cm,width=1.0\textwidth]{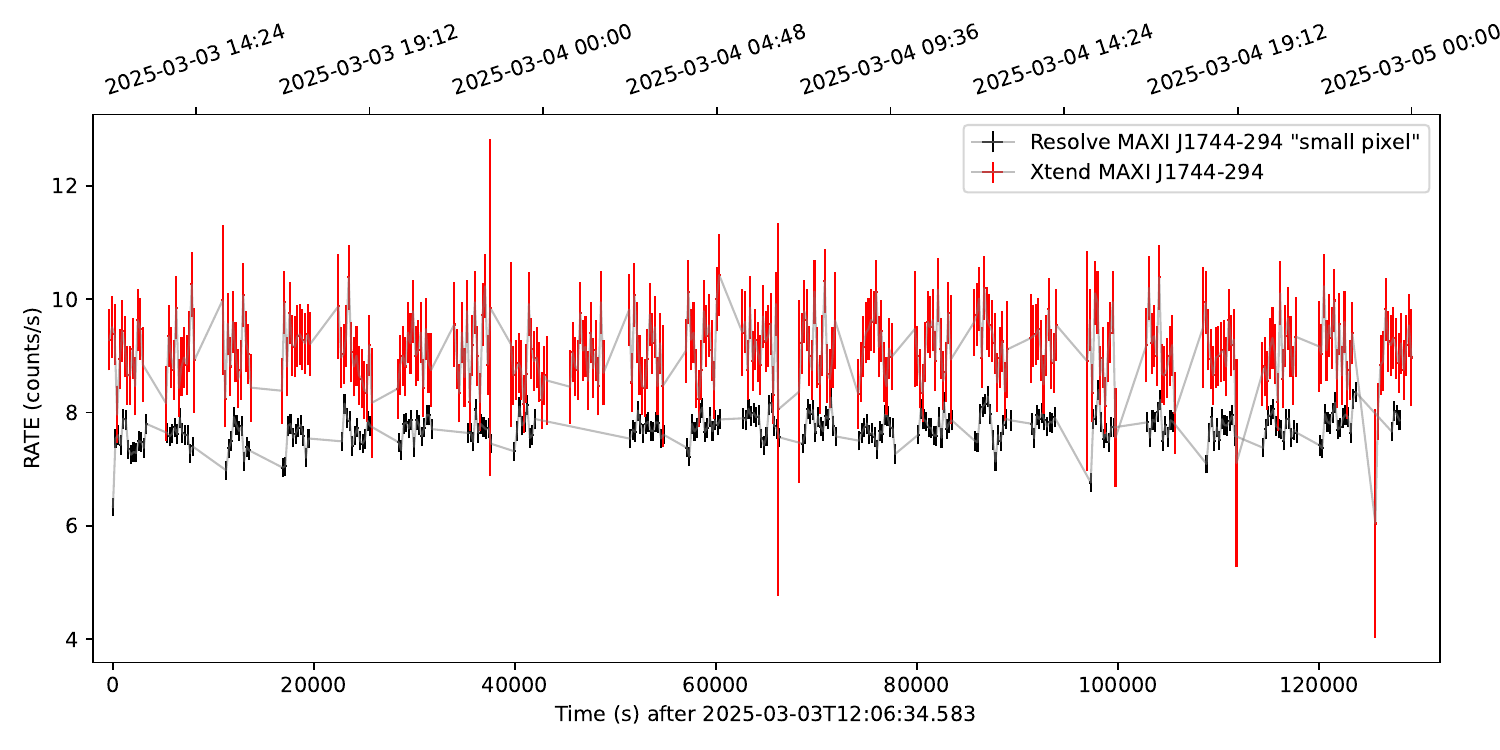}
    \vspace{-2em}
    \caption{2-10 keV Resolve lightcurve computed from the "small" \src{} region, and 0.3-10 keV Xtend lightcurve computed from the \src{} region, both with a 256s binning.}
    \label{fig:XRISM_lc}
\end{figure*}

In this section, we describe our empirical fitting of the \src{} spectra from March 03 2025. We show in Fig.~\ref{fig:XRISM_lc} the detailed lightcurve of the observation using \xrism, the instrument with the highest exposure. As no significant variability is seen across the observation, we perform the entire analysis on time-averaged products. 


After recalling the main spectral parameters of the dust corrected \xmm{} and \nustar{} continuum model, detailed in M26, we first perform a soft X-ray continuum and "mid-resolution" line fitting with Xtend, which, aside from the lack of DSH modeling, has the benefit of a higher spectral resolution and no pile-up compared to the \xmm{} observation, but suffers from calibration uncertainties at low energies \citep{Xrism2025_NGC3783_muli_inst_cal}. Finally, we perform a high-resolution empirical fit on the Resolve data alone, focusing on line features and comparing the results between the two region and background methodologies detailed in Paper I. We apply to the \xrism{} fits the same comptonized disk continuum model as in M26, and freeze the photon index to the value derived in that study, since its value can only be poorly constrained from data below 10 keV, and will be weakly affected by the DSH. Similar considerations are applied to the AXJ model, as will be detailed in future publications.

\subsection{Dust-corrected continuum modeling}\label{sub:empi_xmmnustar}

We first performed an empirical "low-resolution" continuum fitting of M1744 using the \xmm{} and \nustar{} observation simultaneous to \xrism{}. The instruments are complementary and can be fitted together using dust scattering halo (DSH) models tailored to the source. The full list of parameters, residuals, and physical interpretation of that fit are detailed in M26. For our purposes, we simply highlight that this model combined a comptonized thermal disk and a single Gaussian emission line (\texttt{TBabs(thcomp(diskbb)+gaussian)} in XSPEC format), and derived an absorption column of $\mathrm{N}_\mathrm{H}\sim1.72_{-0.02}^{+0.05}\times10^{23}$ cm$^{-2}$, a disk temperature of $kT_{in}=0.61\pm0.01$ keV, and a photon index of $\Gamma\geq2.74_{-0.32}$, for a covering fraction of $f_{cov}=3.3_{-1.8}^{+0.7}\times10^{-3}$. The model also included an unresolved line centered at $6.69_{-0.05}^{+0.06}$ keV, with an Equivalent Width (EW) of $102_{-48}^{+75}$ eV. As the diffuse background contribution and its emission lines are removed in both \xmm{} and \nustar{}, that line is intrinsic to M1744, and corresponds to a blend of some of the features seen by \xrism{}.
The absorption column is, at first order, consistent with that of Sgr A* ($\sim1.6\times10^{23}$ cm$^{-2}$, \citealt{Ponti2017_Sgr_Flare}), and previous transients detected in its direct vicinity, such as Swift J174540.7-290015 ( $\sim1.8\times10^{23}$ cm$^{-2}$, \citealt{Ponti2016_SwiftJ1745407290015}), located 16.6\arcsec{} away from the central BH, and the magnetar SGR J1745-2900 ($\sim1.7\times10^{23}$ cm$^{-2}$, \citealt{Ponti2017_Sgr_Flare}), located 2.4\arcsec{} away. It is however significantly lower than that of the NS-LMXB AX J1745.6-2901 ($\sim3.0\times10^{23}$ cm$^{-2}$, \citealt{Jin2017_DSC}).

\subsection{Xtend continuum and mid resolution modeling}\label{sub:empi_xtend}

\begin{figure*}[t!]
\centering
    \includegraphics[clip,width=0.49\textwidth]{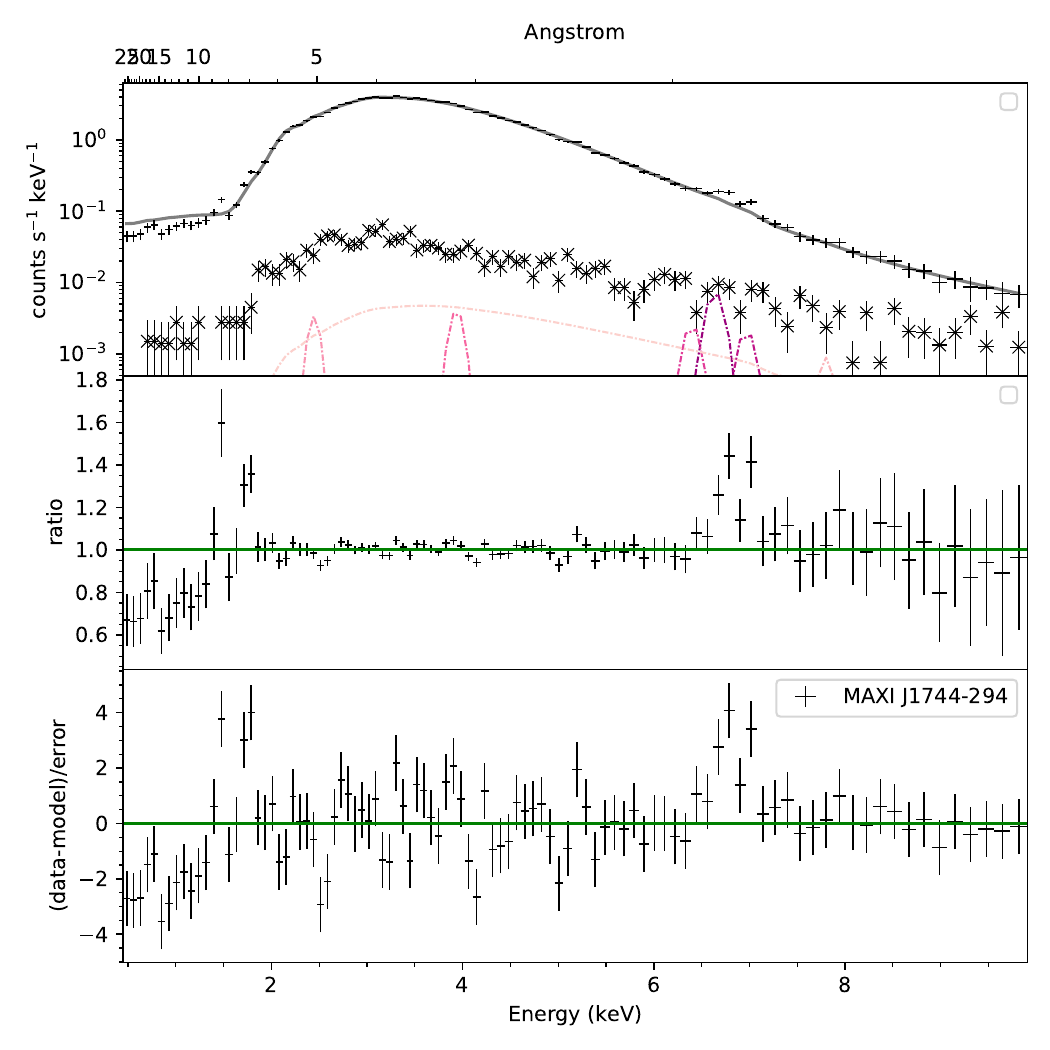}
        \includegraphics[clip,width=0.49\textwidth]{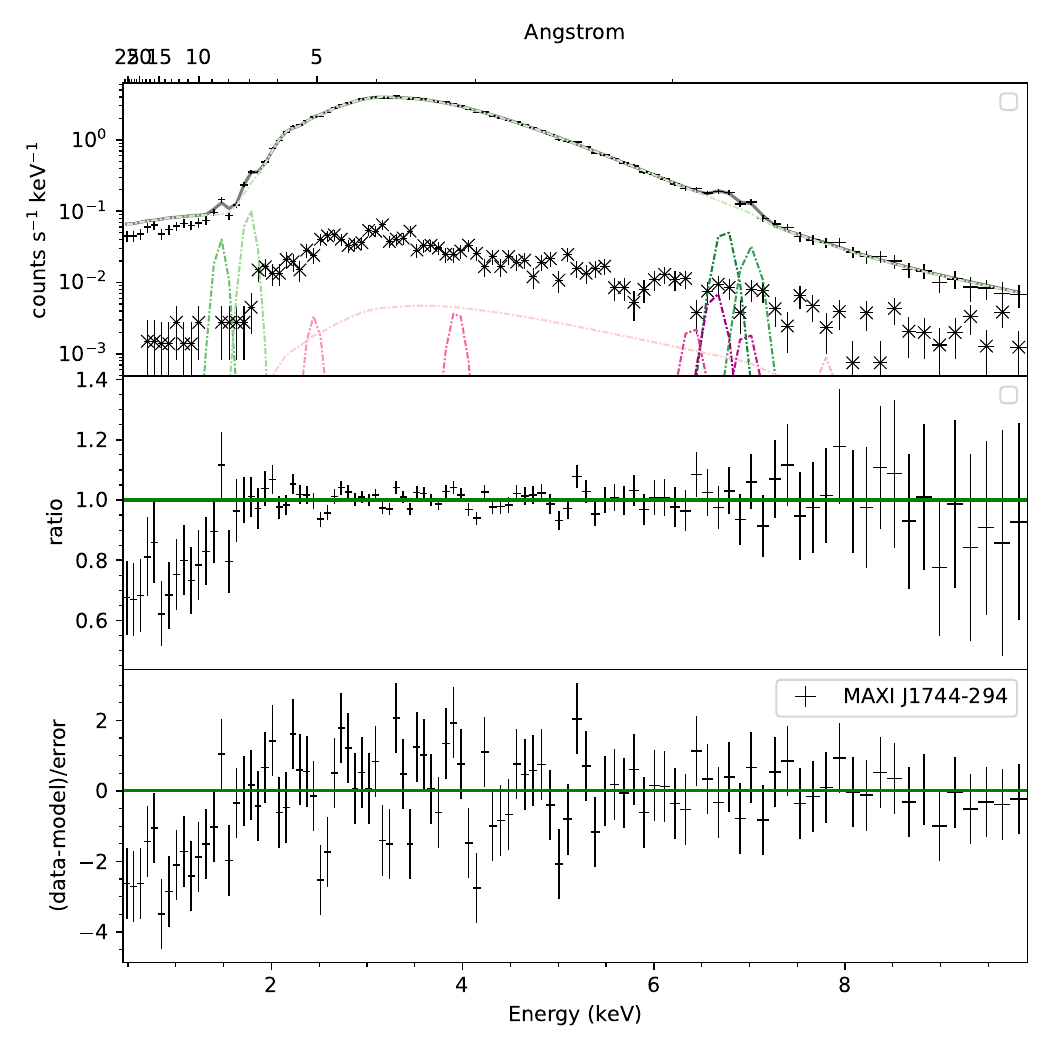}
    \vspace{-1.em}
    \caption{Spectrum, ratio and residuals for the Xtend MAXI J1744-294 region in the DDT observation, in the 0.4-10 keV band, after the empirical continuum modeling described in Section~\ref{sub:empi_xtend}, before (left) and after (right) fitting additional emission lines. Following the colormaps of Paper I, components from MAXI J1744-294 and Sgr A East are shown in shades of green and purple, respectively. The spectrum is rebinned with the optimized scheme of \cite{Kaastra2016_binning_opt}.}
    \label{fig:xtend_empi_resid}
\end{figure*}

For Xtend, our empirical M1744 fit is performed on a single background-subtracted spectrum. It includes one source for M1744 and one for the Sgr A East contribution, which is not considered in the static background. We use the same comptonized disk model as for \xmm{} and \nustar{}, with a comptonization photon index fixed at $2.74$.
We first apply the fit on the 0.4-10keV band, ignoring the 6.4-7.1 keV range (hereafter iron range) to avoid any bias from iron K$\alpha$ emission lines.
We show the resulting fit (after re-noticing the iron range) in Fig.~\ref{fig:xtend_empi_resid}-left. Although the continuum is globally well fitted above 2 keV, there is a clear underfitting at low energies, and several narrow features are evident at both below 2 and above 6 keV, leading to a poor fit C-statistic of $222/94$ d.o.f. 

We thus add the main emission features seen in the initial residuals, using Gaussian lines whose widths are fixed at 0, in accordance with the limited resolution of the instrument. For the two high-energy lines, expected to come from \Fexxv{} and \Fexxvi{} K$\alpha$, we consider a potential velocity shift compared to the averaged rest energies of each complex. The lines at low energies are not tied to a specific complex, but remain compatible with transitions of Magnesium, Aluminum, and Silicon. After adding these 4 lines and refitting the entire continuum, the quality of the fit improves significantly to a C-statistic of $123/89$ d.o.f. We show the residuals of that new fit in Fig.~\ref{fig:xtend_empi_resid}-right, and list the detailed model and line parameters in Tab. A.\ref{tab:comp_param_empi_xtend} in App.~\ref{app:xtend_fit}. The main remaining feature is the overfit of the spectrum at low energies: it can be due to our lack of consideration for the DSH, or to the Xtend calibration discrepancy with \xmm{} at low energies \citep{Xrism2025_NGC3783_muli_inst_cal}. These known calibration uncertainties are nonetheless too weak (max 15\%) and too broad (single broad feature centered on 1.3keV) to explain the low energy emission lines we detect, with ratios of 30 to 60\% of the continuum, and a width constrained to $<0.1$ keV at 3 $\sigma$ for the 1.5keV line. We note that these two lines are detected neither in the simultaneous \xmm{} RGS data nor in the Chandra HETG data taken a few days later, both of which are presented in more detail in the ISM study of Paper III. They are also not seen in the 4 Xtend full-window mode observations of the source from August 2025. Their interpretation as intrinsic M1744 features is thus very uncertain. Additional details on these low-energy features, including comparisons with the calibration of Xtend and alternative models using edges, which provide a more reasonable alternative to emission lines, are presented in Appendix~\ref{app:xtend_lowe}.

\subsection{Resolve High-resolution spectral modeling}\label{sub:empirical_resolve}

Due to the negligible signal-to-noise of M1744 above $\sim7$ keV in the Resolve data, we restrict our analysis to the 2-10keV band. Moreover, our spectral fitting focuses on the characterization of any narrow features, and must thus consider additional effects that become relevant at high resolution. First, to combine models from datasets obtained from different observations, we must correct for the differences in Earth's line-of-sight (LoS) velocity across the Sun. Fortunately, the PV and DDT observations were taken almost exactly one year apart, with LoS velocities of 28 and 29 \kmps, respectively. We thus neglect the (1 \kmps) LoS velocity difference between the two observations. Furthermore, as the LoS velocity itself remains small compared to our derived uncertainties, all values presented in the upcoming subsections are raw, non-LoS-corrected velocities. Finally, any time-averaged analysis of M1744 will include an artificial width increase of all the lines, due to the LoS velocity evolution in its binary system. As we do not yet have information about the optical period, we performed time-resolved analysis on 3 periods of $\sim$40ks, which are the lowest at which the secondary intrinsic features can be detected at high significance. As no significant variation was found in any of the line features with Resolve, all results below will be derived from a time-averaged analysis over the entire period. All Resolve spectra are left ungrouped in the analysis, but our figures will be visually rebinned to highlight the different features of interest in each source. We stress that with this \textit{important} visual rebinning, the significance of the features in the residual plots will be systematically underestimated compared to the actual $\Delta C$ computations in our fits, which use the full spectral resolution of the spectrum. Finally, to highlight the different narrow complexes that would otherwise be blended in the broadband residual plots, we systematically display the model components to a higher resolution than the data itself, using a 3$\sigma$ significance level, which refers to the model components displayed for spectra rebinned at this significance.

\subsubsection{Big MAXI J1744-294 region}\label{subsub:empi_bigpix}

Our first empirical high-resolution fit of M1744, which includes an empirical description of the diffuse emission, is performed on two spectra simultaneously: the  "big" M1744 region shown in cyan in Paper I (Figure 4, left panel), and the AXJ region shown in red in the same figure. Three sources are applied in each of the two spectra: for the empirical diffuse emission "background" contribution, assumed uniform, we directly import and freeze the model derived in Paper I. The contributions of 
M1744 and AXJ are computed with point source ARFs and directly fitted in the observation, starting from the DSH corrected spectral modeling derived in Section~\ref{sub:empi_xmmnustar} and M26. Several parameters are kept frozen in the fit: the column density of the interstellar absorption for AXJ is only weakly constrained due to the very high contamination by M1744 at low energies, and thus kept at $N_H=3\times10^{23}$ cm$^{-2}$. As mentioned in the methodology, we also fix the photon index of the comptonized component in M1744 and the temperature of the hot blackbody in AXJ to the values derived in M26 and our upcoming broadband analysis of AXJ.
Since \texttt{thcomp} requires extending the energy grid, we set the energy grid to \texttt{0.01 0.1 1000 log, 10. 19800 lin, 1000. 1000 log}. This matches the default energy grid of the \xrism\ data between 0.1 and 10 keV, and adds 1000 logarithmic bins above and below. Finally, we let the normalization of the entire M1744 model free in AXJ, to help compensate for any inaccuracies due to the PSF modeling or DSH. The disk temperature of AXJ, which is also poorly constrained due to the lack of signal-to-noise ratio at low energy compared to the contamination by M1744's soft SED, is limited to a maximum value of 2.1 keV, which is already significantly above its dust corrected best fit from \xmm{} and \nustar{}.

\begin{figure*}[t!]
\centering
    \includegraphics[clip,trim=0cm 0.3cm 0cm 0cm,width=1.00\textwidth]{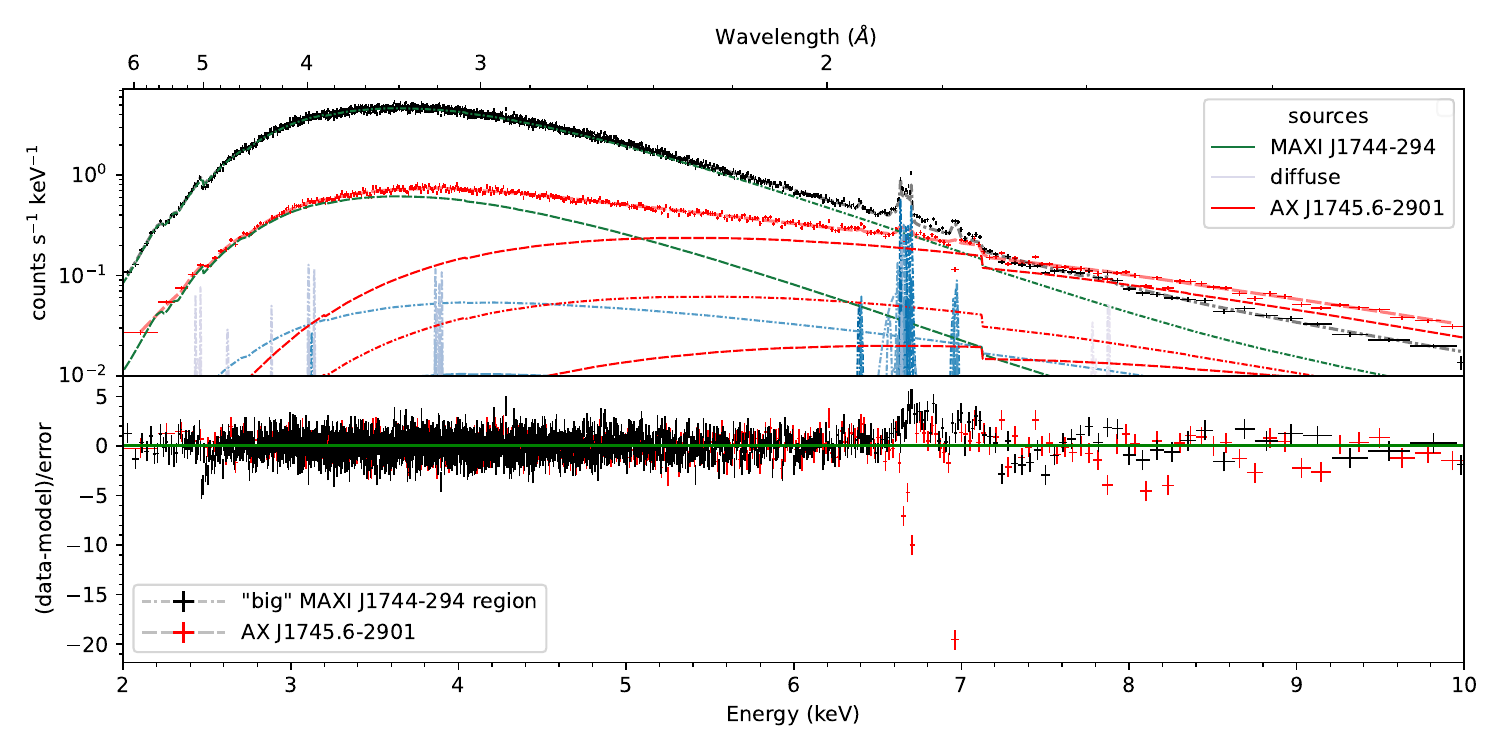}
\vspace{-1.5em}
    \caption{Resolve spectra and residuals for the "big" \src{} region and \axj{} region in the DDT observation, after the first step of their common continuum modeling, and in the entire $2-10$ keV band. Both spectra are visually rebinned at a 20$\sigma$ significance for readability, and model components at a 3$\sigma$ significance level.}
    \label{fig:resolve_bigpix_empi_resid_conti_preline}
\end{figure*}

We start with a "line-free" continuum fit, in which we ignore three specific energy bands where narrow features are apparent in the two spectra: the Sulfur K-edge around 2.5 keV, the iron region between 6.4 and 7.1 keV, and small bands associated with individual strong lines from highly ionized iron and nickel between 7.8 and 8.7 keV. After performing the fit, we notice these bands once again, and obtain a fit C-statistic of 34868/31994 d.o.f. We showcase the residuals of the two spectra in Fig.~\ref{fig:resolve_bigpix_empi_resid_conti_preline}. Several elements are apparent in the residuals, and we thus investigate them iteratively according to their importance for M1744. Zooms of both spectra and their residuals in the relevant energy bands are presented in Fig. A.\ref{fig:resolve_bigpix_empi_resid_conti_preline_zoom} in App.~\ref{app:resolve_fits}.

First, as can be seen in the left panel of Fig. A.\ref{fig:resolve_bigpix_empi_resid_conti_preline_zoom}, the Sulfur edge is clearly incorrectly fitted in both spectra and shows large residuals starting at 2.465 keV. Similar features were already seen in \xrism\ spectra of bright sources with high absorption \citep{Corrales2025_XRISM_sulfur}. Accurate modeling of edges must include new cross-sections and hot ISM phases \citep{Gatuzz2024_sulfur_models}, along with dust, and a yet to be explained energy shift of $\sim8$ eV. The detailed study of the different edges found in the high-resolution M1744 will be presented separately in Paper III. We note that these residuals do not affect the continuum at high energies, and can safely be ignored for the purpose of studying the rest of the features cited below. 

Secondly, a high number of absorption lines are apparent in the AXJ spectrum above $\sim7.8$ keV, as seen in the right panel of Fig. A.\ref{fig:resolve_bigpix_empi_resid_conti_preline_zoom}. Although this is not the focus of our work, these lines must be accurately modeled for the continuum of AXJ and its off-axis contribution in M1744. Since no strong line residuals in emission are seen in the M1744 spectrum, and M1744's off-axis contribution represents less than 3$\%$ of that of \axj{} above 7.8 keV in the \axj{} region, we fit these absorption lines directly, using empirical Gaussians in absorption for the transitions of K$\beta$ and K$\gamma$ transitions of \Fexxv{}, \Fexxvi{}, \Nixxvii{} and \Nixxviii{} detected in the spectrum. This leads to a very significant improvement in the fit ($\Delta$C=500 for 25 d.o.f.), for a final C-statistic of 34368/31969 d.o.f. 

Thirdly, the M1744 spectrum is clearly underfitted between 7.2 and $\gtrsim7.6$ keV. This is a telltale of residuals to the iron edge, which can be affected by non-standard abundances and the presence of dust (see e.g. \citealt{Rogantini2018_ironedge}). Moreover, several recent \xrism{} observations indicate that the edges of lower-Z elements require significant energy shifts (see e.g. \citealt{Corrales2025_XRISM_sulfur}, paper III), and it may thus be the same for iron. After verifying that letting the abundance free to vary with \texttt{TBfeo} did not improve the edge profile, 
we remove this edge from the ISM absorption entirely by setting the iron abundances in \texttt{TBfeo} to 0, and instead add an independent \texttt{edge} component with free energy and optical depth.
This leads to a small but significant improvement in the fit, with $\Delta$C=16 first for 2 d.o.f. when ignoring the 6.4-7.1 keV band, which increases to  $\Delta$C=48 when noticing the entire spectrum. Since this confirms that the emission residuals in the BH described below may be (slightly) influenced by the iron edge fitting, we thus adopt this empirical edge description in the remainder of the "big" M1744 region analysis. As the signal-to-noise spectrum of M1744 is extremely limited beyond $\sim7.6$ keV, we do not focus on potential emission line features beyond this limit.

\begin{figure*}[t!]
\centering
    \includegraphics[clip,width=1.00\textwidth]{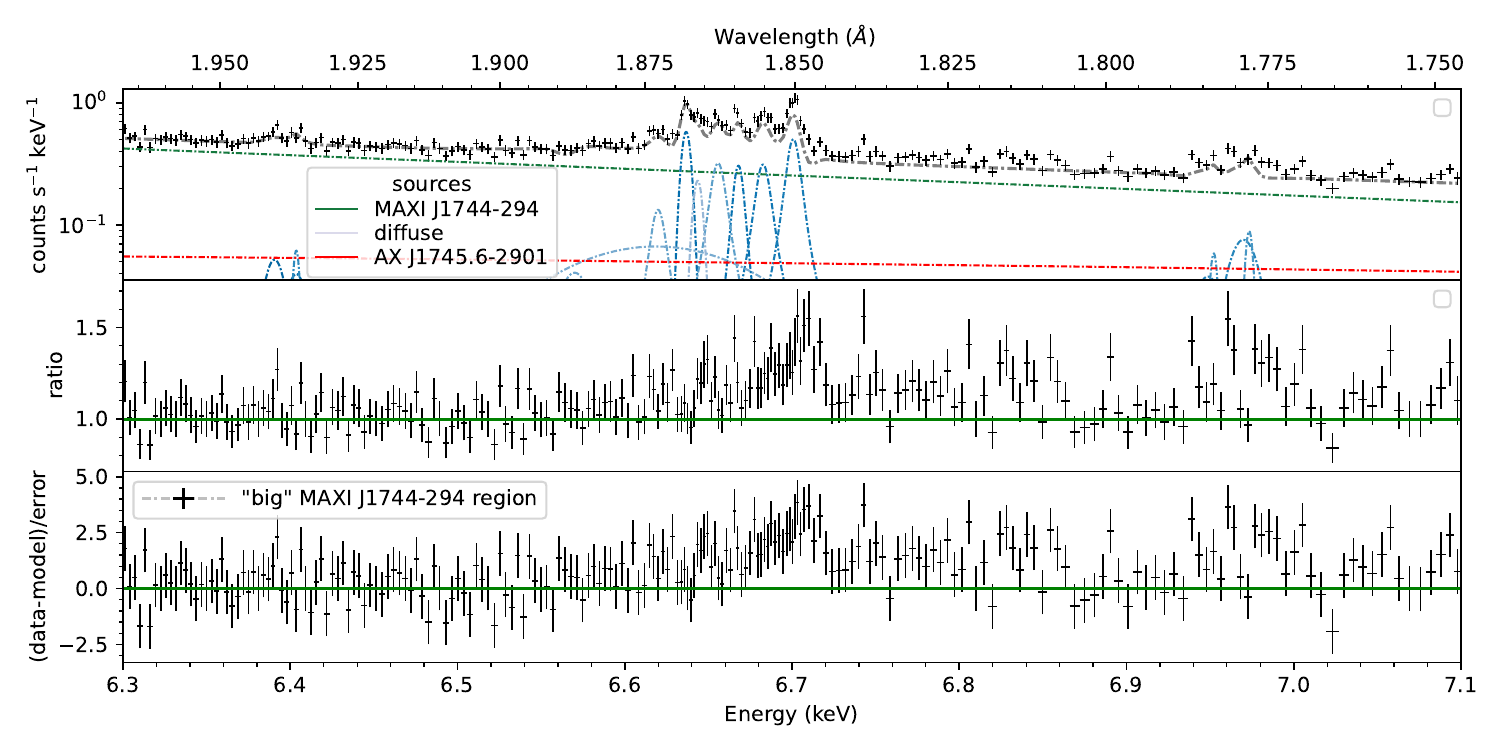}
\vspace{-2em}
    \caption{Zoomed spectrum, ratio, and residuals for the "big" \src{} region in the DDT observation, in the 6.3-7.1 keV band, after the empirical continuum-only modeling, including the edge readjustment and lines above 7.8keV in AX J1745.6-2901. The spectrum is visually rebinned at a 10$\sigma$ significance level for readability, and model components at a 3$\sigma$ significance level.}
    \label{fig:resolve_empi_resid_zoom_BH_big_preline}
\end{figure*}

\begin{figure*}[t!]
\centering
\includegraphics[clip,width=1.00\textwidth]{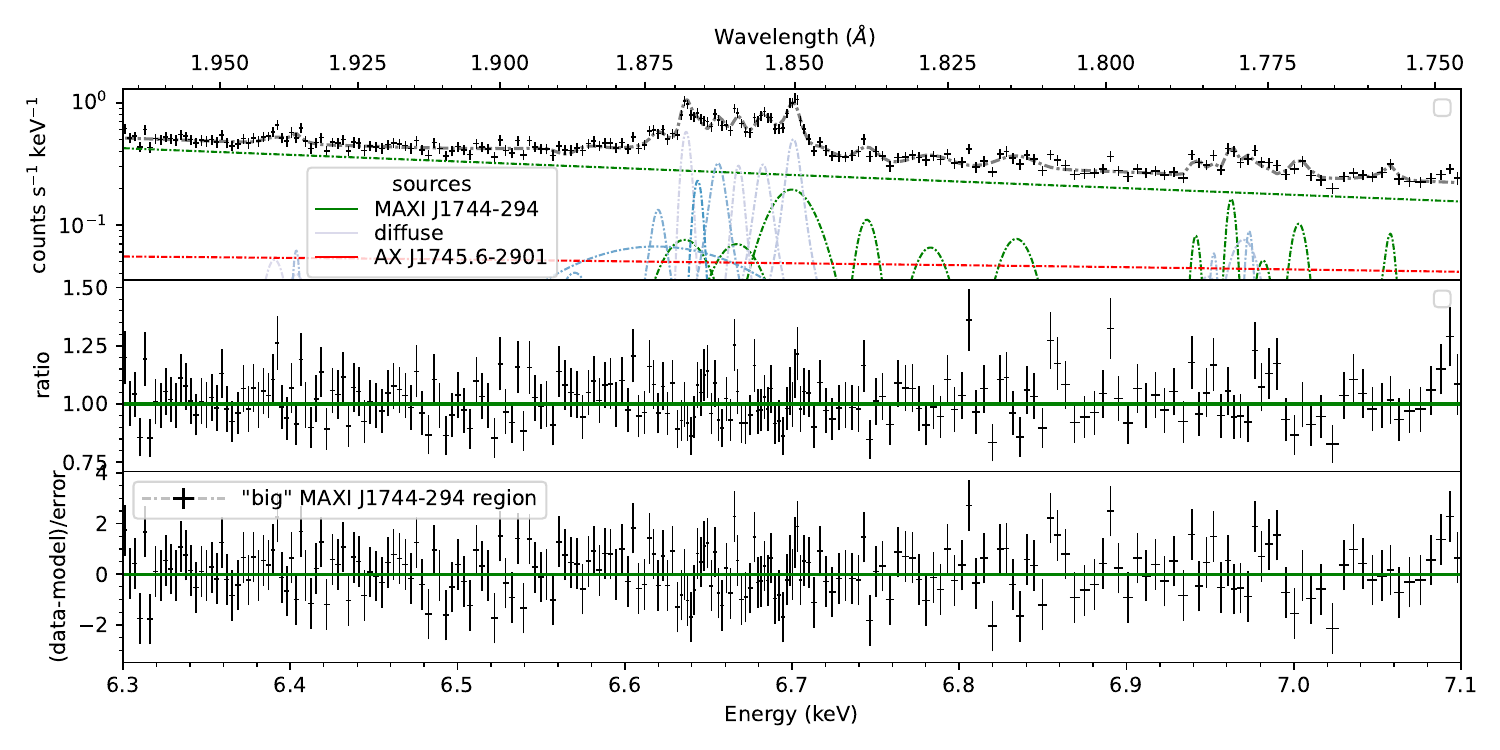}
\vspace{-2em}
    \caption{Zoomed spectrum, ratio, and residuals for the "big" \src{} region in the DDT observation, in the 6.3-7.1 keV band, after the full empirical modeling, including all significant lines. The spectrum is visually rebinned at a 10$\sigma$ significance level for readability, and model components at a 3$\sigma$ significance level.}
    \label{fig:resolve_empi_resid_zoom_BH_big_postline}
\end{figure*}

Fourthly, and most importantly, clear residuals in emission (for M1744) and absorption (for AXJ) can be seen in the iron range, most notably -but not only- for the K$\alpha$ transitions of \Fexxv{} and \Fexxvi{}. The common view of both M1744 and AXJ is shown in the middle panel of Fig. A.\ref{fig:resolve_bigpix_empi_resid_conti_preline_zoom}. To highlight the features in M1744, we show an individual zoom of its spectrum and residuals after the initial continuum fitting in the top left panel of Fig.~\ref{fig:resolve_empi_resid_zoom_BH_big_preline}. 
This prompts us to perform two independent blind searches for narrow absorption features in the 6.3-7.1 keV band for the M1744 and AXJ spectra, in order to benchmark and assess the significance, location, and width of these residuals. We follow the same procedure as in Paper I and \cite{Parra2024_winds_global_BHLMXBs}, and show the output for M1744 in the top-left panel of Fig. A.\ref{fig:blind_search_empi_BH} in App.~\ref{app:resolve_fits}. In the M1744 spectrum, strong components for \Fexxv{} and \Fexxvi{} K$\alpha$ are obvious with marginal velocities, but a series of other narrow emission features are present between 6.7 and 7.1 keV. In the AXJ spectrum, whose detailed analysis is left for Matsunaga et al. (in prep.), we only observe absorption residuals, for the \Fexxv{} and \Fexxvi{} K$\alpha$ lines. For our purposes, all that matters is that they are very well fit by a series of Gaussians in absorption.
In M1744, we use 4 Gaussians with a common width and velocity shift to represent the \Fexxv{} K$\alpha$ complex ($\Delta$C=164 for 6 d.o.f.), and 2 Gaussians with a common width, velocity shift, and a normalization fixed at a 1-2 ratio for the \Fexxvi{} K$\alpha$ complex ($\Delta$C=50 for 3 d.o.f.). Both models are sufficient to represent the residuals at first order, but the fits are biased towards unexpectedly high line widths due to additional emission residuals around the rest energy of the lines. For \Fexxv{}, the separation with another narrow component above 6.74keV is clear, both in the pre-line residuals of Fig.~\ref{fig:resolve_empi_resid_zoom_BH_big_preline} and in the significance maps of the blind searches. 
The Fe XXV K$\alpha$ complex has a high width of $\sigma_{0,25}=14_{-3}^{+6}$ eV, and a velocity of $v_{0,25}=26_{-185}^{+258}$ \kmps{}.
The Fe XXVI K$\alpha$ complex has an even higher width of $\sigma_{0,26}=21_{-6}^{+9}$ eV
 and a velocity of $v_{0,26}=-218_{-342}^{+343}$ \kmps{}, and is thus compatible with zero velocity and the negligible shift of the Fe XXV K$\alpha$ complex within errors.
However, for \Fexxvi{}, a a much better fit and completely different parameters are reached when adding a second, blueshifted line, as we will see below. We show the resulting lines and residuals after the final fit (including the additional components, which will change the shape of the lines) in  Fig.~\ref{fig:resolve_empi_resid_zoom_BH_big_postline}. 

These features can in any case be clearly distinguished from the diffuse emission, having much higher widths and a completely different ratio between the individual transitions of \Fexxv{} K$\alpha$, including a much stronger resonance line at 6.7 keV. Furthermore, we can clearly see in the top-left panel of  Fig.~\ref{fig:resolve_empi_resid_zoom_BH_big_preline} that the forbidden narrow line in the diffuse emission already provided a nearly perfect fit to the data around $\sim6.63-6.642$ keV in the pre-line model. This strongly suggests that the diffuse emission fit is correctly fitted in the DDT observation, and thus that the other residuals originate from a different component, intrinsic to M1744. 

Other features remain apparent, notably in the 6.75-6.85 keV range, and we thus perform a new continuum fit, this time noticing the entire spectrum, before computing two blind searches for narrow line features in the M1744 and AXJ spectra individually. The first, shown in the middle-left panel of  Fig. A.\ref{fig:blind_search_empi_BH} in App.~\ref{app:resolve_fits}, confirms that significant residuals remain, clustered between 6.74 and 6.88 keV. The second, which we show in the top-left panel of Fig. A.\ref{fig:blind_search_empi_NS}, confirms that no significant residuals remain in the AXJ spectrum.

We then fit these unknown line features incrementally with additional Gaussian models. Although we present them in decreasing order of significance, we number them according to their energies, for easier identification and comparison with other fits. We start with the strongest component, around 6.74 keV: it is well fitted with a $\Delta$C=26 (for 3 d.o.f.) narrow ($\sigma_{1}=5.6_{-2.2}^{+4.0}$ eV) Gaussian at an energy of $E_{1}=6741.9_{-2.9}^{+3.1}$ eV. For reference, attributing it to the nearest strong line, namely the resonance ($w$) transition of \Fexxv{} K$\alpha$ at 6.7004 keV, implies a blueshift of $v_{1}=-1987_{-140}^{+135}$ \kmps{}. The second most significant component, which we number as component 3 according to its energy, is wider and found around 6.83 keV: it is well fitted by a "broader" Gaussian, but here letting the width free leads to a degeneracy with the third feature below 6.8 keV, which is partially fitted for $\sigma\gtrsim 20$ eV. For now, we limit the line width to $\sigma_{3}=10$ eV, for a final improvement of $\Delta$C=25 for 3 d.o.f., centered on an energy of $E_{3}=6827.7_{-5.2}^{+8.7}$ eV, and thus a blueshift of $v_{3}=-5931_{-285}^{+337}$ \kmps{} if attributed to \Fexxv{} K$\alpha$-$w$.The third and weakest residual feature, which we number as component 2 due to its intermediate energy, is found around 6.78 keV. With a similar width limit of $\sigma_{3}=10$ eV, its inclusion leads to an improvement of $\Delta$C=17 for 3 d.o.f., for a line centered on an energy of $E_{2}=6783_{-10}^{+13}$ eV, and a potential blueshift of $v_{2}=-3676_{-445}^{+574}$ \kmps{}.
Without constraints on the widths, the two lines around 6.80 keV blend completely: we test this by replacing them with a single line with widths up to 100 eV. This configuration leads to a fit improvement almost identical to that of both lines ($\Delta$C=36 for 3 d.o.f.), although with a much higher width of $\sigma_{2-3}=37_{-11}^{+18}$ eV. This line is centered around $E_{2-3}=6814_{-16}^{+15}$ eV, and thus a potential blueshift of $v_{2-3}=5078_{-701}^{+677}$ \kmps{}. We note that a broader line alone struggles to reproduce all three components, with a $\Delta$C of -13 (worsening) for -3 d.o.f. compared to 2 lines, even at widths of $\gtrsim40$ eV. This will be confirmed in  Section~\ref{subsub:empi_smallpix}, where these complexes are detected at higher significance. To illustrate that we cannot distinguish between two narrower and one broader line, we present the "three lines" configurations in our final fit for the "big" M1744 region. The other configuration will be adopted for the "small" M1744 region.

We then test the effect of fitting a single Gaussian component around 7.00 keV, where a significant narrow feature was detected in the first blind search. The feature is relatively significant ($\Delta$C=12 for 3 d.o.f.) and stronger than seen in the blind search due to its low width of ($\sigma_4=0^{+5}$ eV).
Its energy is very well constrained at $E_{4}=7003.7_{-1.5}^{+1.7}$ eV, or $v_{4}=-1312_{-73}^{+63}$ \kmps{} if attributed to the \Fexxvi{} K$\alpha$-3/2 (6.9732 keV) transition. In this context, its corresponding \Fexxvi{} K$\alpha$-1/2 transition would fall at $\sim$6.983 keV, which is very close to the two transitions of the static component, and could thus explain the uneven residuals between 6.95 and 6.97 keV. We thus perform another fit, this time using a combination of two Gaussian emission lines, with a common width and velocity shift parameter, and a normalization fixed at a 1/2 ratio. This completely changes the parameters of the static \Fexxvi{} K$\alpha$ component, and leads once again to an improvement in the fit ($\Delta C$=7 for no additional d.o.f.), for a total of $\Delta$C=19 for 3 d.o.f, when simultaneously adding both transitions. In this configuration, the parameters of the "static" \Fexxvi{} K$\alpha$ component change completely: its width, previously around 20 eV, is reduced to $\sigma_{0,26}=1.7_{-1.7}^{+2.1}$ eV, and the negligible blueshift turns into a small redshift, now at $v_{0,26}=446_{-53}^{+60}$ \kmps{}. Meanwhile, the blueshifted component, which is now the stronger of the two, sees an increase in width to a well constrained $\sigma_{4}=4.5_{-2.2}^{+2.7}$ eV, with no significant change to its velocity ($v_{4}=-1281_{-99}^{+115}$ \kmps{}). Although this fits perfectly the residuals, this narrow, significantly redshifted component is difficult to reconcile with the well-constrained, broader and slightly blueshifted \Fexxv{} component around 6.7 keV. We thus note that another local minima with $\Delta$C=-6 provides more standard parameters, with a broader "static" component (10 eV)  compatible with 0 velocity shift ($v_{0,26}=273_{-416}^{+324}$ \kmps{}). The parameters of the --still significant-- narrow component remain compatible with the previous fit. Although we adopt the two narrow component configuration in the final best fit used for the plots and tables due to better statistics, we stress that the results for this complex are highly degenerate, and both the width and velocity shift of the static \Fexxvi{} K$\alpha$ component cannot be reliably constrained in this fit.

Finally, we fit the residual feature at 7.06 keV using the newly introduced \texttt{bfekblor} component in XSPEC, which follows the 4 Lorentzian modeling of \cite{Holzer1997_feklor_base}. This results in a $\Delta$C improvement of 14 for 3 d.o.f., and a narrow ($\sigma_{K\beta}<183$ \kmps{}), static ($v_{K\beta}=-6_{-83}^{+108}$\;\kmps) component. Since a detection of the Fe K$\beta$ line without its K$\alpha$ should be impossible, we will investigate this feature using physical models in Section~\ref{sub:FeK_ratios}.  
We show the final residuals in the most "crowded" 6.3--7.1 keV band for M1744 in Fig.~\ref{fig:resolve_empi_resid_zoom_BH_big_postline}. Additional residuals combining the spectra of M1744 and AXJ 
 are shown in Fig. A.\ref{fig:resolve_bigpix_empi_resid_conti_postline} in App.~\ref{app:resolve_fits}. To assess the quality of our final fit, we perform one final round of blind searches for the M1744 and AXJ spectra, which we show in the bottom-left panels of Fig. A.\ref{fig:blind_search_empi_BH} and Fig. A.\ref{fig:blind_search_empi_NS} in App.~\ref{app:resolve_fits}, respectively. Both significance maps confirm that no significant residuals remain in the final spectra, aside from a feature at 7.1 keV, possibly due to our imperfect fitting of the iron edge, and the lack of consideration for dust (see e.g. \citealt{Rogantini2018_ironedge}). As the edge profile is not trivial and has a very limited signal-to-noise, its detailed study is left for future work. 
We list the full M1744 model and line parameters in Tab. A.\ref{tab:comp_param_empi_BH} in App.~\ref{app:resolve_fits}.
The significance of each line is shown with both the raw $\Delta$C differences and corresponding Monte-Carlo (MC) simulations. This MC significance is central to determine the "true" significance of the lines in regard to the look elsewhere effect (see e.g. \citealt{Porquet2004_stat_MC}), and our methodology is detailed in App.~\ref{app:resolve_MC}. All lines are found to be highly significant and above the commonly adopted threshold of 3$\sigma$, except for the \Fexxv{}$_2$ Ly$\alpha$, which we find to be 99.6$\%$ significant. We thus consider this line as marginally significant, but stress that this solution is, in any case, indistinguishable from the wider line blending the two contributions, and that the same configuration ends up much more than 3 sigma significant in the "small" M1744 region thanks to its lower continuum.

\subsubsection{Small MAXI J1744-294 region}\label{subsub:empi_smallpix}

\begin{figure*}[t!]
\centering
    \includegraphics[clip,trim=0cm 0.3cm 0cm 0cm,width=1.00\textwidth]{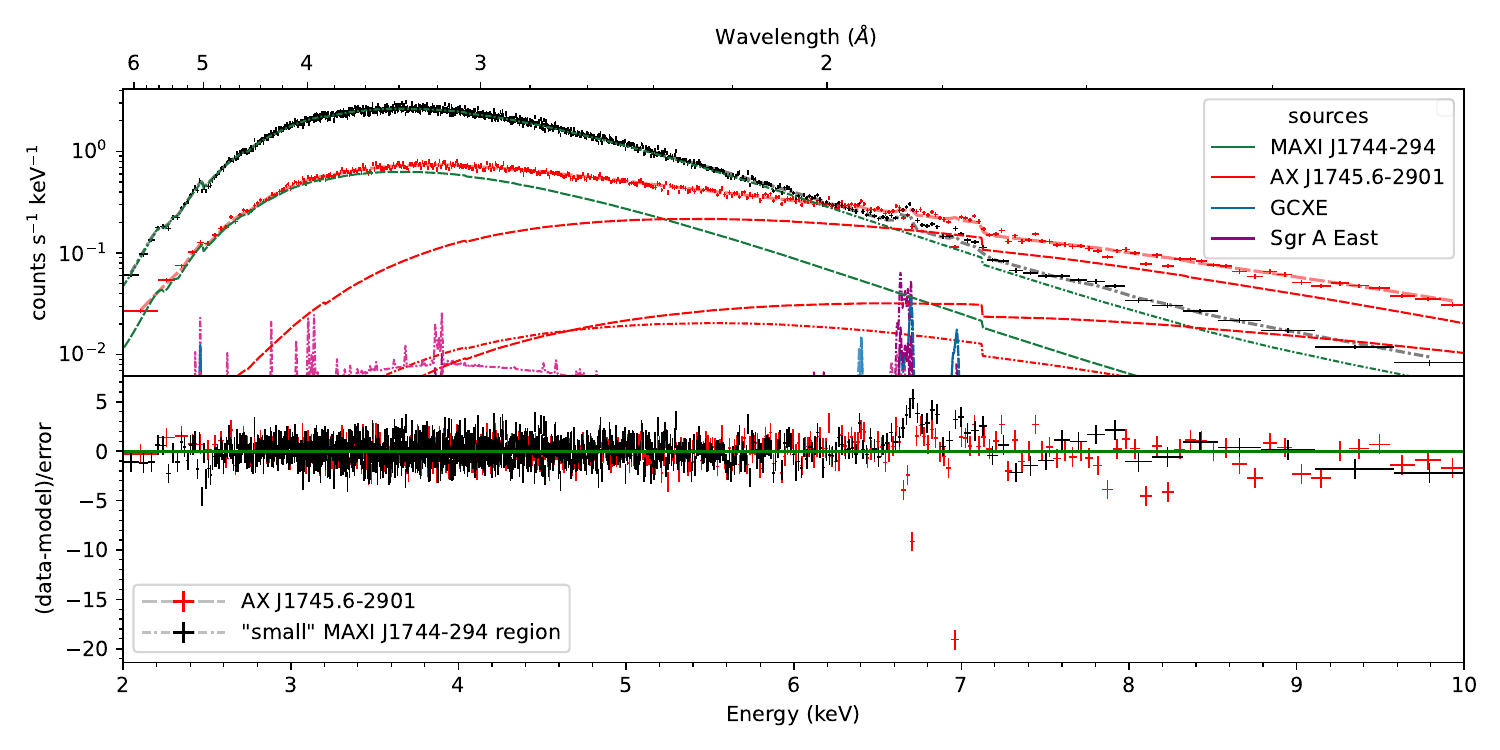}
\vspace{-1.5em}
    \caption{Resolve spectra and residuals for the "small" \src{} region and \axj{} region in the DDT observation, after the first step of their common continuum modeling, and in the entire $2-10$ keV band. Both spectra are visually rebinned at a 20$\sigma$ significance for readability, and model components at a 3$\sigma$ significance level.}
    \label{fig:resolve_smallpix_empi_resid_conti_preline}
\end{figure*}

Our second high-resolution empirical fit of M1744, which includes a physical description of the diffuse emission, is performed on two spectra simultaneously: the "small" M1744 region shown in green in Paper I (Figure 4, left panel), and the AXJ  region shown in red in the same panel. Four sources are applied in each of the two spectra: for the GCXE and Sgr A East diffuse emission sources, we use and freeze the models derived in Paper I, using the angular distributions detailed in that paper and Section~\ref{sec:methodo}.
M1744 and AXJ are computed with point source ARFs, and directly fitted in the observation, starting from the DSH corrected spectral modeling of Section~\ref {sub:empi_xmmnustar} and M26. Our approach to the fit largely follows that of Section~\ref{subsub:empi_bigpix}, with the main difference being the absorption column density of AXJ, which we fix to the value ($N_H=2.47\times10^{23}$ cm$^{-2}$) obtained in the diffuse emission fit in Paper I, for consistency. 

\begin{figure*}[t!]
\centering
     \includegraphics[clip,width=1.00\textwidth]{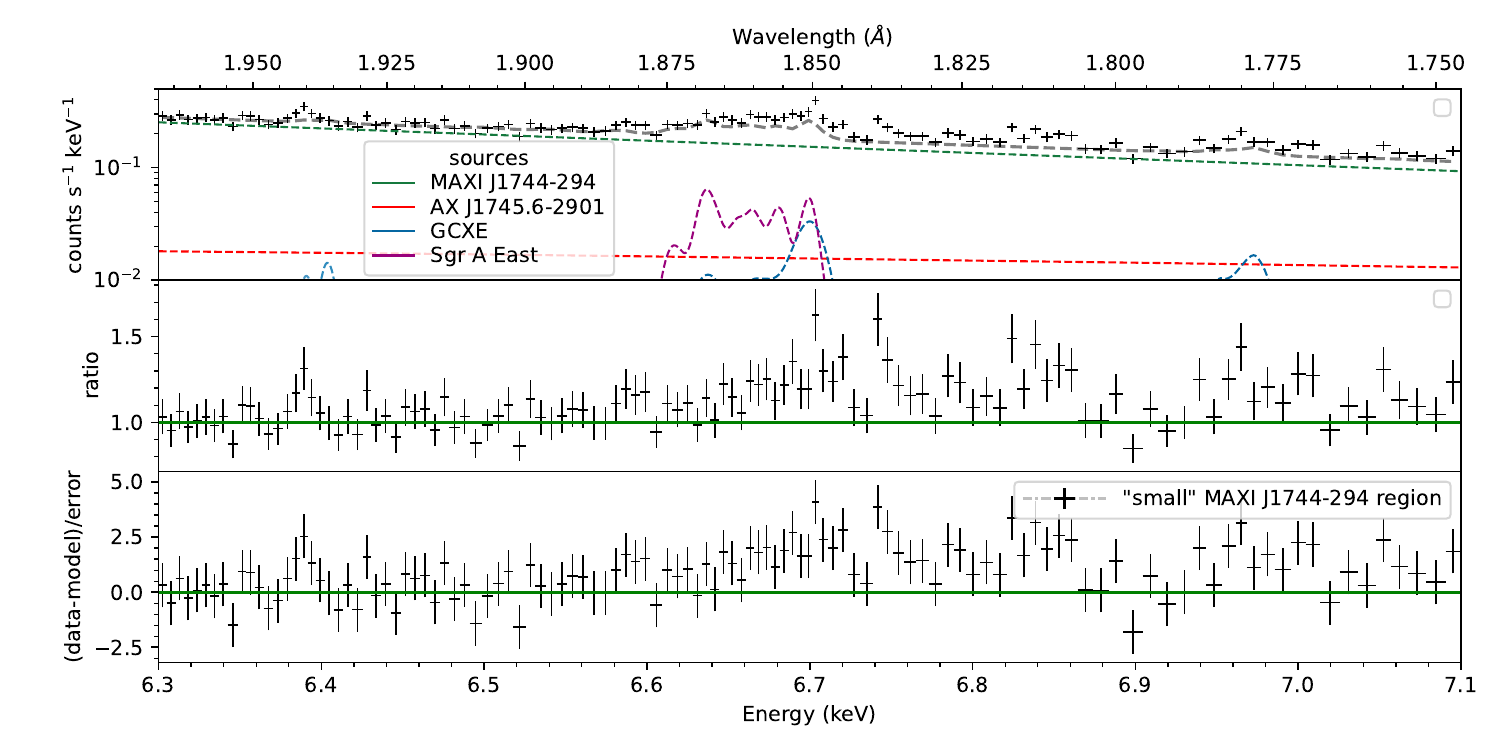}
\vspace{-2em}
    \caption{Zoomed spectrum, ratio, and residuals for the "small" \src{} region in the DDT observation, in the 6.3-7.1 keV band, after the empirical continuum-only modeling. The spectrum is visually rebinned at a 10 $\sigma$ significance level for readability, and model components at a 3$\sigma$ significance level.}
    \label{fig:resolve_empi_resid_zoom_BH_small_preline}
\end{figure*}

We show in Fig.~\ref{fig:resolve_smallpix_empi_resid_conti_preline} the spectrum and residuals after a first fit of the continuum, ignoring the Sulfur edge, 6.4-7.1 keV iron band, and the energies of the high-order absorption lines in AXJ. The Black Hole-only zoom of the 6.3-7.1 keV region is shown in Fig.~\ref{fig:resolve_empi_resid_zoom_BH_small_preline}. The main difference compared to the previous analysis is a much smaller off-axis contribution from other sources, although at the cost of a lower signal-to-noise ratio for the spectrum. As an example, the diffuse emission contribution, which had a maximum in the \Fexxv{} K$\alpha$ complex comparable to the M1744 continuum in the "big" region (see Fig. A.\ref{fig:resolve_bigpix_empi_resid_conti_postline} in App.~\ref{app:resolve_fits}), is less than 3 times lower in the "small" region. This allows additional overlapping features to become apparent, notably a weak Fe I K$\alpha$ emission feature at $\sim$6.39 keV. However, the shapes of the remaining residuals remain similar in both spectra, as can be seen in the zoomed panels of Fig. A.\ref{fig:resolve_smallpix_empi_resid_conti_preline_zoom} in App.~\ref{app:resolve_fits}. Once again, we ignore the residuals around the Sulfur edge, and start by fitting the high-energy lines in the AXJ  spectrum, and refitting the continuum of the two spectra, this time including the entire Resolve band between 7.1 and 10 keV. As the fit statistic does not improve significantly when thawing the energy of the iron edge --perhaps due to a lower signal-to-noise--, we then perform a blind search on the 6.3-7.1 keV band for the BH spectrum, and show its results in Fig. A.\ref{fig:blind_search_empi_BH} in App.~\ref{app:resolve_fits}. The shape of the residuals and their localization remain entirely consistent with the "big" M1744 region fit, and their significance remains high despite the lower signal-to-noise ratio of the spectrum. We note that the normalization of the blind search is computed with respect to the entire set of source models, and changes in the y-axis for the static components reflect only a change in the relative strength of the line compared to the background. Here again, the signal-to-noise ratio of the BH spectrum above 7.5 keV is insufficient to fit higher-energy emission lines.

We thus follow up with a fit of the main static ionized lines in the 6.7-7.1 keV band. Due to the presence of other residuals at higher energies, we here again limit the width of the lines to 10 eV, and obtain very similar parameters than with the "big" M1744 region, with a Fe XXV K$\alpha$ complex dominated by the resonance ($w$) line and with a weak (this time compatible with zero at 2$\sigma$) contribution for the forbidden ($z$) line. The velocity shifts of the \Fexxv{} K$\alpha$ and \Fexxvi{} K$\alpha$ are once again marginal, with the \Fexxv{} K$\alpha$ showing weak blueshifts but remaining compatible with 0 at 2$\sigma$. The \Fexxvi{} component is once again degenerate with a two-component narrower solution, as we will detail below. The \axj{} lines are once again well fit with simple Gaussian absorption components, as will be detailed in Matsunaga et al. (in prep).
As significant residuals are once again apparent after the static highly ionized component fit, we perform a new continuum fit with the entire iron range excluded, before computing a new blind search for each spectrum.
We show the results for M1744 in the middle-right panel of Fig. A.\ref{fig:blind_search_empi_BH} in App.~\ref{app:resolve_fits}. Despite the lower signal-to-noise ratio of the spectrum, the residuals for the unknown features above 6.7 keV are more significant than those for the "big" M1744 region. This is likely due to the lower contribution of other sources in this band for the "small" M1744 region. Moreover, the neutral Fe K$\alpha$ line feature remains significant, albeit weak. We show the results of the line search for AXJ in the upper-right panel of Fig. A.\ref{fig:blind_search_empi_NS} in App.~\ref{app:resolve_fits}. No notable residuals are seen aside from a small feature at 6.6keV, which may result from imperfect modeling of the satellite lines in the diffuse emission (see Paper I). We note that a similar (but weaker) feature was seen in the residuals of AXJ fitted simultaneously to the "big" M1744 region and with an empirical continuum model (upper-left panel of Fig. A.\ref{fig:blind_search_empi_NS}).

We then fit the emission features incrementally with additional Gaussian lines, starting with the new feature seen in this spectrum, for Fe I K$\alpha$. To properly estimate the velocity shift of the line, which appears redshifted in the blind search, we use the \texttt{bfeklor} model in XSPEC, which follows the 7 Lorentzian modeling of \cite{Holzer1997_feklor_base}. The line is well fit by the model but provides only a limited improvement to the data ($\Delta$C=10 for 3 d.o.f.), with for now weakly constrained redshift and width. Then, following the same methodology as per the "big" M1744 region analysis, we use Gaussians for the individual components seen between 6.74 and 6.88 keV. With the approacch of using 3 Gaussians with widths limited to below 10 eV, this leads to $\Delta$C improvements of 31, 16, and 38, for Gaussians centered on $E_1=6746.1_{-2.7}^{+3.1}$ eV, $E_2=6788_{-13}^{+11}$ eV, and $E_3=6835.2_{-5.9}^{+7.8}$ eV, implying velocity shifts of $v_1=-2043_{-138}^{+120}$ \kmps{}, $v_2=-3915_{-477}^{+574}$ \kmps{}, and $v_3=-6032_{-351}^{+264}$ \kmps{} under the assumption of a blueshifted \Fexxv{} K$\alpha$ (w) line. Using a common, broader line for the two highest energy features leads to a slightly better fit than a single line ($\Delta$C=2 for -3 d.o.f.), with similar constraints on its parameters compared to the "big" M1744 region: a width of $\sigma_{2-3}=36_{-13}^{+21}$ eV, and an energy of $E_{2-3}=6826_{-20}^{+12}$ eV, equivalent to a potential blueshift of $v_{2-3}=-5621_{-556}^{+877}$ \kmps{}. We also confirm that fitting all features with a single Gaussian line leads to an unambiguously worse fit ($\Delta$C=-20 for -3 d.o.f.) even for a line width of $>50$ eV, which ensures that at least 2 narrow emission lines are required in this energy range. We present the "two line" configuration in the final fit, one narrow for the 6.74 keV feature and one broader for the 6.83 keV feature. 

\begin{figure*}[t!]
\centering
     \includegraphics[clip,width=1.00\textwidth]{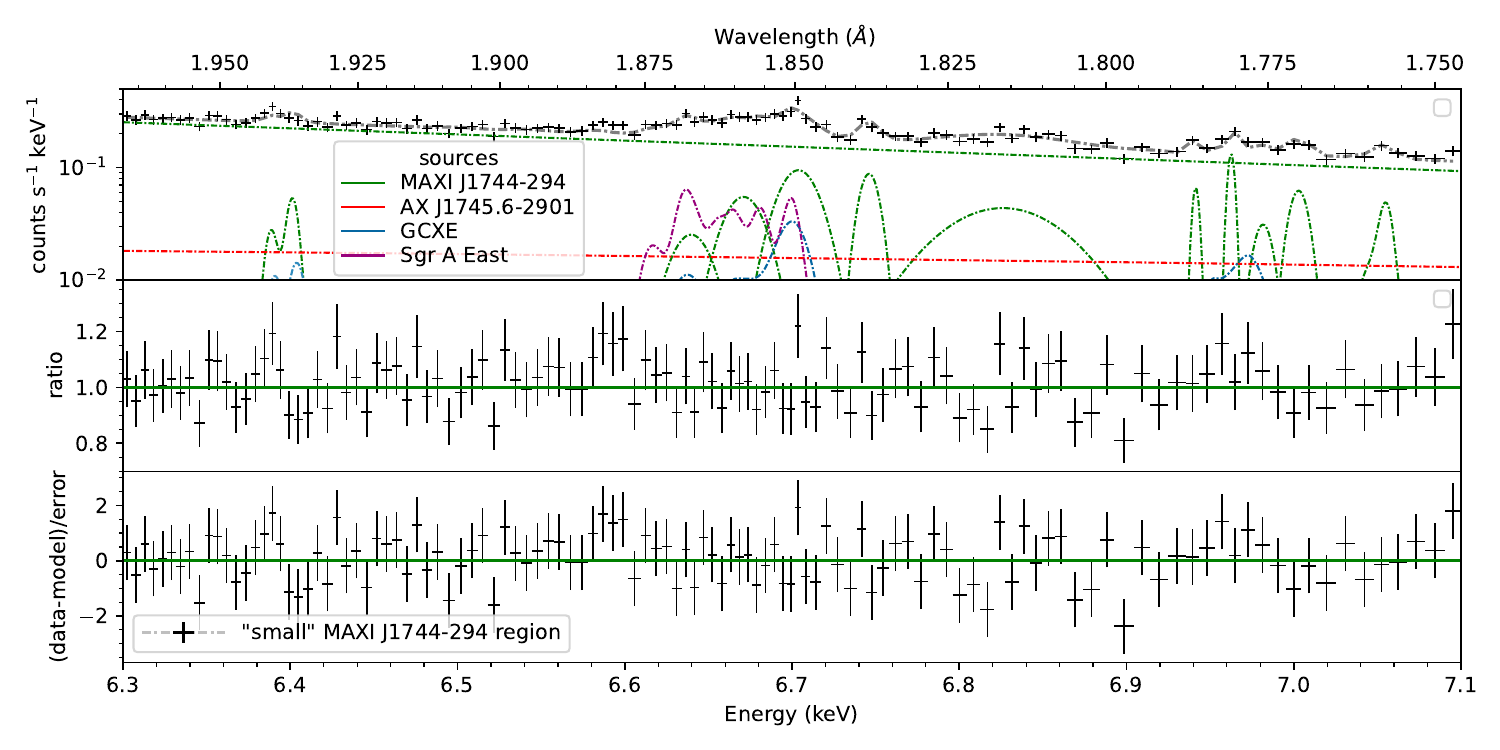}
\vspace{-2em}
    \caption{Zoomed spectrum, ratio, and residuals for the "small" \src{} region in the DDT observation, in the 6.3-7.1 keV band, after the full empirical modeling, including all significant lines. The spectrum is visually rebinned at a 10 $\sigma$ significance level for readability, and model components at a 3$\sigma$ significance level.}
    \label{fig:resolve_empi_resid_zoom_BH_small_postline_zoom}
\end{figure*}

We then test the effect of fitting the residuals around 7.00 keV, first using a Gaussian component around 7.00 keV. The feature is significant ($\Delta$C=12 for 3 d.o.f.), and, once again, stronger than seen in the blind search, due to its low width ($\sigma_4=0^{+6}$ eV). 
Similarly to the "big" M1744 region, its energy is well constrained at $E_{4}=7003.7_{-1.1}^{+1.2}$ keV, or $v_{4}=-1313_{-51}^{+48}$ \kmps{} if attributed to the \Fexxvi{} K$\alpha$-3/2  transition at 6.9732 keV. We then test a fit using both components of \Fexxvi{}, with tied widths and blueshifts and a 1/2 normalization ratio. 
This leads to an adjustment of the static \Fexxvi{} K$\alpha$, with a small improvement in the fit ($\Delta C$=7, for a total of $\Delta$C=19 for 3 d.o.f.), but the parameters remain poorly constrained: this fit becomes once again dominated by the blueshifted component, which remains at a high velocity of $v_{4}-1277_{-131}^{+177}$ \kmps{}, with a higher width of $\sigma_{4}=5.4_{-5.4}^{+3.9}$ eV, while the static component once again lowers to a reduced width of $\sigma_{0}=0.8_{-0.8}^{+2.2}$, and a stronger redshift of $v_{0,26}=457_{-42}^{+36}$ \kmps{}. However, another local minimum exists at a $\Delta C$ of -3, where the "static" component has a much less constrained redshift at $v_{0,26}=472_{-200}^{+110}$ \kmps{}. Thus, while we adopt the global minimum as our final solution, we once again stress that the results for the "static" component are highly degenerate, and the width and velocity shift of the static \Fexxvi{} K$\alpha$ component cannot be reliably constrained.

Our last added component is a \texttt{bfekblor}, used to represent the Fe I K$\beta$ neutral iron feature at 7.06 keV, and whose width and velocity shift are tied to those of the Fe I K$\alpha$ feature. This leads to a fit improvement of $\Delta$C=6 for 1 additional d.o.f. ($\Delta$C=17 for 4 d.o.f. for the two Fe I complexes together), with weakly constrained common widths and velocities of  $\sigma_K=94_{-94}^{+1181}$ \kmps{} and $v_K=101_{-707}^{+452}$ \kmps{}. The abnormal values of the iron line ratio are discussed in Section~\ref{sub:FeK_ratios}. 

We show the final residuals in the 6.3-7.1 keV band for M1744 in Fig.~\ref{fig:resolve_empi_resid_zoom_BH_small_postline_zoom}, and additional residuals combining the spectra of M1744 and AXJ in Fig. A.\ref{fig:resolve_smallpix_empi_resid_conti_postline} in App.~\ref{app:resolve_fits}. We perform one final round of blind searches for each spectrum, as shown in the bottom-right panels of Fig. A.\ref{fig:blind_search_empi_BH} and Fig. A.\ref{fig:blind_search_empi_NS}. Here again, the significance maps do not show any significant residuals, aside from the emission feature at 7.1 keV, which may be attributable to the iron edge \citep{Rogantini2018_ironedge}.
We list the full M1744 model and line parameters in Tab. A.\ref{tab:comp_param_empi_BH} of App.~\ref{app:resolve_fits}. 
In the table, the significance of each line is shown both with raw $\Delta$C differences and corresponding MC simulations.  This MC significance is central to determine the "true" significance of the lines in regard of the look elsewhere effect (see e.g. \citealt{Porquet2004_stat_MC}), and our methodology is detailed in App.~\ref{app:resolve_MC}. Aside from the static \Fexxv{} Ly$\alpha$-z transition, for which we only have an upper limit at $\sim80\%$ significance, and the individual transitions of the Fe I K$\alpha$ line, which are both lower than $99\%$, all transitions are significantly detected above the 3 $\sigma$ level. Furthermore, even if we present the solution with the "broad" composite highly blueshifted line above 6.8keV  in Tab. A.\ref{tab:comp_param_empi_BH}, we stress that the configuration with two individual lines has a much higher significance in the "small" M1744 region, with a total $\Delta$C of 54 for both lines, and each detected above the $3\sigma$ confidence level. Thus, while the two solutions are degenerate and indistinguishable \textit{in the fit}, the "small" M1744 region analysis confirms that they are both highly significant with regards to photon noise.

\section{Physical modeling of the line features} \label{sec:anal_phys}

The line features derived in the previous section are not only significant, but also highly consistent between the different regions and background modeling methodologies, which cements their status are "real", intrinsic features from M1744. They can be summarized as such:
\begin{itemize}
    \item A group of resolved, narrow static emission lines for \Fexxv{} K$\alpha$, with negligible velocity shift and a width around $\sim15$ eV.
    \item One narrow emission line at $E_{1}\sim6.745$ keV, which can be interpreted as a blueshifted \Fexxv{} K$\alpha$-$w$ transition at a blueshift of $v_{2}\sim-2000$\kmps{}.
    \item A duo of emission features around $E_{2}\sim6.78$ and $E_{3}\sim6.83$ keV (or the same transition at $v_{2}\sim-3500$ \kmps{} and $v_{3}\sim-6000$\kmps{}), which can be equally well represented by two narrow ($\sigma_2<10$, $\sigma_3<10$ eV) lines or one broader ($\sigma_{2-3}\sim30$ eV) line centered around $E_{2-3}\sim6.80$ keV (or $v_{2-3}\sim-5500$ \kmps{}).
    \item A duo of resolved, narrow, static or redshifted ($\sim400$ \kmps{}) emission lines for \Fexxvi{} K$\alpha$, with weakly constrained parameters due to the degeneracy with the complex below.
    \item A narrow emission feature around $E_{4}\sim7.00$ keV, most significant when interpreted as a duo of \Fexxvi{} K$\alpha$ blueshifted emission features with $v_{4}\sim-1300$ \kmps{}.
\end{itemize}

In addition, the "big" M1744 region analysis hints at a weak Fe K$\beta$ emission line, and the "small" M1744 region at a duo of Fe K$\alpha$ and Fe K$\beta$ lines. We also see a well-resolved Sulfur edge profile at $\sim2.65$ keV, which will be analyzed in depth in Paper III, and a potential mismatch to the Fe edge at $\sim7.2$ keV in the "big" M1744 region spectrum. 

In the following section, we investigate several plausible candidates for the origin of the highly ionized emission lines, assuming an intrinsic origin and either photoionization or collisional ionization. In addition, we use a physically motivated assessment of the neutral Fe K$\beta$/K$\alpha$ flux ratios to assess whether the intrinsic neutral Fe emission features seen in the previous are significant and warrant the presence of additional features. A common analysis of the neutral and ionized emission lines, along with fully self-consistent reflection models, is left for future work.

\subsection{Intrinsic photoionization}\label{sub:phys_photo}

The first and (arguably) most straightforward approach is to assume that the multiple components detected in our empirical analysis are due to photoionization of the different phases of an outflow. 
We thus computed photoionization tables using \texttt{pion} in SPEX. These tables were extracted from SPEX and then implemented in XSPEC in the form of additive table models (\texttt{atable}). The illuminating SED was fixed to the dust-corrected, unabsorbed model derived in Section~\ref{sub:empi_xmmnustar}. We assume a source distance of 8.178kpc, following the most up-to-date measurements for  Sgr A* \citep{Abuter2019_SgrAstar_distance}.
Our grid of photoionization tables used solar abundances from \citep{Lodders2009_abundances}, across a range of column densities ($N_H$), ionization parameter (log$\xi$), and turbulent velocity ($v_{turb}$). Considering the limited influence of the density with our weak statistics, and to reduce the size of the tables, we assumed a fiducial value of $n_{p}=10^{12}$cm$^{-3}$ across all simulations. We performed 14976 simulations, with 20$\leq$log $N_H$$\leq$25 in 26 steps of 0.2, 1$\leq$log $\xi$$\leq$8 in 36 steps of 0.2, and 3$\leq$$v_{turb}$$\leq$3000 \kmps{} in 16 (logarithmic) steps of 0.2. 

\begin{figure*}[t!]
\centering
    \includegraphics[clip,trim=0cm 0.3cm 0cm 0cm,width=0.99\textwidth]{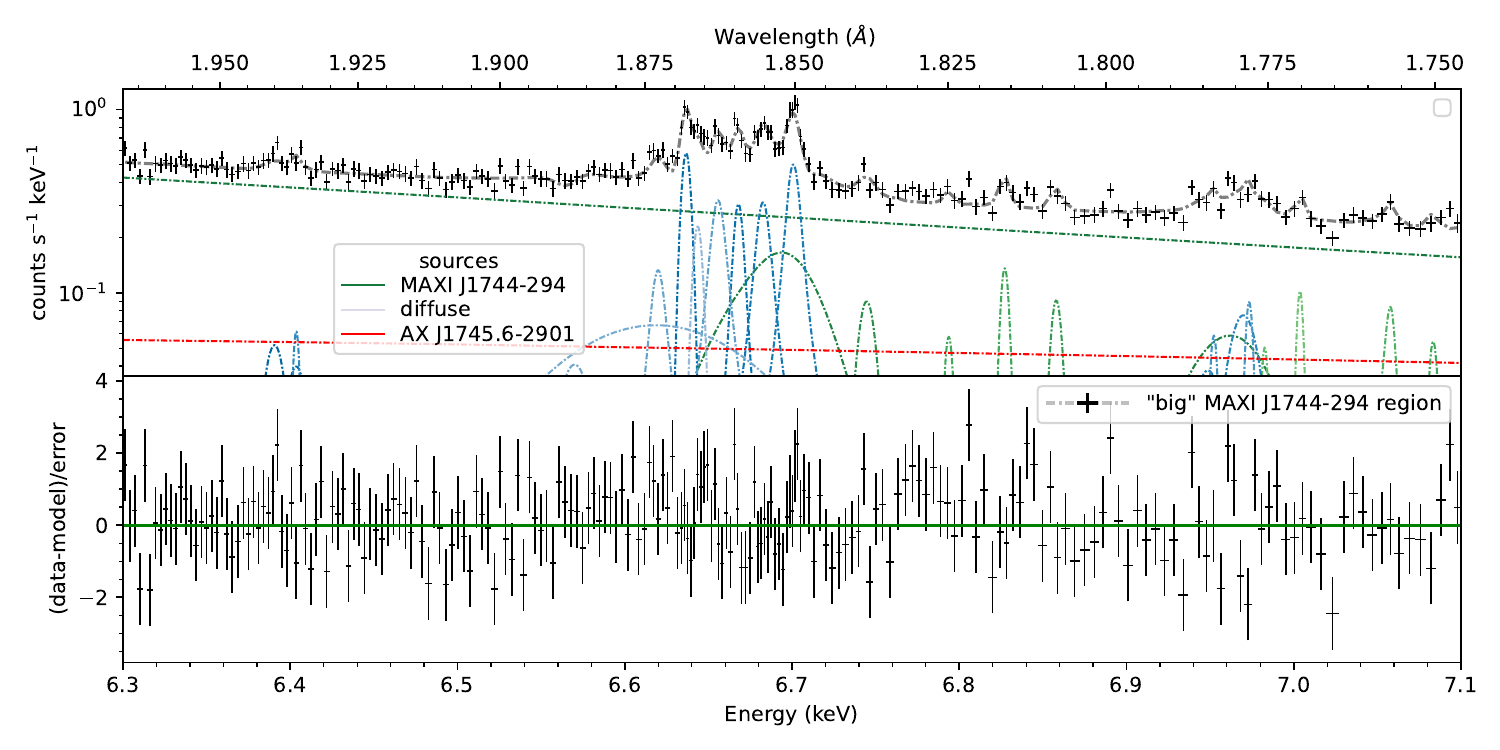}
    \includegraphics[clip,trim=0cm 0.3cm 0cm 0cm,width=0.99\textwidth]{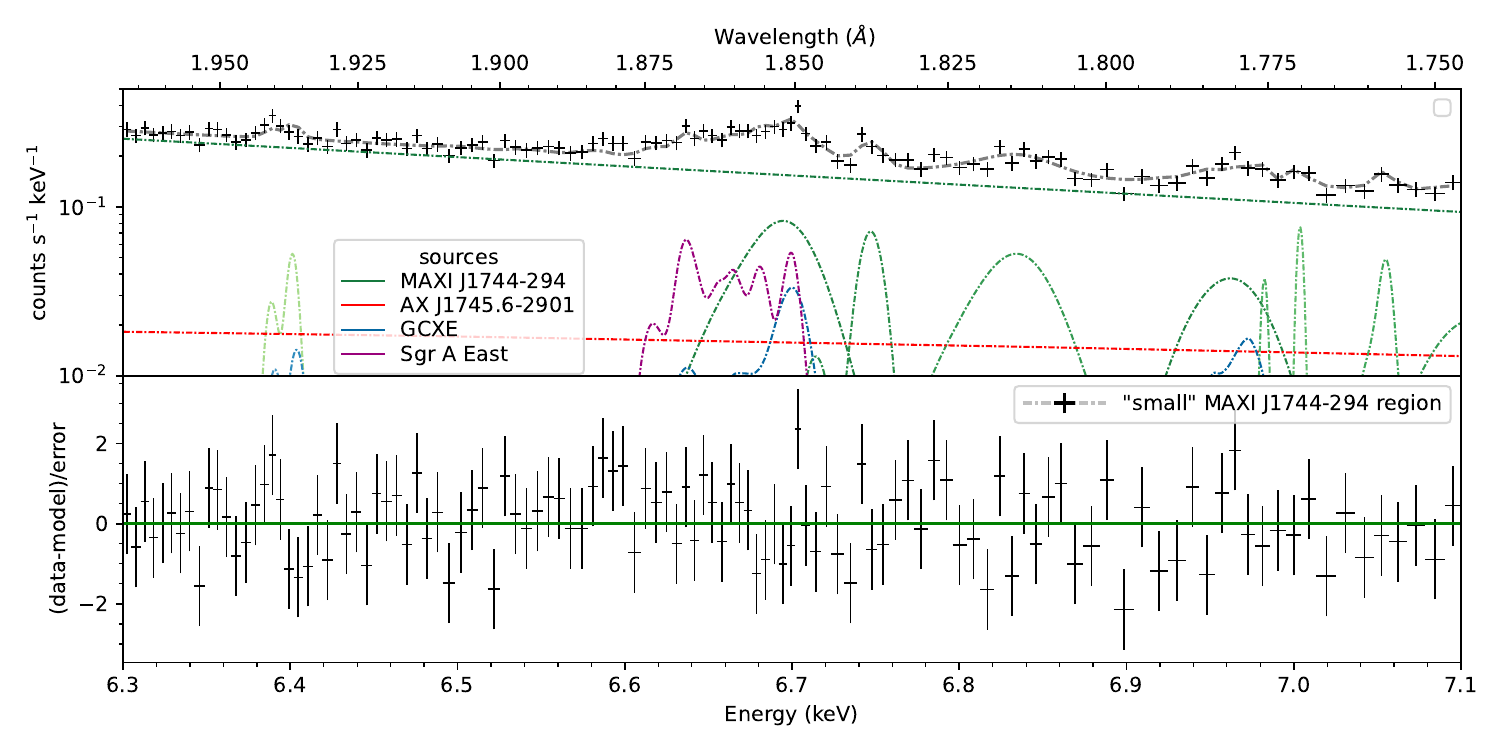}
\vspace{-1.em}
    \caption{Full residuals \textbf{(left)} and zoom in the 6.3-7.1 keV band \textbf{(right)} of the photoionization modeling of the line features in the "big" \textbf{(top)} and "small" \textbf{(bottom)} MAXI J1744-294 region. The residuals for the AX J1745.6-2901 spectrum, which are largely unaffected by the switch to physical models, are omitted for legibility. The spectra are visually rebinned at 20$\sigma$ and 10$\sigma$ in the left and right panels, respectively, and model components at a 3$\sigma$ significance level.}
    \label{fig:resolve_photo_resid}
\end{figure*}

Following the choice of components adopted in the "big" and "small" M1744 region fits, we incrementally replaced each set of lines with a single photoionization component and assessed the parameter constraints. For the sake of simplicity, and due to the lack of constraints in detailed line profiles, we fix the normalization of each pion component to match a covering fraction $\Omega$ of 1, considering the distance highlighted above. Finally, to avoid degeneracy with continuum fitting, we fix an upper limit of $v_{turb}=1000$ \kmps{} ($\sigma=22$ eV at 6.7 keV) to all photoionization layers except the blueshifted composite layer in the "small" M1744 region, whose proposed width was measured above this value in the empirical fit.
Section~\ref{sub:empirical_resolve} has shown that the shape of the residuals between 6.7 and 7.1 keV is only weakly dependent on the region and methodology. Thus, although we use an additional component for the "big" M1744 owing to our choice of two individual features to represent the residuals at 6.78 and 6.83 keV, the results are qualitatively similar for the two fits. We show the zoomed residuals in the 6.3-7.1keV range in Fig.~\ref{fig:resolve_photo_resid}, and a more detailed view of the contribution of the different intrinsic components in the middle panels of Fig.~\ref{fig:mod_comp_ratio}. The broadband Resolve residuals are presented in Fig. B.\ref{fig:resolve_photo_resid_full} in App.~\ref{app:phys_mod}, and we list the parameters derived for the different photoionized layers in Tab. B.\ref{tab:comp_param_phys_BH} . 

The most important constraint is that most emission layers are highly ionized, with log$\xi\sim5-6$ except for the highly degenerate "blue 3" layer in the "big" M1744 region fit. Given the limited statistics on the line shapes themselves, this stems from the balance between two elements. The first is that log$\xi$ must be high enough to avoid producing strong emission lines at lower energies, since none are observed in the spectrum. This constraint is far stronger for the layers with the highest column density, and thus most notable for the static component, where it translates to a requirement of log$\xi\gtrsim5.6$.  However, we note that the requirement of a highly ionized static layer is strongly dependent on the abundances of lower elements, and subsolar abundances of e.g., S, Ca, and Cr, would likely allow for a wider parameter space. This can be tested at first order by restricting the fit to energies above 6.3 keV, which effectively removes the influence of any lower-energy emission ions on the fit. In that case, in both the "small" and "big" M1744 region spectra, the fit converges to a narrower zone with log$\xi\sim$3.5, and can be significantly improved by adding a second narrow zone with higher broadening and log$\xi\gtrsim5.5$, similar to our main zone in the broad band fit, albeit much weaker. However, only the higher $\xi$ solution survives the requirement of a lack of low-$\xi$ lines, and the limited statistics do not allow for pinpointing the need for a second static emission component.

In parallel, the "static" component is the only one where both \Fexxv{} and \Fexxvi{} are identified. All other components are seen through a single line with a given blueshift ($\sim-2000$, $\sim-3700$ and $\sim-6000$ \kmps{}), and thus any photoionization component at the velocity shift of their closest complex will include not only this line but also another emission line from the other highly ionized Fe transition in the vicinity. As the ionization potential of \Fexxv{} is lower than that of \Fexxvi{}, the layers extrapolated by the \Fexxv{} lines must have an ionization parameter low enough to avoid creating significant \Fexxvi{} features. The combination of the two effects provides decent ionization constraints, even for the weaker layers.

One notable exception is the narrow component seen at 7.000 keV, which is the only non-static component interpreted as a  \Fexxvi{} transition ($-1300$ \kmps{} \Fexxvi{} K$\alpha$). Here, the corresponding \Fexxv{}(w) transitions fall between the \Fexxv{}(w) transition of the main static component (6.700 keV) and the first blueshifted emission feature (6.74 keV), in a region where the lack of a narrow feature is very well constrained. The only way to reproduce the \Fexxvi{} line with photoionization is thus to reach negligible contributions of \Fexxv{}, which is only possible at the edge of the parameter space: the \Fexxv{}(w)/\Fexxvi{} K$\alpha$ only reaches $\lesssim10\%$ at log$\xi$=8, and in turn, due to the lower \Fexxvi{} fraction, requires a column density of $N_H\gtrsim2\times10^{23}$ cm$^{-2}$. Even then, the component's fit improvement remains negligible ($\Delta C$=6). This is because in the empirical fit, the width of the \Fexxvi{} complex was significantly smaller than that of \Fexxv{}, while here the main component is driven by the higher statistics of the \Fexxv{} residuals, and its higher width for \Fexxvi{} is less compatible with an additional, narrow, blueshifted component. We thus choose not to fit this component with a photoionized layer. As we cannot exclude a more complicated configuration for the intrinsic layer, and to avoid biasing the static component's width by the 7.00 keV residuals, in both the "big" and "small" M1744 region fits, we instead maintain the empirical double-peaked \Fexxvi{}-only emission line used previously. This very broad \Fexxvi{} profile leads to a weaker significance for the empirical blue \Fexxvi{} component, down to $\sim99.4\%$ in both fits, but we stress that this may very well be due to the imprecision of our single-component modeling for the "static" layer, where the \Fexxvi{} K$\alpha$ line is impacted by the much more significant (and broader) \Fexxv{} complex.  

In parallel, we also note that the main static layer contributes to fitting a weak excess in the Ni and Fe lines between 7.8 and 7.9 keV. Although this may represent a real need for intrinsic emission at higher energy, we stress that it is difficult to assess both the continuum and the contribution from diffuse emission lines in this range, and thus we cannot provide any firm conclusion on the presence of highly ionized Nickel. Similarly, the so-called "blue 2" emission component benefits from fitting the 7.1 keV emission residuals, which we previously interpreted as a potential pre-edge residual but could also be the \Fexxvi{} pendant of the feature seen at $\sim6.83$ keV in both spectra. In the "big" M1744 region, where we explore a more "narrow" configuration with more components, and force $v_{turb}<500$\kmps{} for the blueshifted zones, the lines of the "blue 2" layer cover residuals both below and above 6.8keV with a single component, which allows for zone 3 to shift to higher energies ($\sim6900$\kmps{}). However, we stress that this configuration is indistinguishable from a single, broader line for zones 2 and 3 (as exemplified by the "small" M1744 region fit) with the current statistics.

As per usual, we computed blind searches for residual line features in the 6.3-7.1 keV range following the final fits, including all photoionization layers and the empirical description for the -1300 \kmps{} \Fexxvi{} complex, as well as the empirical models for the \Fei{} K$\alpha$ and K$\beta$ lines. The results, which we show in Fig. B.\ref{fig:blind_search_photo_BH} in App.~\ref{app:phys_mod}, corroborate the lack of significant difference in the fit statistic, compared to the empirical modeling, and confirm that no remaining significant residual feature is seen within each of the highly ionized line profiles. We note some imperfection in the \Fexxv{} K$\alpha$ fit for the big-pixel region, consistent with the lower statistical improvement of that photoionized component compared to our previous empirical fitting.

\subsection{Intrinsic Collisional ionization}\label{sub:phys_colli}

\begin{figure*}[t!]
\centering
    \includegraphics[clip,trim=0cm 0.3cm 0cm 0cm,width=0.99\textwidth]{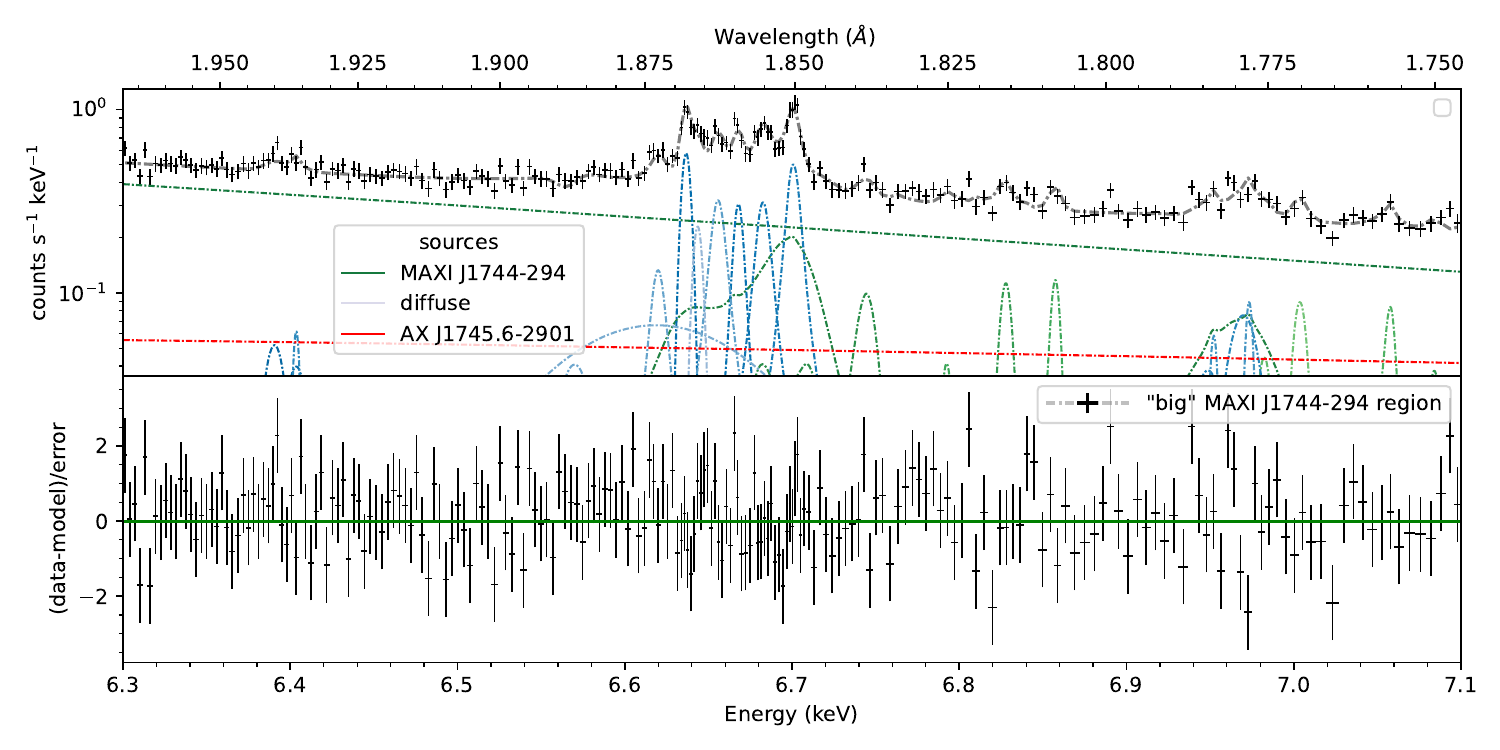}
    \includegraphics[clip,trim=0cm 0.3cm 0cm 0cm,width=0.99\textwidth]{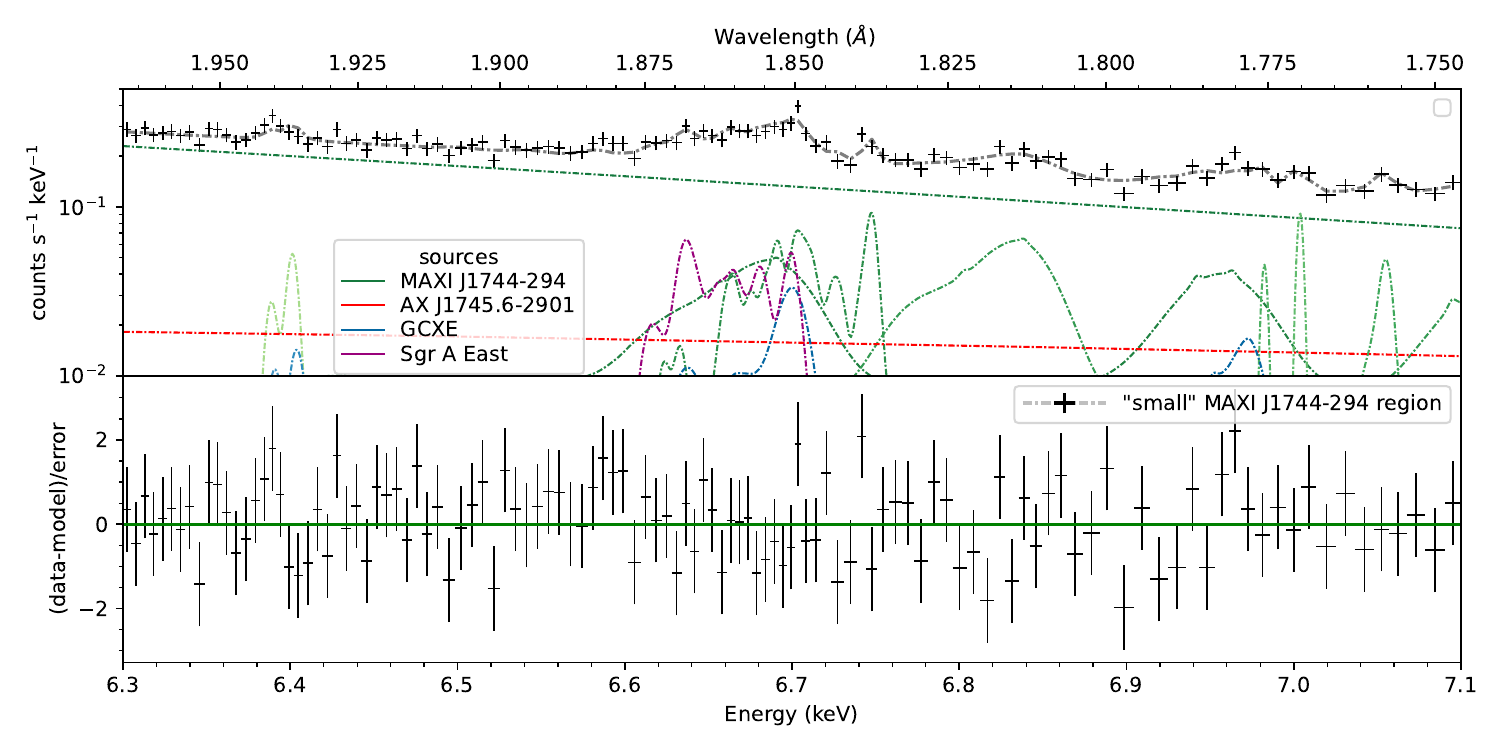}
\vspace{-1.em}
    \caption{Zoom in the 6.3-7.1 keV band of the collisional ionization modeling of the line features in the "big" \textbf{(top)} and "small" \textbf{(bottom)} MAXI J1744-294 region. The residuals for the AX J1745.6-2901 spectrum, which are largely unaffected by the switch to physical models, are omitted for legibility. The spectra are visually rebinned at a 10$\sigma$ significance level, and model components at 3$\sigma$.}
    \label{fig:resolve_CIE_resid}
\end{figure*}
The second approach is to assume that the line emission arises from different phases of a collisionally ionized plasma. For this, we model the same components as in the previous sections, this time using the $\texttt{btapec}$ model in XSPEC (once again with ATOMDB v3.13), which represents a collisionally ionized plasma in equilibrium, with independent temperatures for the continuum and lines, and includes both the effect of intrinsic thermal broadening (with \texttt{APECTHERMAL} set to yes) and an additional source of velocity broadening. All parameters of the different components are left free to vary, aside from the abundances, which are set to 1, and the continuum temperature, which we initially set to a negligible value to fit the line features independently, and then link to the temperatures derived for the lines before re-adjusting the continuum. 

We show the residuals for each fit zoomed in the 6.3-7.1 keV in Fig.~\ref{fig:resolve_CIE_resid}, and a more detailed view of the contribution of the different intrinsic components in the bottom panels of Fig.~\ref{fig:mod_comp_ratio}. The broadband Resolve residuals are presented in Fig. B.\ref{fig:resolve_CIE_resid_full} in App.~\ref{app:phys_mod}, and we list the parameters derived for the different collisional layers, along with the updated continuum, in Tab. B.\ref{tab:comp_param_phys_BH}. 

The models reproduce the data very well with a series of layers with $kT\sim3-8$ keV, and similar widths and blueshifts to the photoionization models. Here again, the blueshifted FeXXVI K$\alpha$ feature cannot be fitted and is thus kept as an empirical component. The 7.1 keV emission residuals, which may be due to the proximity of the iron edge, significantly affect the fitting of the blue zone 2. For the "big" M1744 region, although negligible broadening is preferred for the blue zones 2 and 3, the resulting turbulent velocity still considers the $\sim100$ \kmps{} from the purely thermal contribution. 
This approach introduces important differences compared to the photoionization layers: first, a much more resonance-dominated FeXXV K$\alpha$ line for all components, which, along with the additional continuum contributions at high energies, leads to a better fit in both regions ($\Delta C=38$ for the "big" M1744 region, and $\Delta C=20$ for the "small" M1744 region). However, with the many degeneracies and our limited signal-to-noise, this is not enough to conclude that the CIE solutions are favored. Secondly, with this choice of model, the empirical blueshifted \Fexxvi{} line remains more significant than with the photoionization layers, and above $99.9\%$ MC significance in both the "big" and "small" M1744 regions. 
Finally, the CIE layers introduce a very significant contribution to the continuum at high energies, which dominates over the Comptonized component. In the \xrism\ spectrum alone, the spectral range and contamination by other sources are insufficient to provide meaningful constraints to the high-energy spectral shape, although the continuum fit including all CIE layers requires a comptonization fraction below $f_{cov}\lesssim10^{-3}$, compared to $f_{cov}=7.3_{-0.5}^{+0.5}$ in the empirical model. We thus tested whether the \nustar{} continuum was compatible with a model including the 3 bapec components. We found that the \nustar{} spectrum could not be reproduced by a continuum including all bapec components (C-stat/d.o.f.$>$2.5), as they significantly overfit the 10-30keV band of the spectrum even without the inclusion of any intrinsic hard tail. Moreover, even the static component by itself led to a significant decrease in fit quality, with $\Delta C=$-16 when forcing $\Gamma\geq$ 2, along with clear underfitting at higher energies. As the photon indexes of BH-XRB hard tails in the soft state are generally at or above 2 (see e.g. \citealt{Cangemi2021_hardtail_CygnusX-3,Cangemi2021_hardtail_CygnusX-1,Cangemi2023_hardtail_multi,Parra2025_4U1630}
), the \nustar{} data disfavors a CIE origin, at least within this parameter space. We stress that this is not a definitive argument, as the comptonized continuum we used to perform this comparison is itself imperfect at high energies \citep{Mandel2026_ApJ}. A complete comparison with a more complex hard tail, e.g., including reflection or returning radiation, is left for future work.

We once again computed blind searches for residual line features in the 6.3-7.1 keV range following the final fits, including all collisional ionization layers, the empirical Gaussians for the -1300 \kmps{} \Fexxvi{} feature, and the empirical models for the \Fei{} K$\alpha$ and K$\beta$ lines. The results, shown in Fig. B.\ref{fig:blind_search_CIE_BH} in App.~\ref{app:phys_mod}, help visualize the quality of the fit, the lack of residual features, and differences compared to the photoionization fit, whose blind search is shown in Fig. B.\ref{fig:blind_search_photo_BH}. 

\subsection{Neutral iron line ratios}\label{sub:FeK_ratios}

\begin{figure*}[t!]
\centering
    \includegraphics[clip,trim=0cm 0cm 1.9cm 0cm,width=0.304\textwidth]{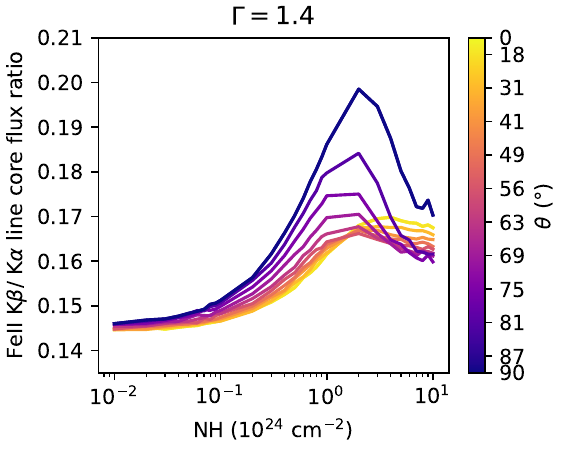}
    \includegraphics[clip,trim=0cm 0cm 1.9cm 0cm,width=0.304\textwidth]{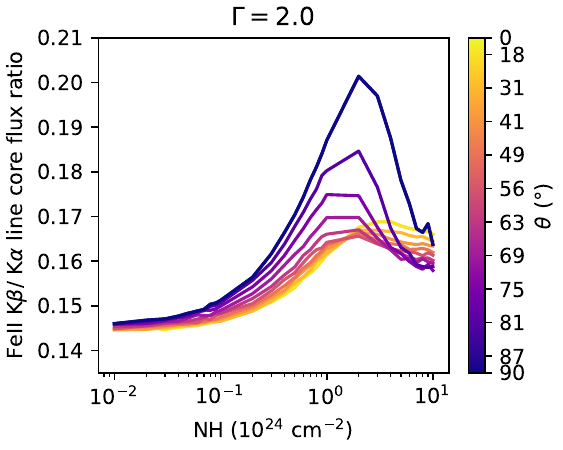}
    \includegraphics[clip,width=0.377\textwidth]{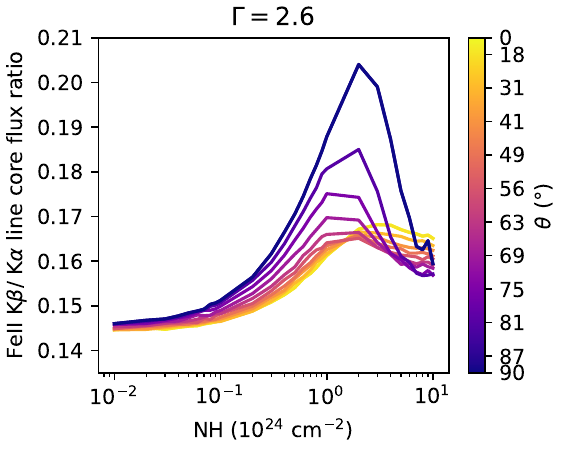}
    \caption{Ratios of the Fe I K$\beta$/Fe I K$\alpha$ line core fluxes across the parameter space of MYTORUS, for 3 representative illuminating power-law photon indexes. Inclination angles follow the angle bins of the model, with 0$^\circ$ for a face-on BH with the torus perpendicular to the line of sight, and 90$^\circ$ for an edge-on BH, where the entire width of the torus crosses the line of sight. See the MYTORUS manual and Section~\ref{sub:FeK_ratios} for details.}
    \label{fig:FeK_ratios}
\end{figure*}

The neutral iron features which we fitted in Section~\ref{sec:empirical}, being both weak and very narrow, can be treated independently from the other highly ionized components. Although the parameters of each line are very standard, the Fe I K$\beta$/$\alpha$ ratios are extremely high. In the "small" M1744 region, the only one where both features are seen in the residuals, the flux ratio of the lines is $R=1.0_{-0.8}^{+4.0}$. In the "big" M1744 region, where the Fe I K$\alpha$ line is not even significantly detected, we can set an upper limit of $R\gtrsim0.43$. These values, although weakly constrained, are much higher than the representative theoretical and experimental value of 0.135 (see \citealt{Murphy&Yaqoob2009_MYTORUS} and references therein). However, in realistic configurations, this ratio can be significantly modified by many physical (illuminating SED, column density, ionization) and observational (inclination angle) properties, none of which can be properly assessed in our case with our marginal detections. We can, however, derive an approximate upper limit to this flux ratio across the possible parameter space using existing physical models and assess the significance of the spectral residuals. For this, we use the updated version of MYTORUS \citep{Murphy&Yaqoob2009_MYTORUS,Yaqoob2024_MYTORUS_2}, which is one of the few available physical models for neutral iron with micro-calorimeter level (2eV) resolution. This approach has two important caveats. Firstly, fluorescence yields and K line energies both evolve with ionization level, which drastically influences the line shape even for low ionization parameters \citep{Kallman2004_photo_iron}, and this effect has recently been detected in one XRB with strong "neutral" iron emission lines \citep{Nagai2026_CenX-3_FeK_shift}. MYTORUS only treats cold matter, and our approach is thus limited to the (very strong) assumption that the Fe I features come from entirely neutral gas. 
Secondly, MYTORUS is suited to eponymous geometries in Active Galactic Nuclei (AGNs) and power-law illuminating SEDs. We note that only the previous version of MYTORUS is presently available for disk-dominated SEDs and  corresponding geometries. In parallel, relxillNS and xillverNS \citep{Garcia2022_relxillNS} are often used in such cases for CCD-level spectra, but as of the writing of this paper, their limited spectral resolution prevents any application to narrow lines in micro-calorimeter datasets. 

We thus computed the Fe I K$\beta$/$\alpha$ ratio over the entire parameter range of $N_H$, $\Gamma$, and $\theta$ in MYTORUS. Since the ratios we derived in our observational analysis were based on the fit of the "narrow" core of the line, we purposefully disregard any contribution from the Compton shoulder, restricting our flux computations to a 10 eV band centered on the average energy of each line.  We show representative values for the lowest, middle, and highest available values of $\Gamma$ in MYTORUS in Fig.~\ref{fig:FeK_ratios}. We note that our ratios are systematically higher by $\sim0.01$ compared to those derived in the MYTORUS manual\footnote{available at \url{https://mytorus.com/mytorus-manual-v0p0.pdf} and on the wayback machine}, likely due to our voluntary choice of computing the line flux in narrow energy bands. More importantly, a notable deviation from the theoretical ratio can be seen in configurations where the medium becomes optically thick to either (and eventually both) emission lines, and thus has different opacities to each. The highest deviations are seen for mildly Compton thick column densities and a 90$^\circ$ angle, which corresponds to the most absorbed line of sight, parallel to the torus plane. These deviations slightly increase with $\Gamma$, reaching a ratio of R$\sim$0.205 at the steepest photon index available. We cannot exclude this geometry as we do not know the geometrical origin of the hard tail, and if it is the one generating the reflection, it could be obscured while the disk emission remains unabsorbed. However, if we assume that the reflection is due to the disk, this configuration is unrealistic as such a geometry would require an overwhelmingly absorbed disk continuum, which we do not see in M1744, or an emitting material in the innermost disk regions, which we can exclude due to the weak line widths. For a non-obscured line of sight, the upper limit becomes R$\sim$0.17, this time for completely face-on configurations and the lowest available $\Gamma$. Meanwhile, in the diffuse emission, the lack of constraints on the physical origin of the Fe I K lines prevents any conclusion on the geometry.

\begin{figure*}[t!]
\centering
    \includegraphics[clip,trim=0.4cm 0.4cm 0.5cm 0.2cm,width=0.495\textwidth]{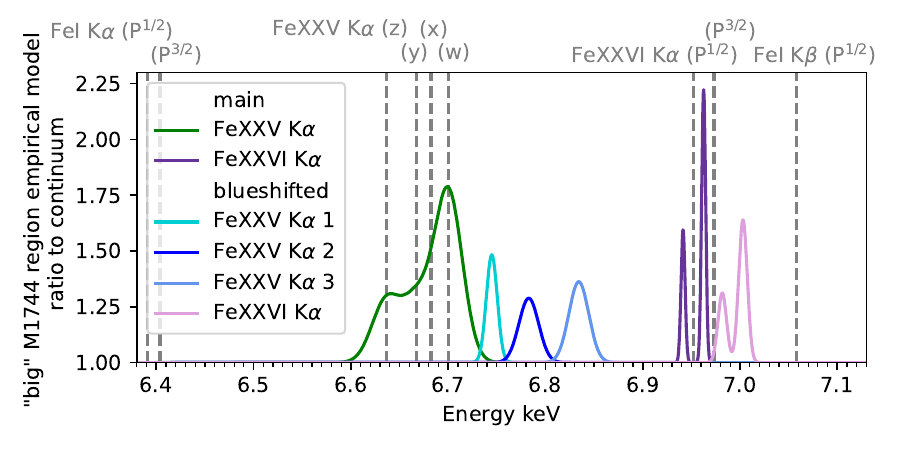}
    \includegraphics[clip,trim=0.4cm 0.4cm 0.5cm 0.2cm,width=0.495\textwidth]{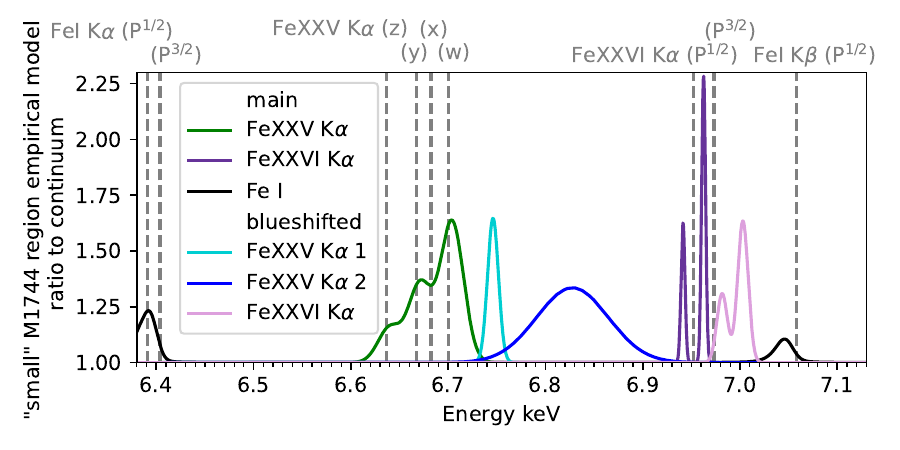}
    \includegraphics[clip,trim=0.4cm 0.4cm 0.5cm 0.2cm,width=0.495\textwidth]{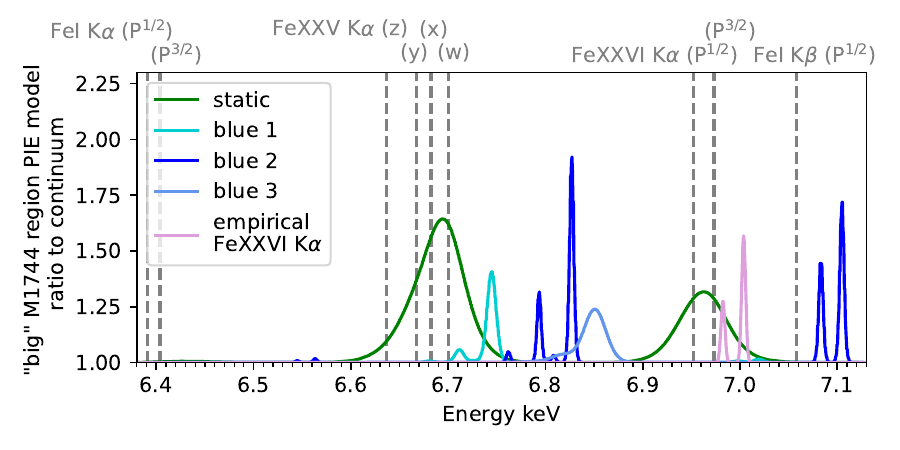}
    \includegraphics[clip,trim=0.4cm 0.4cm 0.5cm 0.2cm,width=0.495\textwidth]{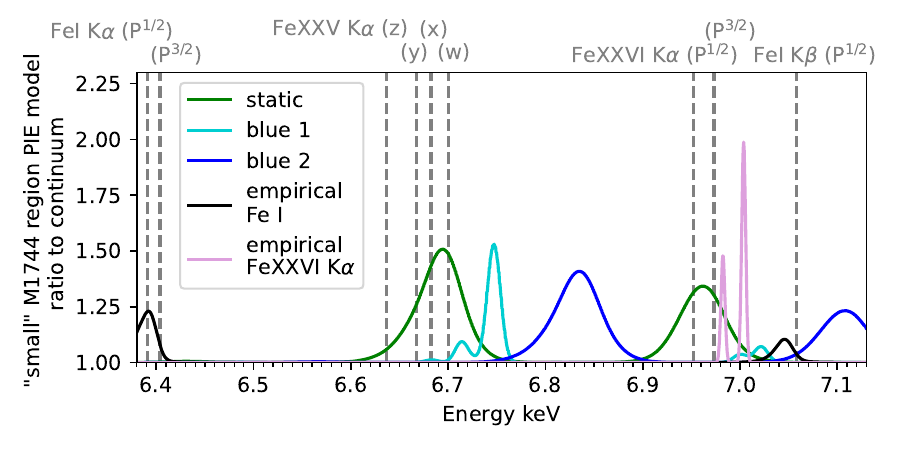}
    \includegraphics[clip,trim=0.4cm 0.4cm 0.5cm 0.2cm,width=0.495\textwidth]{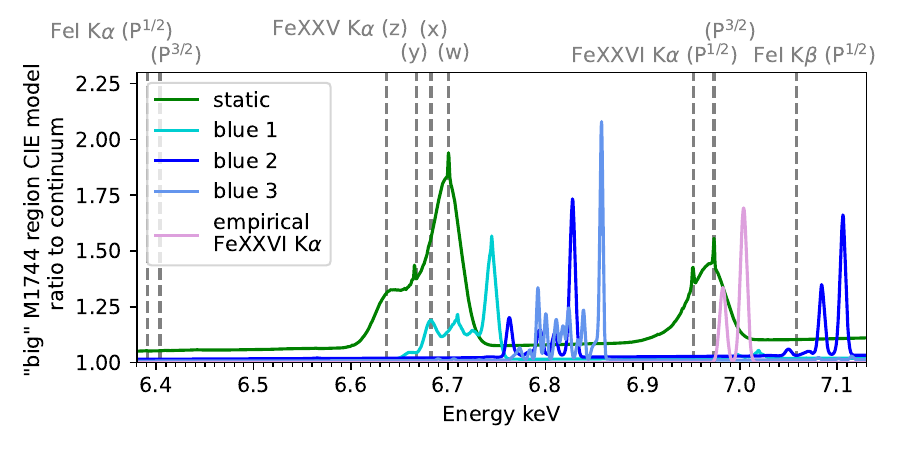}
    \includegraphics[clip,trim=0.4cm 0.4cm 0.5cm 0.2cm,width=0.495\textwidth]{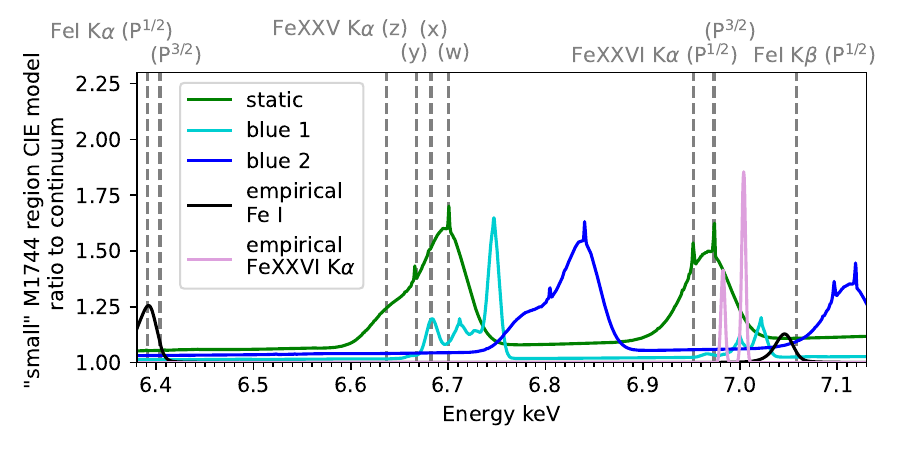}
    \vspace{-1.5em}
    \caption{Flux ratios of the \src{} line components to the deabsorbed continuum of the source, for the empirical \textbf{(top)}, photoionization \textbf{(middle)}, and collisional ionization \textbf{(bottom)} models, for the "big" \textbf{(left)} and "small" \textbf{(right)} MAXI J1744-294 region, from the results of Section~\ref{sec:empirical} and Section~\ref{sec:anal_phys}, including the constrained Fe I line ratios from Section~\ref{sub:FeK_ratios}. Empirical and photoionization components with width upper limits are put at a fiducial width of $\sigma=$2 eV ($\sim90$ km $^{-1}$) for visualization, which is within their allowed parameter space. }
    \label{fig:mod_comp_ratio}
\end{figure*}

Initially, only the physical diffuse emission models of the "small" M1744 region included physically consistent iron line ratios. Meanwhile, for the diffuse emission in the "big" M1744 region, we did not include a Fe I K$\beta$ line because there were no significant residuals. We can, however, assess whether the data remains compatible with a Fe I K$\beta$/K$\alpha$ ratio at either end of the parameter space. The initial diffuse emission empirical fit of Paper I included two different Fe I K$\alpha$ line complexes with different widths to reproduce potential emission from both Sgr A East and the GCXE. Replacing them with a single \texttt{bfeklor} (the relevant XSPEC component for Fe I K$\alpha$) leads to an almost similar fit, with $\Delta$C=+5 for +4 d.o.f. There is thus no need for a second component for the Fe I K$\alpha$ line with this more precise model. We then added a \texttt{bfekblor} (the relevant XSPEC component for Fe I K$\beta$), and refitted the model with a fixed normalization ratio between the two. The result is a very marginal improvement in the fit, with a $\Delta C$ between -2 and -3 for normalization ratios between 0.145 and 0.205. While this would not be sufficient to justify the addition of the Fe K$\beta$ independently, it ensures that the data are compatible with the entire range of normalization ratios. Meanwhile, in the physical fit of the "small" M1744 region diffuse emission fit of Paper I, only the GCXE included a Fe I K line contribution, and the normalization ratio was initially frozen at 0.17. Due to the weakness of the Fe K$\beta$ line and the even more limited signal-to-noise ratio at 7 keV, the data are again compatible with the entire range of normalization ratios, with values of 0.145 to 0.205 leading to only a $\Delta C$ of $\sim$1.

We then revisited the fits of Section~\ref{sub:empirical_resolve}, forcing the diffuse emission ratio to 0.205 in both modeling approaches, in order to get the most conservative estimates of potential residuals. In the "big" M1744 region, the addition of the Fe I K$\beta$ diffuse component lowers the improvement of an additional component at 7.06keV from $\Delta C=13$ to $\Delta C$=7, much more compatible with spurious statistical fluctuations. We further confirmed that adding a physically consistent Fe I K$\alpha$ and K$\beta$ intrinsic to M1744 did not lead to any improvement to the fit, due to the lack of strong residuals at 6.4 keV, which are likely dominated by the GCXE. We confirm this with an additional blind search of the 6.3-7.1keV range with the diffuse emission Fe K$\beta$ flux ratio at 0.205, and no intrinsic Fe I component, shown in the left panel of Fig. B.\ref{fig:blind_search_phys_fekratio} in App.~\ref{app:phys_mod}.

Meanwhile, in the "small" M1744 region, freezing the normalization ratio between the two components significantly changes the solution, which now predominantly fits the slightly redshifted, broader feature around 6.37 keV, with only a marginal contribution at 7.06 keV, regardless of the chosen normalization ratio. With a conservative ratio of 0.205, the intrinsic Fe I contribution results in a fit improvement of $\Delta C$=14 for 3 d.o.f., confirming the detection of the complex, but narrow residuals remain at 7.06 keV. Adding an additional narrow line at 7.06 keV leads to an improvement of $\Delta C$=9 for 3 d.o.f. One physical interpretation could be yet another narrow blueshifted component of Fe XXVI K$\alpha$ at $\sim-3500$ \kmps{}, matching weak residuals observed at a similar blueshift for the Fe XXV K$\alpha$ line at around 6.78 keV. However, we stress that the statistics remain too limited to distinguish this potential line from photon noise, and that the influence of a complex Fe edge on the continuum modeling (which we cannot constrain within our observation) adds additional uncertainty to these results. We showcase these residuals with a blind search in the 6.3-7.1 keV range in the right panel of Fig. B.\ref{fig:blind_search_phys_fekratio} in App.~\ref{app:phys_mod}. Its base model includes both the diffuse emission and the intrinsic Fe I K flux ratios at 0.205, but no additional 7.06 keV component. The new Fe I K$\alpha$ parameters are reported in Tab. B.\ref{tab:comp_param_empi_BH}, and their component contribution is shown in the right panels of Fig.~\ref{fig:mod_comp_ratio}. We note that while this newly adjusted component is now exactly 3$\sigma$ significant w.r.t. MC simulations, it remains degenerate with the unknown emission feature observed around $\sim7.06$ keV.

\section{Discussion} \label{sec:discu}

In this section, we detail the physical interpretations of the different emission components detected in \src{}, their relevance with respect to the literature, and their discrepancies with the signatures from the diffuse emission. 
In Section~\ref{sub:discu_origin}, we explore the possible wind and jet scenarios able to explain the presence of the static and blueshifted phases of highly ionized plasma. In Section~\ref{sub:line_attrib}, we question and justify our interpretation of blueshifted phases for the line features seen at atypical energies in the \src{} spectrum. We compare our results with the literature of compact objects at all mass scales in Section~\ref{sub:compa_lit}. Finally, in Section~\ref{sub:discu_diffuse_origin}, we quantitatively confirm that none of the intrinsic emission features can be imputed to an incorrect estimate of the diffuse emission contribution.

\subsection{origin of the highly ionized phases} \label{sub:discu_origin}

 Pure narrow emission residuals for highly ionized iron lines are extremely rare in BH-LMXBs, having only been detected in the exotic states of 3 systems to date: V404 Cygni \citep{King2015_GS2023+338_wind_x}, V4641 Sgr \citep{Shaw2022_SAXJ1819_wind_emission_2020}, and GRS 1915+105 \citep{Miller2025_GRS1915+105}, all of which are during strongly obscured phases, with many elements pointing towards previous or simultaneous Super-Eddington accretion. This is not the case here, as M1744 shows relatively normal spectral states, luminosity, and long-term evolution for a BH-XRB candidate. The addition of several individual components with blueshifts of thousands of \kmps{} is even rarer, and to date has only been detected in SS 443 \citep{Medvedev2019_SS433_Chandra,Shidatsu2025_SS433}, but at relativistic speeds, and is interpreted as the signature of precessing jets. At lower velocities, only a single tentative report of non-relativistic speeds above $\gtrsim1000$\kmps{} in a BH-LMXB exists with high-resolution instruments, for V4641 Sgr \citep{Parra2025_V4641Sgr_XRISM}. 
However, distinguishing several narrow emission features from a broad emission line is virtually impossible without micro-calorimeters, or very favorable conditions for grating instruments to leverage their resolution despite a limited effective area. This is exemplified by the lack of constraints on the presence of multiple lines in the many observations obtained with CCDs for this very source \citep{Mandel2026_ApJ}.

\subsubsection{wind origin}

For the static phase specifically, if such highly ionized lines are a standard characteristic of soft state sources, then we could expect them to be linked to the main source of narrow lines in this accretion state, namely, disk winds. Absorption lines are a staple of disk winds seen at \textit{relatively} high ($\sim55-80^\circ$) inclination, where the line of sight crosses the bulk of the outflow. At even higher inclination angles, few systems called Accretion Disk Coronae (despite no link with Black Hole Coronae) show strong emission lines and a continuum dominated by reprocessed components and emission lines, and may be compatible with the geometry of thermal winds \citep{Tomaru2023_wind_ADC}. This, along with the confirmation of non-negligible covering fractions in recent \xrism\ observations of multiple NS-LMXBs \citep[][Caruso et al. in prep.,...]{XRISM2025_GX13+1,Tsujimoto2025_CirX-1,XRISM2024_CygnusX-3,Miller2025_4U1630-47}, confirms the presence of a scattered component across a range of accretion states and inclination. In lower-inclination systems, this scattered component should naturally be observed without being convolved with a highly ionized absorber. 

A more precise interpretation remains difficult: due to the lack of detections of such residuals with the previous generation of instruments, studies of physical wind-launching mechanisms have overwhelmingly focused on absorption profiles, particularly for MHD winds. For thermal winds, the few existing reports of emission lines show a variety of complex profiles \citep{Tomaru2020_H1743-322_wind_model_thermalradiative_2,Tomaru2023_wind_ADC}, and may be able to reproduce the static emission phase. One contrasting element is that in absorption, even Super-Eddington wind signatures such as GX 13+1 only show turbulent velocities of a few hundred \kmps{} \citep{XRISM2025_GX13+1}. These are generally lower than our results for the main wind phase, but a disk wind or atmosphere spread across the entire disk would lead to a much higher velocity spread if viewed face-on. Regarding the line shape, the combination of a dominating \Fexxv{} line and a very high ionization parameter, although very rare, is likely to be the consequence of the extremely soft SED of M1744, with not only a pure disk but a very low disk temperature compared to other bright BH-XRB soft states. This apparent temperature is likely a byproduct of relativistic effects in the vicinity of the Black Hole and points to a low-inclination source \citep{Munoz-Darias2013_HID_i}.

On the other hand, the highly blueshifted phases do not match any of the current wind launching mechanisms. The velocities are purely too high for thermal winds, as confirmed by the few hundred \kmps{} blueshifts seen in the Super-Eddington state of GX 13+1 \citep{XRISM2025_GX13+1}. Line driving, seldom proposed in BH-XRB sources but able to reach higher velocities, would be completely suppressed in such a bright soft state. This leaves two possible interpretations for a wind origin. First, magnetic driving, which can easily reach $10^4$ \kmps{} under the assumption of a wind starting in the inner disk regions. A magnetic origin would nonetheless face several issues when considering the radial evolution of its density, ionization, and velocities within the framework of current self-similar prescriptions (see e.g. \citealt{Fukumura2010_MHD_wind_AGNs,Jacquemin-Ide2019_wind_weak_magnetic_JEDSAD_modeling}). First, high velocity regions are traditionally expected to be completely overionized \citep{Fukumura2017GROJ1655-40_wind_magnetic,Datta2024_WED_signatures}, and thus would not imprint the X-ray spectrum with emission lines. One possible mitigation is the extremely soft spectrum of this source, which allows very high ionization parameters to retain high ionic column densities, but the extent has yet to be quantified. The second and most important issue is that the continuous nature of these solutions forces single-peak asymmetric line profiles with a blueshifted tail \citep{Fukumura2017GROJ1655-40_wind_magnetic,Tomaru2022_GROJ1655-40_wind_thermal,Datta2024_WED_signatures}. Such signatures are completely incompatible with the stratification of multiple components seen in our dataset. As recent studies for AGNs \citep{XRISM2025_PDS456,Mehdipour2025_XRISM_NGC3783} and XRBs \citep{Miller2025_4U1630-47,XRISM2025_GX13+1} alike show similar multi-component narrow profiles, winds may be systematically clumpy at the macroscopic level, or at least strongly stratified, no matter their launching mechanisms. Such solutions have yet to be formally explored for MHD winds -or any other physical origin-, in part due to the extreme complexity and computational power required to properly consider and resolve clumps. We note that self-similar MHD solutions naturally allow a "clump-free" wind stratification, as the solutions strongly depend on the number of magnetic field oscillations in the disk, which is a quantified number \citep{Jacquemin-Ide2019_wind_weak_magnetic_JEDSAD_modeling}. This phenomenon is governed by the magnetization $\mu$, hitherto assumed constant across the entire disk; a more realistic scenario, with $\mu$ decreasing at larger radii, would naturally lead to a succession of different "classes" of solutions. This would imply different degrees of bending for successive magnetic field lines, and thus different projected velocities across any line of sight. 

The last and most exotic wind launching mechanism available is radiation driving from the continuum of a strongly Super-Eddington source. Indeed, radiation driving is the most likely candidate to explain the Ultra-Fast Outflow (UFOs) absorption and emission features seen in soft X-rays for the brightest Ultra-Luminous X-ray Sources \citep{Kosec2021_ULX,PintoKosec2023_ULX}. ULXs exhibit a wide variety of wind signatures, with strong state dependence and hints of stratification between UFOs and non-relativistic components \citep{Pinto2020_ULX,Pinto2021_ULX,Pinto2025_ULX} that are compatible with the signatures seen in M1744. However, this directly contradicts the continuum properties of the source, whose apparent luminosity is of the order of $L_{X}\sim10^{37}$ erg/s under the reasonable assumption of an object in the Galactic center, much below the Eddington limit of any compact object. One possibility is that the inner flow of the source is obscured by a fully ionized, Compton-thick absorber, decreasing the source's flux while having a negligible influence on its spectral shape. This would provide an intermediate, perhaps more face-on middle ground between standard galactic XRBs and the extreme cases of obscured high-inclined sources thought to be accreting at very high rates, such as GRS 1915+105 \citep{Neilsen2020_GRS1915+105_obscured_NICER_winds_hard,Miller2025_GRS1915+105} and V4641 Sgr \citep{Shaw2022_SAXJ1819_wind_emission_2020,Parra2025_V4641Sgr_XRISM}.In the case of a highly obscured, potentially Super-Eddington episode, the low disk temperature could then be interpreted as the signature of a so-called "soft" Ultra-Luminous state \citep{Gurpide2021_ULX}. The source would nonetheless have to mimic the spectral-timing evolution of a standard outbursting XRB, making this scenario very convoluted. 

\subsubsection{jet origin}

As none of the current wind frameworks provide satisfactory answers, it may be that the emission lines have a completely different origin. The only other well-established physical scenario is collisional ionization, which in BH-XRBs is historically linked to baryonic ejections. The only recorded detections of this phenomenon are the precessing relativistic jets of SS 433 (see e.g. \citealt{Fabrika2004_SS433_review,Marti2018_SS433_mm,Medvedev2019_SS433_X-rays}). First and foremost, we stress that with the right angle, a jet origin is perfectly compatible with blueshifts of a few $\sim-1000$ km s$^{-1}$. Moreover, despite intrinsic jet speeds of $\sim0.26$c, the maximum blueshifts seen in the X-ray spectra of SS 433 are only $\lesssim24000$ km s$^{-1}$ ($\sim0.08$c), which is only 3-4 times above our fastest component. Both of these aspects are the consequence of the same relativistic effects, which we will detail below. Furthermore, as SS 433 is a system of incredible complexity, several elements can be connected to our observational case. 

The primary culprit is the direct X-ray emission from the precessing baryonic jets themselves, but both the physical scenario and observables must be adjusted significantly to match our situation. The adaptability of the physical scenario can be questioned considering the difference between the two sources: SS 433 is a supercritical accretor with near persistent, highly mass loaded jets, whose velocity is tightly constrained and can vary between $\sim0.26c$ \citep{Fabrika2004_SS433_review,Blundel2007_jet_speed_optical,Sakai2026_SS433_jet_X-ray_vs_OIR} and up to 0.30c during flares \citep{Blundel2011_SS433_jet_sped_optical_flare,Jeffrey2016_SS433_jet_speed_radio_flare,Sakai2026_SS433_jet_X-ray_vs_OIR}. This may favor a radiative acceleration through so-called "line-locking" (see e.g. \citealt{Milgrom1979_SS433_line-locking,Shapiro1986_SS433_line-locking}, although see \citealt{Gravity2017_SS433} for recent observational constraints). M1744, on the other hand, appears to be a standard BH-XRB candidate, as mentioned in M26, and was in the soft state during the \xrism\ observation. The presence of transient jets this "deep" in the soft state (very low HR, no QPOs, negligible hard tail) would be atypical -compact jets notwithstanding-, but M1744 nonetheless shows peculiar radio properties, which are currently investigated and will be presented in a forthcoming paper (Grollimund et al. in prep.). 

While theoretical investigations of transient ejecta as a source of X-ray CIE emission are beyond the scope of this work, we can at least explore how current and future observables could constrain physical parameters of the system under the guise of special relativity. We refer to \cite{Urry1995a_AGN_jets} for details of the relativistic derivations. Here, the main difference compared to SS 433 is the lack of redshifted jet detection, which prevents any direct determination of the intrinsic jet velocity $\beta_{intr}$. We recall that the light emitted from particles in the jet is blueshifted by a doppler factor which we will dub $\delta_{line}$ (from its observational constraints), which is a function of both $\beta_{intr}$ and the inclination of the jet $\theta$, through:

\begin{equation}\label{eq:delta}
    \delta_{line}=\frac{1}{\gamma_{intr}\;(1-\beta_{intr}\;\mathrm{cos}\;\theta)}
\end{equation}

With $\theta=0$ for a jet directly in the line of sight and $\gamma_{intr}$ the intrinsic Lorentz factor of the jet. In SS 433, since both $\delta_{line}$ and $\beta_{intr}$ are known, this formula can be inverted to retrieve the inclination of the jet and its evolution with precession, but not here. However, the same parameters intervene in the formula defining the apparent transverse velocity of transient ejecta $\beta_{app}$:

\begin{equation}\label{eq:beta_app}
    \beta_{app}=\frac{\beta_{intr}\;\mathrm{sin}\;\theta}{1-\beta_{intr}\;\mathrm{cos}\;\theta}
\end{equation}

Thus, by assuming a common origin, and that the jet angle and velocity are similar in the inner, X-ray emitting regions, and outer, radio emitting regions, we can uniquely solve $\beta_{intr}$ and $\theta$ with a single transient ejecta, using the two observables $\delta_{line}$ and $\beta_{app}$. We note that this assumption is at first order valid for the jets in SS 433, the only system where it can be tested \citep{Jeffrey2016_SS433_jet_speed_radio_flare,Shidatsu2025_SS433}. Analytically, this leads to:

\begin{equation}\label{eq:beta_intr}
    \beta_{intr}=\sqrt{\frac{\beta_{app}^4+2\beta_{app}^2(\delta_{line}^2+1)+(\delta_{line}^2-1)^2}{(\beta_{app}^2+\delta_{line}^2+1)^2}}
\end{equation}

\begin{equation}\label{eq:theta}
    \theta=\mathrm{tan}^{-1}\bigg(\frac{2\beta_{app}}{\beta_{app}^2+\delta_{line}^2-1}\bigg)
\end{equation}

\begin{figure*}[t!]
\centering
    \includegraphics[clip,trim=1.9cm 0cm 1.3cm 1.3cm,width=0.49\textwidth]{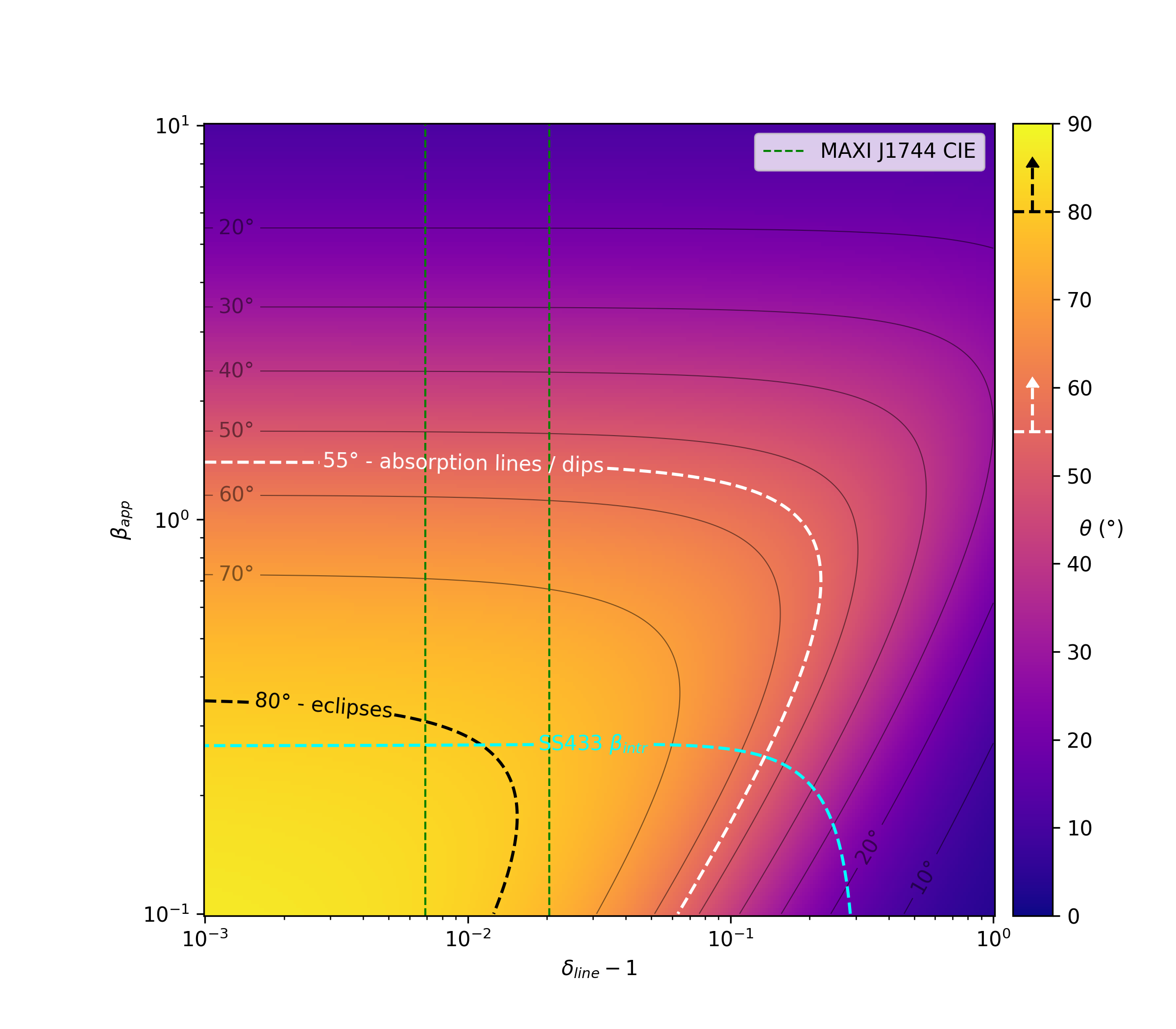}
    \includegraphics[clip,trim=1.9cm 0cm 1.3cm 1.3cm,width=0.49\textwidth]{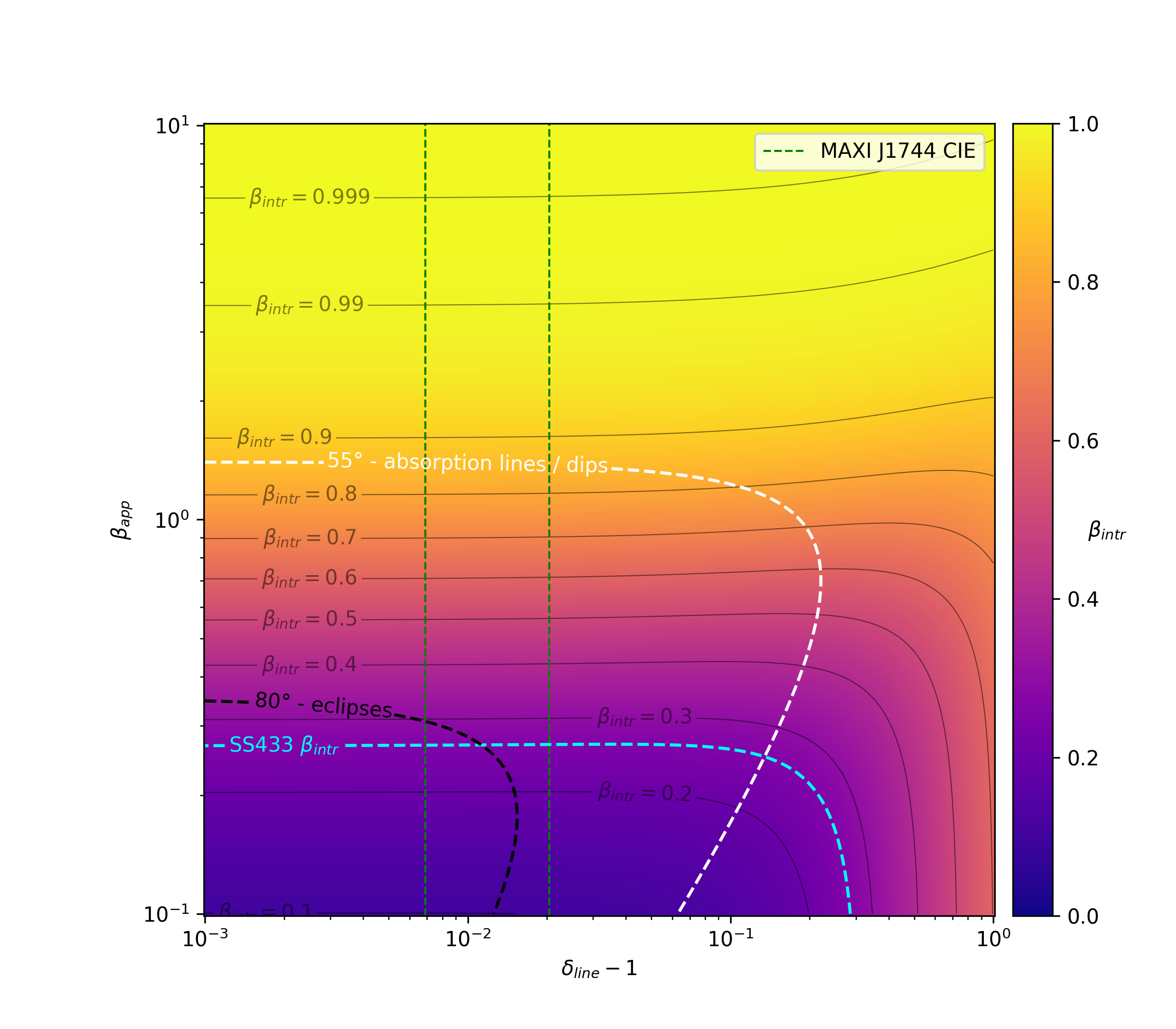}
\vspace{-1.em}
    \caption{Jet inclination angle \textbf{(left)} and intrinsic velocity \textbf{(right)} as a function of two observables: the Doppler factor measured from the X-ray emission lines $\delta_{line}$, and transverse velocity of radio ejecta $\beta_{app}$, assuming a common physical origin. The x-axis shows $\delta_{line}-1$ to allow a logarithmic scale. We highlight canonical inclination thresholds for wind absorption lines and dips in white, and for eclipsing systems in black. The intrinsic velocity of SS 433's jet is shown in cyan for reference. Vertical green dashed lines highlight the Doppler factors of the two blueshifted components in the CIE models of Section~\ref{sub:phys_colli}.}
    \label{fig:jet_par_2D}
\end{figure*}

We computed the allowed parameter space for $\beta_{intr}$  and $\theta$ for $1.001\leq\delta_{intr}\leq2$ and  $0.1\leq\beta_{app}\leq10$ (we recall that the \textit{apparent} transverse velocity can become larger than c for $\beta_{intr}\gtrsim0.7$ under the right inclinations). We show the results in Fig.~\ref{fig:jet_par_2D}. In both panels, we highlight several angle thresholds and the intrinsic jet velocity of SS 433 for reference, along with the $\delta_{line}$ values for the two main blueshifted CIE components derived for the "small" M1744 region in Fig.~\ref{fig:jet_par_2D}. We note that the third, marginally faster CIE component, proposed for the "big" M1744 region fit, would lead to identical interpretations. 

Following the $\beta_{intr}$ of SS 433 (cyan line) is a good way to visualize the influence of the angle on the observables: at high inclination angle (edge-on), the apparent transverse velocity is maximized, while the Doppler blueshift of the line is minimized (for a forward jet). The opposite happens for low inclination angle (face-on). In reality, the precession and nutation of the jets in SS 433 restricts $\theta$ to a range of high-inclination angles, strongly reducing the maximum $\delta_{line}$ to $\sim1.08$ \citep{Medvedev2019_SS433_X-rays,Shidatsu2025_SS433}. In the general case, a single $\delta_{line}$ only admits solutions for $\theta<\theta_{lim}=\mathrm{sin}^{-1}(\delta^{-1})$. Thus, in M1744, if a jet did create the blueshifted lines, then the inclination of the system can be constrained to $\lesssim 78^{\circ}$. This matches the lack of eclipses throughout the outburst, which already constrained the system to $\lesssim 80^{\circ}$. 

Now, aside from the limit case, which gives a single solution, every angle admits one "slow" and one "fast" solution due to the competing effects of $\gamma_{intr}$ and $1-\beta_{intr}\; \mathrm{cos}\;\theta$ in Eq.~\ref{eq:delta}. This can also impose constraints on our observational case: namely, for the small $\delta_{line}$ observed across all our components, lower inclination angles will almost always have a "slow" solution with $\beta_{intr}<0.1$. Such jet velocities are lower than any measurements for BH and NS-XRBs (see e.g. \citealt{Mirabel1999_XRB_jets,Carotenuto2024_jet_BHXRB}) and can thus be excluded. The remaining "fast" solution has a monotonic link between $\theta$ and both $\beta_{intr}$ and $\beta_{app}$, allowing us to divide two classes of solutions: if $\theta\gtrsim60^\circ$, then $\beta_{app}\lesssim1$, and the source is likely to be a Neutron Star (see \citealt{Mirabel1999_XRB_jets}, although see \citealt{Stephens2026_ScoX-1_fast_jet} for a sign of potential fast jets in a NS). Meanwhile, if $\theta\lesssim60^\circ$, then $\beta_{app}\gtrsim1$, and the source is likely to be a Black Hole. The remaining high inclination interval with "slow" solutions above $\beta_{app}\sim0.1$, being also restricted to $\beta_{intr}\lesssim0.4$, gives the same Neutron Star interpretation. Consequently, the lack of dips or wind signatures in absorption, the low disk temperature, low polarization degree \citep{Marra2025_MAXIJ1744-294_IXPE}, and the timing properties, all of which favor a low-inclination source in M1744, strongly point towards a fast jet and a BH origin. Of course, this is only valid under the assumption that the jet is at the origin of these blueshifted lines. The measurement of transient ejecta can disprove this scenario, as low velocities, particularly below $\beta_{app}\sim0.3$, would require eclipsing inclinations that we can exclude. Meanwhile, a high $\beta_{app}$ would yield a very precise inclination measurement for this physical configuration, to be compared with other estimates. We note that the very low measured $\delta_{line}$ of our different components implies that they can be obtained with marginal variations of $\theta$ or $\beta_{app}$. This means that a very small opening angle along the jet, or a change in velocity due to variations in medium density (e.g., exiting a static atmosphere around the source), could be enough to create different X-ray components. The process must be at least partly different from SS 433, where velocity variations of up to $\sim0.05c$ are observed, but correspond to a gradual variation of a unique jet component over time. Another possibility is that different "clumps" of transient ejecta are stacked along the line of sight, each with distinct properties.

We recall, nonetheless, that the inner jet regions are only one of the possible regions where blueshifted high-temperature CIE emission is expected, and the comparison of SS 433 already provides a second scenario. Indeed, recent reports of spatially resolved X-ray spectroscopy within SS 433's nebula W50 have identified several jet knots with proper motions of at most a few 1000 km $^{-1}$ within the lobes of the Nebula \citep{Tsuji2025_SS433_W50_knots_X-ray}. Although the configuration of these knots and the true jet velocities at distances of several parsecs are very difficult to determine, the diversity of X-ray spectra across knots \citep{Kayama2025_SS43_knots}, along with indications of significantly slower jet velocities  \citep{Kayama2022_SS43_knots} and UHE emissions from the lobes \citep{LHAASO2025_UHE_microquasar} shows the possibility of another source of collisional ionization due to the interaction of the jet with the outer nebula. In SS 433, these emission sources appear spatially distinct only because the source is both very close to us and seen edge-on. If M1744 is seen face-on, then its jet could shock the edge of a neighboring nebula or equivalently dense structure in the ISM, and produce CIE profiles that would be very close to the angular position of the source. 

Several candidates for such an interaction are known to be in the vicinity. The first is the Sgr A East SNR, which is superposed to M1744 within our line of sight \citep{Mandel2026_ApJ}. We stress that the BH would not need to be at the origin of the SNR: the gravitational well of Sgr A* favors, in any case, a scenario where M1744 is inside Sgr A East, since the BH is also 18\arcsec{} away from the supermassive BH, equivalent to less than $\sim0.7$ pc of projected distance. However, a similar interaction with the SNR could also happen if M1744 was behind the remnant. A combined X-ray view at both high spectral and angular resolution, combined with an independent measurement of the jet inclination, could help constrain this scenario. The main limit remains the distance between the source and the possible edge of the nebula, which would need to be at most $\sim0.1-0.2$ pc, to allow a jet to reach the nebula in the span of a few months since the start of M1744's outburst. To circumvent this, another candidate is the dense circumnuclear disk around Sgr A* (see e.g. \citealt{Hsieh2021_circumnuclear_disk}). Its south-east edge is coincident with M1744, and should sit much closer to Sgr A* (and thus M1744) along the axis of the line of sight. Although interactions between a jet and this structure are difficult to quantify, we note that this configuration would naturally explain the lack of a redshifted feature due to a single collision front. 

Finally, another possibility is that a remnant of core jet emission, relativistic or not, leads to this blueshifted CIE emission. High-cadence radio monitoring has recently shown that an optically thin core component may remain active along the soft-intermediate state for at least some BH-XRBs \citep{Hughes2025_radio_1,Hughes2025_radio_2,Zhang2026_BHXRB_wind-jet_4U1630}, but its origin remains for now unexplained. Although such a hypothesis could be tested with a more comprehensive sampling of low-inclined soft state BH-XRBs, it will be difficult to discriminate from a face-on wind, unless its properties can be tied to those of the radio emission.

\subsection{attribution of the atypical line components}\label{sub:line_attrib}

In the previous sections, we have followed the assumption that the many additional emission residuals found separately from the main Fe XXV K$\alpha$ and Fe XXVI K$\beta$ transitions were exclusively from blueshifted transitions of the nearmost iron complex. Although these assumptions cannot be fully verified, they can be considered "reasonable" given our atomistic and physical expectations for the system.

First, the choice of linking all components to the main iron transitions stems from the current state of atomistic databases and abundance estimates. For instance, the latest version of NIST, as of the writing of this paper (v 5.12), lists 19 possible transitions between 6.705 and 6.900 keV when considering all possible elements. They can be sorted between weak transitions of iron and transitions of any other elements. For the former, all listed transitions are dielectric recombination satellites of \Fexxvi{}\footnote{these transitions result from a complex double ion configuration and are thus sometimes identified as a \Fexxv{} transition, for instance in NIST and AtomDB}. Although some of them could be compatible with our detections -notably one satellite feature at 6.845 keV-, they remain far too weak. Indeed, the corresponding theoretical computations cited in NIST \citep{Dubau1981_satellite_FeXXVI} feature an analytical formula to derive the evolution of the intensity ratios between the satellite and main lines with temperature for N $\lesssim10^{17}$ cm$^{-3}$ (which is a very reasonable assumption in our case). The satellite intensity ratios are proportional to $1/T$ at first order, and become completely negligible for temperatures of a few keV. This rules out this configuration for collisional ionization, but not necessarily for photoionization, where the plasma temperature is proportionally much lower.
In addition, the ratios between the satellites themselves are almost independent of the temperature, and several satellite features are more than an order of magnitude stronger than the one at 6.846 keV. Since neither of these two elements is seen in our data, this completely rules out this satellite as a possible origin for the "blue 2" component, outside of uncharted physical conditions that would completely change the line ratios compared to \cite{Dubau1981_satellite_FeXXVI}. 
We note that the only reported study of these satellite lines in laboratory experiments is in \cite{Gu2012_diaelectronic_FeXXV_FeXXVI_EBIT}, and the 6.846 keV feature is neither seen nor explicitly mentioned, as it is far too weak to be detected. As the more recent efforts have, for now, focused on characterizing transitions up to \Fexxv{} \citep{Shah2025a_diaelectronic_FeXXV_EBIT}, there are no experimental confirmations of the strength of this line. To the best of our knowledge, the only claim of such detection in any extra-solar X-ray observations is a recent tentative detection of a very weak feature in a collisional plasma in a binary star \citep{Kurihara2025_DRlines}. In addition, few similar high-resolution studies of the sun highlight the presence of weak dielectric recombination \Fexxvi{} features in the sun, albeit only above $\sim6.90$ keV \citep{Watanabe2024_sun_X-ray_highres}. This physical process may thus be of weak relevance when compared to the parameter spaces of astrophysical plasmas. 

Meanwhile, the remaining non-Fe line complexes can be ruled out as they are all secondary transitions of rare elements (e.g., Vanadium, Cesium, Cerium). Their detection would require both extraordinary non-solar abundances and additional, much stronger detections at other energies where no lines are seen. Similar considerations can be used to rule out the presence of other lines above 6.97 keV. One notable candidate for the feature at 7.1 keV (which we conservatively attributed to edge residuals) is a partly ionized K$\beta$ emission feature. The spectral templates shown in \cite{Kallman2004_photo_iron} show that around log$\xi\sim0.2$, the Fe K line emission profile is dominated by transitions of \Fexi{} and \Fexii{}, with a K$\beta$ complex centered on 7.1keV. However, our lack of detection of a strong Fe K$\alpha$ component would require a Fe K$\alpha$/Fe K$\beta$ ratio much higher than what is possible physically, even with extreme obscuration conditions, as discussed in Section~\ref{sub:FeK_ratios}.

Even if we expect that all non-standard features are indeed from Fe lines, the diversity of transitions in the vicinity prevents any definitive interpretation. Our initial argument was to assume a highly ionized plasma. We then chose to minimize the "absolute" velocity shift of each component, which in practice means only blueshifts. 
This interpretation is reasonable with respect to a face-on wind origin, where no redshifted profiles should be detected since the entirety of the wind will be either static or directed towards us. We note that the additional components above 6.97 keV could be \Fexxv{} components with even higher blueshifts, which would allow the 7.00 keV line to be fit by both photoionization and collisional ionization. However, this would imply even higher velocities ($\sim-15000$\kmps{}), and thus an even more atypical configuration. In the same vein, relativistically blueshifted neutral Fe K$\alpha$ lines have not yet been unambiguously reported in astrophysical systems, but this may stem from the commonly adopted approach of attributing unknown transitions to the neighboring strong line. In this observation, we do not have the signal-to-noise to rule out the scenario of $\sim-16000--30000$ \kmps{} neutral Fe K$\alpha$ components.
Meanwhile, a combination of blueshift and redshifts is expected for a SS 433-like jet scenario, unless the source is face-on enough to hide the inner regions of its receding jet. However, as the luminosity of the jet emission scales with $\delta_{line}^4$, the two profiles at 6.74 and 6.84 keV, should they be redshifted, would be at least $\sim15\%$ weaker than their blueshifted counterparts. Our data show the opposite, at least for the features we detect, with the emission lines between 6.7 and 6.9 keV being much stronger than those above 6.97 keV, even when including the 7.1 keV residuals. However, we cannot exclude that stronger blueshifted features would exist beyond $\sim7.1-7.2$keV. A study of this more exotic scenario is made very difficult due to the lack of signal-to-noise ratio in that range, and would likely require additional observations.

\subsection{comparison with other high-resolution observations}\label{sub:compa_lit}

To search for alternative physical explanations and links to other sources, we compared our residuals with the literature. We first focus on other \xrism\ observations: for AGNs, the diversity of UFOs and blueshifted profiles has, for now, remained restricted to absorption, without meaningful constraints on the covering factor due to the inherent faintness of these sources. The only system with a report of highly ionized emission lines is NGC 7213 \citep{Kammoun2025_NGC7213}, and interestingly possesses both a static highly ionized phase and two unexplained (and statistically significant) narrow emission features, one of which at $\sim6.75$ keV. However, they could not assess whether these features were the consequence of inflows or outflows. In X-ray Binaries, two obscured BH systems have shown XRISM spectra dominated by highly ionized emission lines: GRS 1915+105 \citep{Miller2025_GRS1915} and V4641 Sgr \citep{Parra2025_V4641Sgr_XRISM}. While the former only had components with negligible velocity, part of the blueshifted profiles seen in M1744 are remarkably similar to some of the tentative components seen for V4641 Sgr, which included both a $\sim-1300$\kmps{} \Fexxvi{} K$\alpha$ line and weaker residuals around 6.8keV. However, V4641 Sgr was both heavily obscured and highly inclined, making it difficult to identify an outflow mechanism capable of producing the lines in both systems. Higher signal-to-noise-ratio observations for both systems will be required to draw clearer parallels. 

Among other binaries, GX 340+0 is by far the most similar to our present case, as it features an emission line-dominated spectrum, which was interpreted as a combination of relativistic reflection and highly ionized zones, in both absorption and emission \citep{Chakraborty2025_GX340+0_XRISM}, two of which with a significant blueshift of $\sim-2700$ \kmps{}. While that spectrum clearly shows narrow residuals around the \Fexxv{} energies and around $\sim6.75$keV,  the lack of empirical modeling of individual profiles makes it difficult to compare with our results: since their absorption and emission zones self-compensate and affect multiple line features across the entire 6.4-7.0 keV range, the significance they report for each component cannot be directly translated to the significance of a single line, and it is difficult to assess the true nature of each feature. We could likely reproduce the feature at 6.74 keV with a similar combination of broad emission and absorption, but we do not see any sign of the \Caxx{} P-Cygni profile that prompted the introduction of an absorption component in GX 340+0. Part of the highly ionized features in M1744 may correspond to a more face-on (and thus without absorption) view of what is seen in that system, but we stress that our spectra do not show any signs of additional features around 6.8, 7.00, or 7.10 keV. We note that the obscured NS system Circinus X-1, which features either pure emission or P-cygni profiles for highly ionized ions depending on the phase of the system \citep{Tsujimoto2025_CirX-1}, does not feature blueshifted emission, or highly blueshifted contributions of any kind. 

Finally, to the best of our knowledge, only a handful of detections of blueshifted X-ray emission lines have been made before the era of microcalorimeters. For AGNs, tentative detections of emission lines with very high velocities have been reported for quasars, with a potential redshift in \cite{Yaqoob1998_emission_PKS0637} and a high blueshift in \cite{Yaqoob1999_emission_PKS2149}.
In X-ray Binaries, 4U 1630-47, which has been observed by \xrism\ and showed strong absorption profiles with non-negligible covering fraction \citep{Miller2025_4U1630-47}, also has a single claim of highly ionized baryonic ejections from blueshifted, highly ionized X-ray emission lines \citep{DiazTrigo2013_4U1630-47_baryonic_ejection} simultaneous with radio detections. This conclusion remains debated, in part due to the lack of any X-ray line detections during stronger follower radio flares \citep{Neilsen2014_4U1630-47_2012-13emjetdebate}, showing that this was at best a highly anomalous event. Direct comparisons with our results are limited, as this observation was obtained with the limited spectral resolution of \xmm{}, but we note that their spectral modeling shows even more ionized features, with no apparent contribution from \Fexxv{} K$\alpha$.

\subsection{Discrepancy with a diffuse emission origin}\label{sub:discu_diffuse_origin}

Although heavy spatial or time variability of the diffuse emission beyond what we tested in Paper I is heavily unlikely, as a sanity check, we tested how many of the residual features in the \xrism\ spectrum could be retrieved by rescaling some of the components of the diffuse emission, and how much rescaling would be needed, within the framework of the models currently available for the GC diffuse emission. We focused on the "small" M1744 region analysis because of its more detailed background modeling. For this, we simply performed a fit over the entire \xrism\ band, with additional constants $c$ left free to vary for each diffuse emission model. Moreover, due to large differences in ARFs, the AXJ absorption features would be significantly affected by switching the emission features from M1744 to a diffuse source, and would need to be refit. For simplicity, we thus ignored the 6.5-7.0 keV range where the main K$\alpha$ emission and absorption lines overlap for AXJ. 

We first tested rescaling the GCXE emission exclusively: this resulted in a normalization factor of $c_{GCXE}=3.7$, and an improvement of the fit statistic of $\Delta C$=115. We show the corresponding blind search for narrow features in the top panel of Fig. C.\ref{fig:resid+blind_rescale_BH} in App.~\ref{app:bkg_rescale}. Both panels show that, after rescaling, the GCXE lines qualitatively reproduce the static component. Meanwhile, since the GCXE model doesn't include any blueshifted component, none of the additional features between 6.7 and 7 keV are reproduced. We stress that an underestimation of the contribution of the GCXE by almost a factor 4 is a much higher discrepancy compared to any of the remaining systematic effects (DSH, PSF, non-uniformity of the GCXE) that may remain following our analysis, and the only possibility would thus be for the GCXE to have increased very significantly since 2024, and only at the position of M1744. Although we cannot fully rule out a spurious contribution by a background transient in our observation alone, the extensive monitoring presented in M26 clearly shows that the long-term variation of the iron lines was (expectedly) linked to the continuum of M1744 along its entire outburst, which heavily disfavors transient contamination.

In addition to the GCXE, we also tested rescaling the Sgr A East emission exclusively. This time, the normalization increased by a factor of 1.9, with a much more limited improvement of the fit statistic ($\Delta C$=61). This is because the Sgr A East model has a much higher FeXXV K$\alpha$ / FeXXVI K$\alpha$ ratio and a lower width, compared to the GCXE or the residuals in our spectrum, and thus the model cannot reproduce both lines at once, no matter its normalization. This is very apparent in the middle panel of Fig. C.\ref{fig:resid+blind_rescale_BH}. We do note that the spatial distribution of Sgr A East being very well known, along with its nature as an SNR, would in any case forbid any strong spatial or temporal evolution of its spectrum in the span of a single year.

Finally, to test a composite effect, we used a configuration in which both the GCXE and Sgr A East normalizations were left free to vary. As shown in the bottom panel of Fig. C.\ref{fig:resid+blind_rescale_BH}, this led to a very similar situation to a pure GCXE rescaling, with a GCXE normalization factor of 3.5, and Sgr a East normalization factor of 1.2. This is expected since any significant increase of Sgr A East would affect the FeXXV K$\alpha$ / FeXXVI K$\alpha$ ratio of the remaining residuals, making them impossible to fit with a rescaled GCXE model.

\section{Conclusions} \label{sec:conclu}

In this work, we presented a detailed analysis of the first \xrism\ spectrum of MAXI J1744-294 (M1744), a new BH-XRB candidate observed in March 2025 during its discovery outburst, building on the data analysis and background modeling presented in Parra et al. (submitted to ApJ, Paper I).
This paper focuses on the intrinsic spectral properties of the Black Hole, most particularly the various narrow emission lines arising from highly ionized gas and, in part, unexpected energies. The simultaneous coverage from \xmm{} and \nustar{} provides a complementary view to the broad-band X-ray continuum. Indeed, since we do not model the Dust Scattering Halo (DSH) around our sources with \xrism{}, and considering the difficulty of combining instruments with different angular resolutions and Spatial-Spectral Mixing (SSM), we split our analysis into 3 groups. 
\begin{enumerate}
\item First, the low-resolution, DSH corrected broadband continuum of \xmm{}-PN and \nustar{}, which we modeled with a simple Comptonized disk in Section~\ref{sub:empi_xmmnustar}, reveals a very low temperature at such an Eddington ratio, along with an extremely weak but firmly detected hard tail. Here, a single broad Gaussian emission line is sufficient to reproduce a strong emission feature around 6.9keV. This continuum is used as the basis for subsequent physical models, reducing bias from Dust Scattering and SSM. The contextualization of these observations within the entire monitoring campaign of the source, along with a more holistic approach to continuum modeling, can be found in M26. 
\item Secondly, the non-DSH corrected soft X-ray "low-resolution" Xtend spectrum, modeled in Section~\ref{sub:empi_xtend}. Its view of the continuum is expectedly very similar to the previous analysis, but its better spectral resolution unambiguously requires two separate narrow mission lines to fit the large feature seen at $\sim$6.7-6.9 keV in the previous spectra. The better absolute energy calibration of Xtend already shows that these lines are blueshifted, but the extent of the blueshift remains uncertain due to the instrument's limited spectral resolution. Two additional features are observed below 2 keV and can be equally well represented by emission lines or absorption edges, whose origin remains uncertain. The former scenario implies an extremely high equivalent width and is thus disfavored. The second may be explained by non-solar abundances and/or a complex ISM medium with dust and hot absorbers, but the uncharacteristic energy of one of the edges remains puzzling.
\item Thirdly, the non-DSH corrected Resolve spectrum, modeled in Section~\ref{sub:empirical_resolve}, and for which we must model the MAXI J1744-294 (Black Hole), AX J1745.6-2901 (Neutron Star), and diffuse emission contributions simultaneously due to their overlap in the Resolve regions. After subtracting these other contributions, several narrow emission features remain in the M1744 spectra, whose consistency across our background subtraction methodologies and region choices ensures they can be attributed unambiguously to M1744, as we later quantitatively confirm in Section~\ref{sub:discu_diffuse_origin}. These lines can be separated between a very weak neutral iron (Fe I) K line, narrow ($\sim5-15$eV) static components from \Fexxv{} and \Fexxvi{} K$\alpha$, and three narrow features at 6.74, $\sim 6.84$ (which can also be fit with two narrower components depending on our region choice), and 7.00 keV. All lines are significant at $\gtrsim3\sigma$ confidence after MC simulations, and none of the non-standard energies match strong or expected transitions. The most straightforward scenario is thus that these lines are highly Doppler-shifted versions of the nearby, highly ionized Fe lines. To minimize the velocity shifts required, we tentatively associate them with blueshifted \Fexxv{} and \Fexxvi{} K$\alpha$ features with blueshifts of $\sim-2000$, $\sim-5500$ (or $\sim-3500$ and $\sim-6000$), and $-1300$ \kmps{} respectively. 
\end{enumerate}

While the detection of static highly ionized emission features is already a first for non-obscured BH-XRB systems, the high velocity features (whose attribution is discussed in Section~\ref{sub:line_attrib}) are even more atypical. We thus attempted to model the different highly ionized features with photoionization (Section~\ref{sub:phys_photo}) and collisional ionization (Section~\ref{sub:phys_colli}) models. We restrict ourselves to the single combination of velocities highlighted above: in that case, both types of ionization can reproduce the majority of the line features, with the notable exception of the $-1300$\kmps{} \Fexxvi{} K$\alpha$ component, due to the lack of \Fexxv{} counterpart. For photoionization, most components require particularly high ionization parameters due to a soft ionizing spectrum and the lack of line features from lower-ionization ions. For collisional ionization, the fit requires less extreme parameters, but the non-negligible continuum contribution significantly worsens the high-energy fit with \nustar{}. Unless a more complex continuum (e.g., including reflection or returning radiation) can alleviate this tension despite the very soft spectral state, this disfavors the theory that all ionized phases arise from collisional ionization. A hybrid configuration with some phases from photoionized emission and others from collisional emission may be investigated with further broadband spectral analysis. 

To explain the physically abnormal Fe I K$\beta$/K$\alpha$ line flux ratios seen in the data, in Section~\ref{sub:FeK_ratios}, we explored the physical parameters of neutral iron emitters. After deriving their possible parameter range using MYTORUS, we find that with the introduction of additional K$\beta$ lines, which are not significantly detected in the diffuse emission spectra, using physically motivated line ratios, we still obtain an intrinsic Fe I K$\alpha$ line with expected line ratio values in the M1744 spectrum. Remaining weak 7.06 keV residuals may trace an additional, weaker Fe XXVI K$\alpha$ component at -3500\kmps{}. 

We discuss the origin of the highly ionized phases as winds or jets in Section~\ref{sub:discu_origin}. For winds, emission features are expected for face-on spectra, but physical models have yet to focus on this inclination range. The presence of highly blueshifted phases would require magnetic driving, but within an inhomogeneous medium, which is very different from the continuous magnetic wind prescriptions historically developed. For jets, there is already precedent for highly shifted X-ray emission lines in SS 433. With M1744 being in a particularly soft state during the observation, the presence of transient ejecta in radio should be investigated. With relativistic computations, we show that a transient ejecta seen in radio and X-ray would allow us to derive unambiguously the intrinsic lorentz factor of the jet and the inclination of the system. Moreover, the doppler shifts of the X-ray lines link by themselves the inclination angle and intrinsic Lorentz factor of the source, and imply that if the source is indeed low-inclination, as expected from the bulk of its X-ray properties, then a jet origin would require the high Lorentz factors ($\gamma>1$) historically associated with BHs.

For now, the limited signal-to-noise ratio of each abnormal emission component, combined with a lack of dynamical information about M1744, prevents any definitive conclusion on the physical origin of these features. In the future, a large number of observations of BH transients with microcalorimeters, sampling a wide range of accretion states and inclinations, will be the key to determining whether the lines seen in this observation are a normal characteristic of soft state BH-XRBs, or the telltale of an anomalous system that does not represent the behavior of the typical outbursting BH population.

\begin{acknowledgments}
We thank the XRISM operation team for accepting our DDT proposal and conducting the observation, along with the XRISM Science Data Center, help desk, and calibration teams for their continued assistance. This research has made use of software provided by the High Energy Astrophysics Science Archive Research 
Center (HEASARC), which is a service of the Astrophysics Science Division at NASA/GSFC. MP acknowledges support from the JSPS Postdoctoral Fellowship for Research in Japan, grant number P24712, as well as the JSPS Grants-in-Aid for Scientific Research-KAKENHI, grant number J24KF0244.
Support for SM, KM and the Columbia University team was provided by \nustar\ AO-10 (80NSSC25K0653), \chandra\ AO-26 (SAO GO5-26016X) and \xmm\ AO-23  (80NSSC25K0651) programs. 
SM acknowledges support by the National Science Foundation Graduate Research Fellowship under Grant No. DGE 2036197 and the Columbia University Provost Fellows Program. 
Part of this work was financially supported by Grants-in-Aid for Scientific Research 19K14762, 23K03459, 24H01812 (MS) from the Ministry of Education, Culture, Sports, Science and Technology (MEXT) of Japan. TY acknowledges support by NASA under award number 80GSFC24M0006.
\end{acknowledgments}




%
\facilities{\xrism{} (Resolve and Xtend), \xmm{}(EPIC), \nustar{}}


\bibliography{ref,bibliocompl}{}
\bibliographystyle{aasjournalv7}

\appendix

\section{Empirical modeling}

\subsection{Xtend}

\subsubsection{empirical model description}\label{app:xtend_fit}
\begin{table*}[h!]
\centering
\caption{Model parameters for the empirical MAXI J1744-294 Xtend fit in the DDT observation. For the unknown lines -which can also be represented with edges, see below-, we showcase their velocities with respect to the fluorescence K$\alpha$ lines of Aluminum and Silicon.}
\label{tab:comp_param_empi_xtend}
\begin{tabular}{lccccccc}
\cline{1-8}
\multicolumn{8}{c}{Continuum} \T\B\\
Parameter & $N_H$ & $\Gamma$ & $kT_{in}$ & $f_{cov}$ & $kT_{e}$ &F$_{abs,2-10}$$^\star$ & C-stat/d.o.f.$^\star$ \T\B\\
Unit  & 10$^{22}$ cm$^{-2}$ & &  & 10$^{-3}$ & & 10$^{-11}$ cgs & \T\B\\
\cline{1-8}
Value &$18.4\pm 0.2$ & 2.74$^\dagger$ & $0.62_{-0.01}^{+0.01}$ & $11.2_{-1.4}^{+1.4}$ &150$^\dagger$ & $88_{-2}^{+12}$ & 123/89\T\B\\
\cline{1-8}
\cline{1-8}
\multicolumn{8}{c}{Spectral lines}\T\B\\
\multicolumn{2}{c}{line complex}  & $E_{\rm rest}$ (keV) & $v_{raw}$ (\kmps{})& $\sigma$ (eV) & norm ($10^{-5}$) & EW (eV) & $\Delta$ C-stat \T\B \\
\cline{1-8}
\multicolumn{2}{c}{Al K$\alpha$ calibration ?} & 1.4865$^\dagger$ & $3152_{-5035}^{+5045}$ & 0$^\dagger$ & $(8.5_{-4.8}^{+6.6})\times10^{5}$ & / & 14 \T\B \\
\multicolumn{2}{c}{ Si K$\alpha$ calibration ?} & 1.7397$^\dagger$ &  $-5801_{-3369}^{+3369}$ & 0$^\dagger$ & $(0.24_{-
0.24}^{+0.76\dagger})\times10^5$ & / & 38 \T\B \\
\multicolumn{2}{c}{\Fexxv{} He$\alpha$} & 6.7$^\dagger$ & $-1792_{-1857}^{+1793}$ & 0$^\dagger$ & $25_{-8}^{+10}$ & $88_{-46}^{+62}$ & 28 \T\B \\
\multicolumn{2}{c}{\Fexxvi{} Ly$\alpha$} & 6.97$^\dagger$ & $-1821_{-2383}^{+2352}$ & 0$^\dagger$ & $14_{-6.8}^{+8}$ & $65_{-65}^{+62}$ & 12\T\B \\
\cline{1-8}
\cline{1-8}
\end{tabular}\\
\raggedright
$\star$ computed from the full model, including the line components.
$\dagger$ frozen or at the limit of the parameter space. 
\end{table*}

\subsubsection{origin of the low-energy features}\label{app:xtend_lowe}

\begin{figure*}[t!]
\centering
\vspace{-2.5em}
    \includegraphics[clip,trim=0cm 0cm 0cm 0.8cm,width=0.95\textwidth]{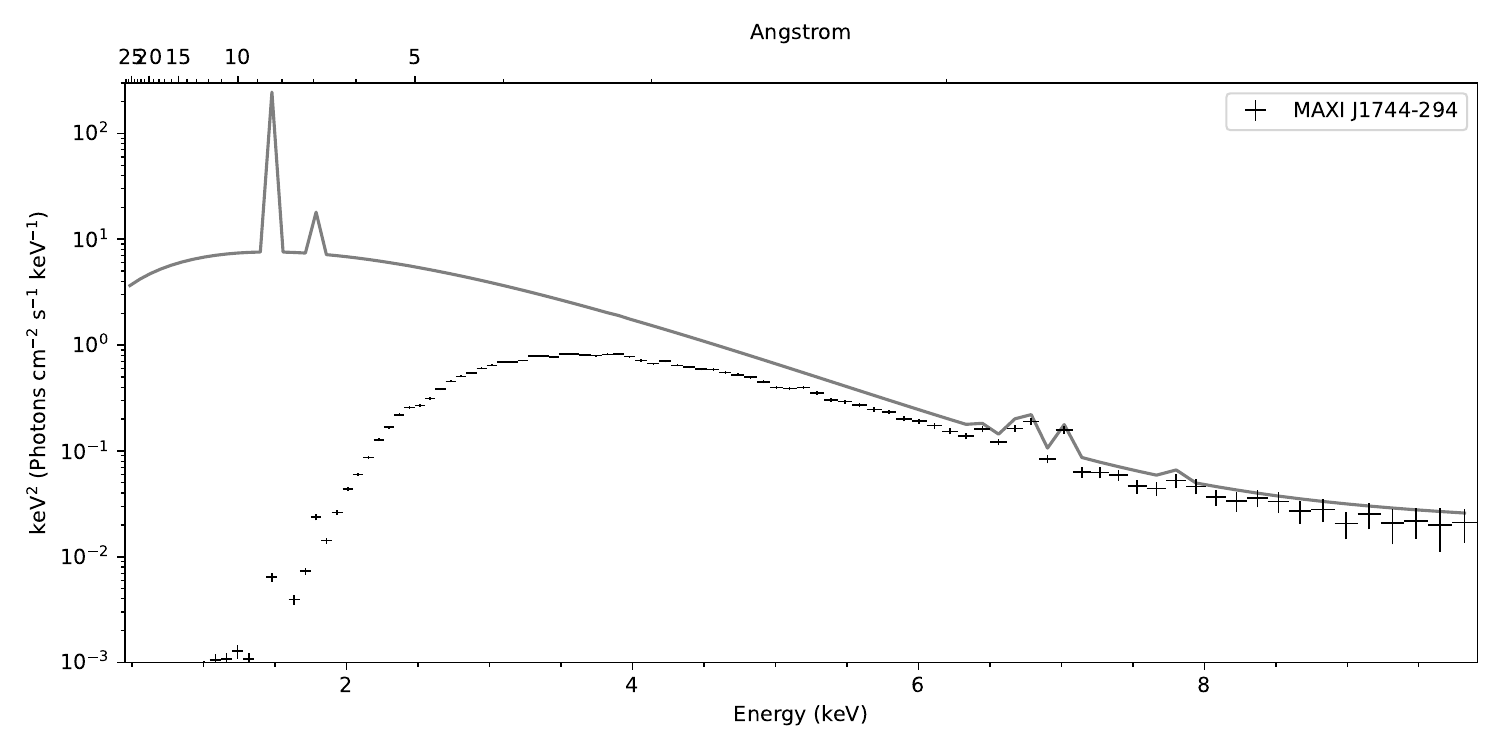}
    \caption{Unfolded spectrum and unabsorbed MAXI J1744-294 spectrum, following the Xtend fit including the two abnormal low-energy lines.}
    \label{fig:empi_Xtend_eeuf}
\end{figure*}

Here, we investigate the origin of the low-energy Xtend emission features seen in Fig.~\ref{fig:xtend_empi_resid}.
The most straightforward transitions are the fluorescence K$\alpha$ lines of Aluminum (1.49 keV) and Silicon (1.74 keV), or the H-like K$\alpha$ lines from Magnesium (averaged at 1.49 keV) and Aluminum (1.72 keV). Although the second line would appear significantly blueshifted, all of these lines are compatible with being at rest with their respective transition at 1 $\sigma$, when considering the limited spectral resolution and signal-to-noise of Xtend. However, any of these lines would have to be extremely strong to be detected close to the peak of the disk component, and despite an absorption column of $\mathrm{N}_\mathrm{H}\sim1.7\times10^{23} \; \mathrm{cm}^{-2}$. This can be seen in the unfolded spectrum of the unabsorbed model, which we show in Fig. A.\ref{fig:empi_Xtend_eeuf}, and reflects in the EW of the lines: 2.4 keV for the 1.5 keV feature, and 100 eV for the one at 1.8 keV. This magnitude of emission lines has never been seen in this category of sources, even in heavily obscured, super-Eddington accretors such as V404 Cyg \citep{King2015_GS2023+338_wind_x}, making this interpretation extremely unlikely. We further note that both the AXJ static background (shown in Fig. A.\ref{fig:xtend_empi_resid}) and the local diffuse emission (whose analysis was presented in detail in Paper I) are several orders of magnitude too weak to explain any of the low-energy features.

\begin{figure*}[t!]
    \includegraphics[clip,trim=4.5cm 2cm 5.4cm 1.5cm,width=0.5\textwidth]{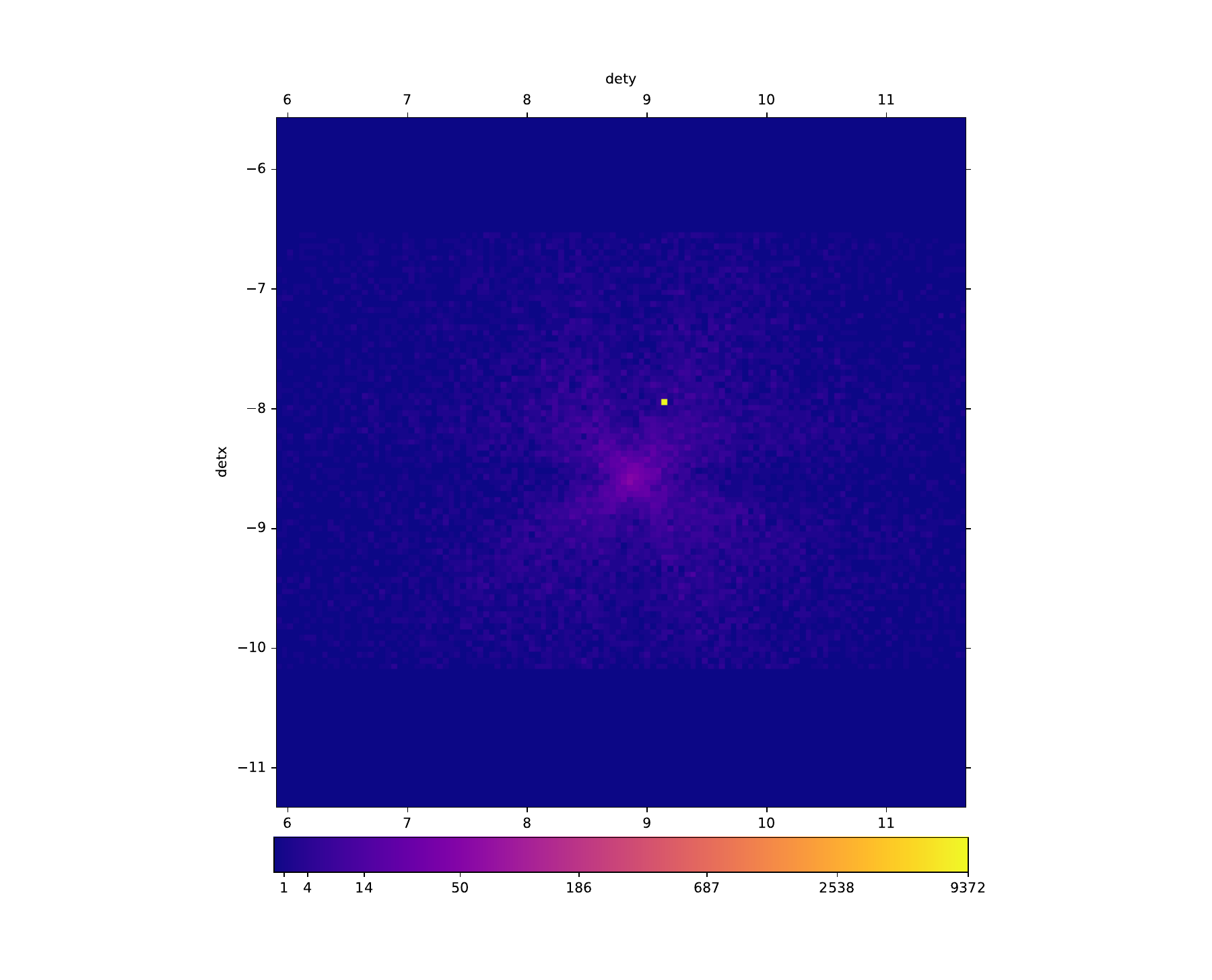}
    \includegraphics[clip,trim=4.5cm 2cm 5.4cm 1.5cm,width=0.5\textwidth]{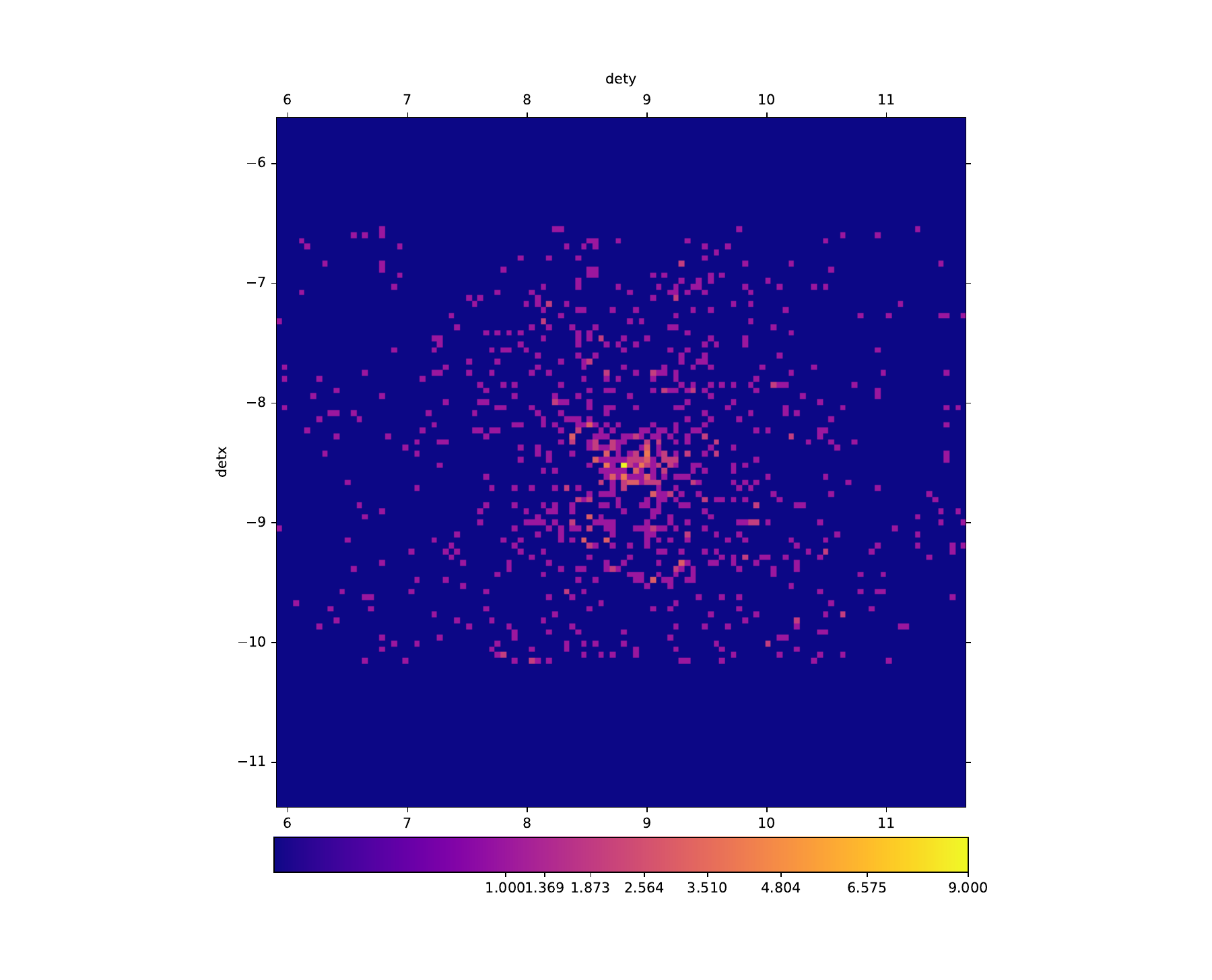}
    \caption{2.35-2.5 keV band \textbf{(left)} and 1.4-1.55 keV band \textbf{(right)} Xtend detector images for the DDT observation, centered on MAXI J1744-294 and computed prior to the SFP filtering.}
    \label{fig:Xtend_SFP_images}
\end{figure*}

We thus investigate in more detail an origin independent from M1744. The first possibility is that the emission lines are due to the non-X-ray background (NXB) or flickering pixels. However, our background subtraction should get rid of any spatially constant NXB, and the current NXB model threads\footnote{\href{https://heasarc.gsfc.nasa.gov/docs/xrism/analysis/nxb/xtend_nxb_db.html}{https://heasarc.gsfc.nasa.gov/docs/xrism/analysis/nxb/xtend\_nxb\_db.html}} do not mention any NXB line below 2 keV. We have also filtered for flickering pixels during the Xtend analysis detailed in Paper I. To better assess the origin of the line and compare it to potential flickering pixel contamination, we thus computed Xtend images before the flickering pixel filtering, both in the 2.35-2.5 keV band, where the anomalous pixel's emission peaks, and in the 1.4-1.55 keV band, to isolate the strongest of our two unknown lines. We compute both images in detector coordinates to avoid spreading the event counts across several pixels after the coordinate conversion, and show them in Fig. A.\ref{fig:Xtend_SFP_images}. The left panel highlights that a single anomalous pixel is present in the outer edge of the M1744 PSF, but comparisons with the Xtend images presented in Paper I and computed from cleaned event files confirm that this pixel was correctly removed, and in any case, outside of our source region. Meanwhile, the right panel confirms that the count rates at the energy of the 1.5 keV line originate from a standard distribution, peaking at the center of the M1744 PSF. 

\begin{figure*}[h!]
    \includegraphics[clip,width=0.5\textwidth]{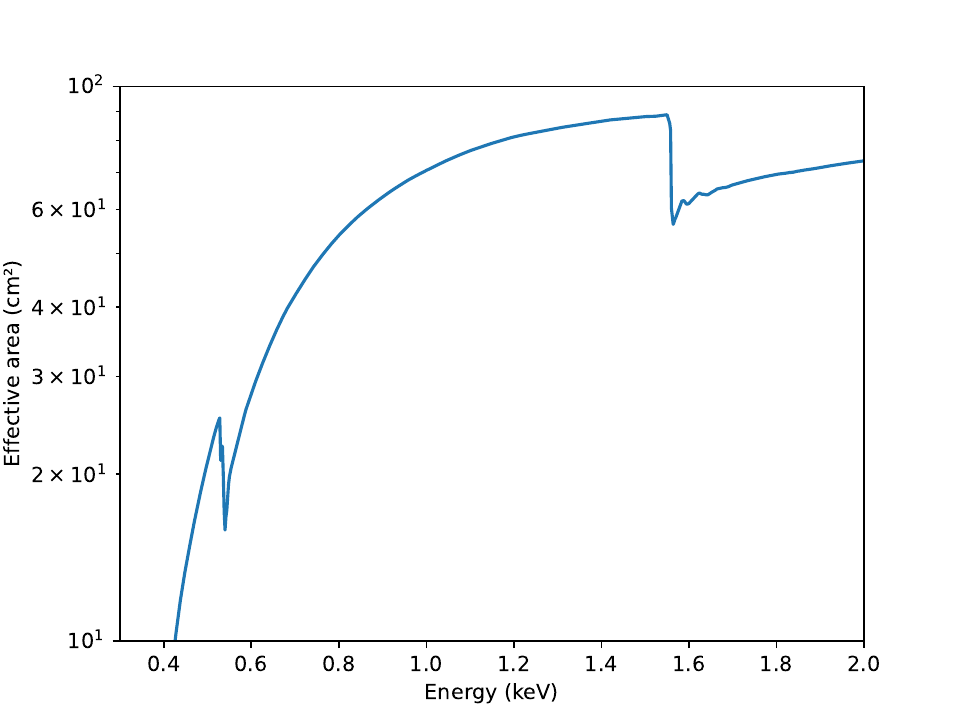}
    \caption{$0.3-2.0$ keV zoom of the Xtend ARF, computed for the MAXI J1744-294 source region in the DDT observation, using Heasoft 6.35.1 and \xrism\ CALDB11}
    \label{fig:Xtend_arf_zoom}
\end{figure*}

Another scenario involves residuals from edges in the instrument calibration. We thus plot the ARF of the instrument computed for this observation in Fig.~\ref{fig:Xtend_arf_zoom}. Since a strong Aluminum K$\alpha$ edge is already present around 1.5keV, an imperfect consideration of that edge could be at the origin of the first emission line seen in our dataset. We thus tested replacing the 1.5 keV emission line with an edge, with its energy left free to vary within the relevant region. Although the artificial edge perfectly reproduces the line, its best-fit energy is very well constrained around 1.50 keV, and remains at 4$\sigma$ from the theoretical edge energy ($1.559$ keV) in the fit, but could be interpreted as a part of a more complex edge profile, which was not fully considered in Xtend's calibration. Meanwhile, if no strong edges are present around 1.83keV in the ARF, where our second line is detected, there might be uncertainties due to a possible gain discontinuity near the Si K$\alpha$ edge at 1.839 keV, similar to Suzaku XIS \citep{Koyama2007_Suzaku_XIS}. Introducing this second edge into the fit can reproduce the higher-energy line feature. It is thus possible for at least one of the two lines to have an instrumental origin. 
We note that additional changes to the Xtend energy calibration were introduced in CALDB 12\footnote{\href{https://heasarc.gsfc.nasa.gov/docs/xrism/calib/releases/XRISM\_Xtend\_CalDB\_12\_Release\_Note.txt} {https://heasarc.gsfc.nasa.gov/docs/xrism/calib/releases/XRISM\_Xtend\_CalDB\_12\_Release\_Note.txt}} and may have an influence on these emission lines.

A final possibility is for these edges to be due to ISM absorption, for instance, due to non-standard astrophysical edge profiles, arising from non-solar abundances, dust, or hot ISM phases in the GC region. This is already strongly hinted at by another edge residual, seen with a secondary feature at 2.45 keV in the Xtend residuals \ref{fig:xtend_empi_resid}, and much more apparent in the Resolve spectrum (see Section~\ref{sub:empirical_resolve}). However, the lack of counterparts in the HETG and RGS observations, presented in the dedicated ISM analysis of Paper III on our observations, strongly disfavors interpreting these features as ISM edges.

\clearpage
\subsection{Resolve}

\subsubsection{Spectral fitting results}\label{app:resolve_fits}

\begin{figure*}[h!]
\centering
        \includegraphics[clip,trim=0.1cm 0cm 0.36cm 0.3cm,width=0.318\textwidth]{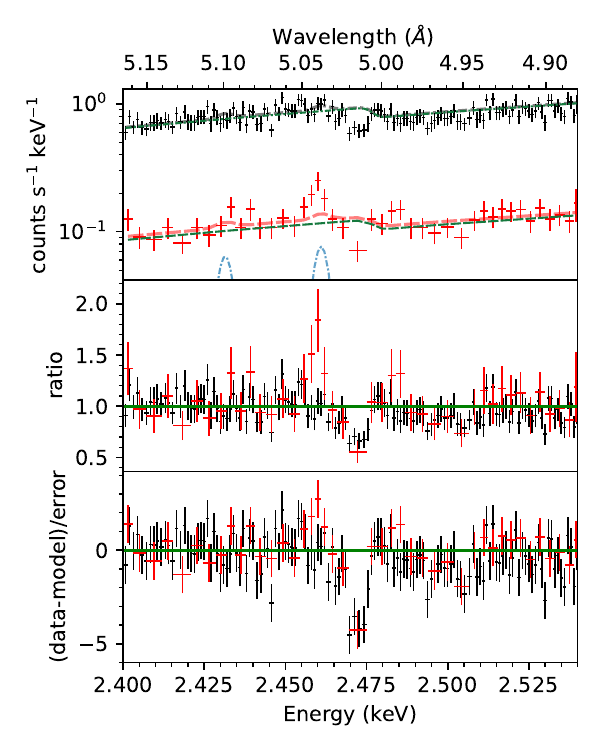}
    \includegraphics[clip,trim=1.85cm 0cm 0.4cm 0.3cm,width=0.426\textwidth]{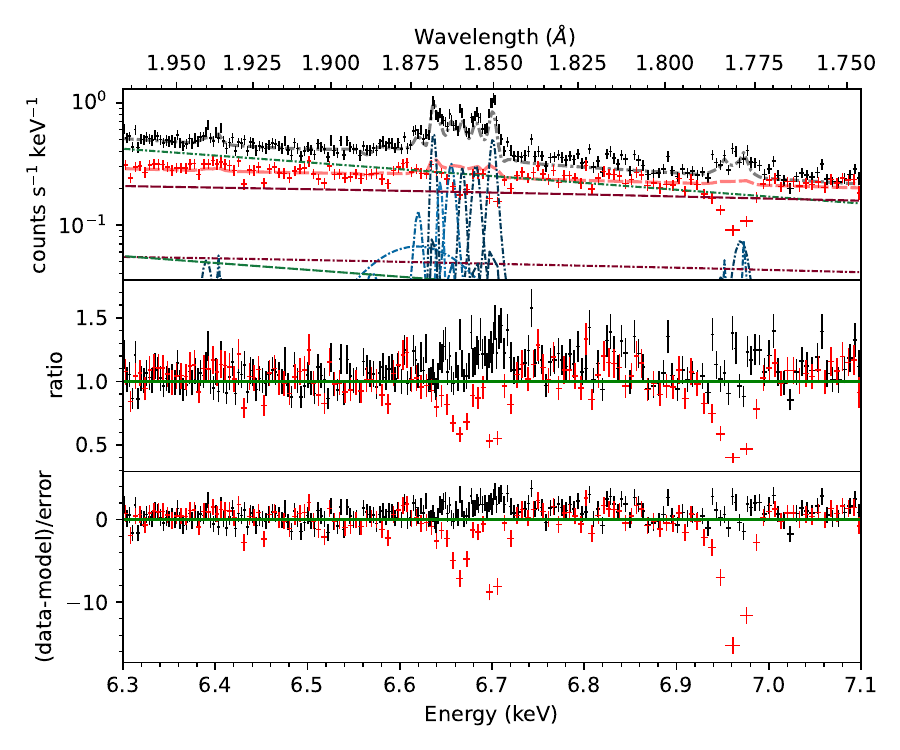}
    \includegraphics[clip,trim=2.55cm 0cm 0.3cm 0.3cm,width=0.2397\textwidth]{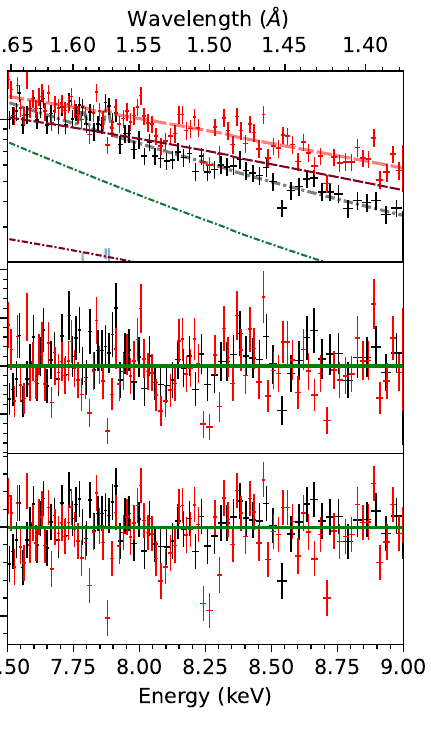}
\vspace{-1.em}
    \caption{Zooms of the Resolve spectra, ratio, and residuals for the "big" \src{} region and \axj{} region in the DDT observation, after the first step of their common continuum modeling, focusing on regions with narrow features. The spectra are visually rebinned in each panel to highlight the different lines, and model components are always shown at a 3$\sigma$ significance level.}
    \label{fig:resolve_bigpix_empi_resid_conti_preline_zoom}
\end{figure*}

\begin{figure*}[h!]
\centering
    \includegraphics[clip,trim=0.1cm 0cm 0.36cm 0.3cm,width=0.318\textwidth]{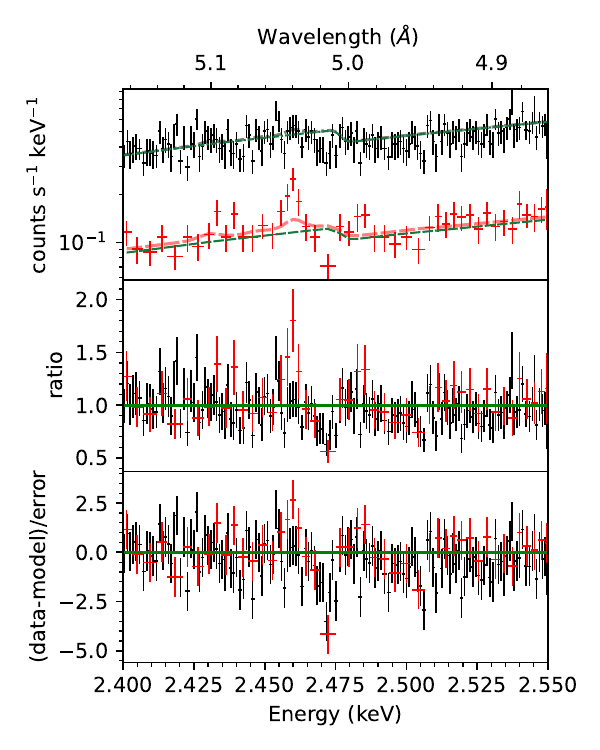}
    \includegraphics[clip,trim=1.85cm 0cm 0.4cm 0.3cm,width=0.426\textwidth]{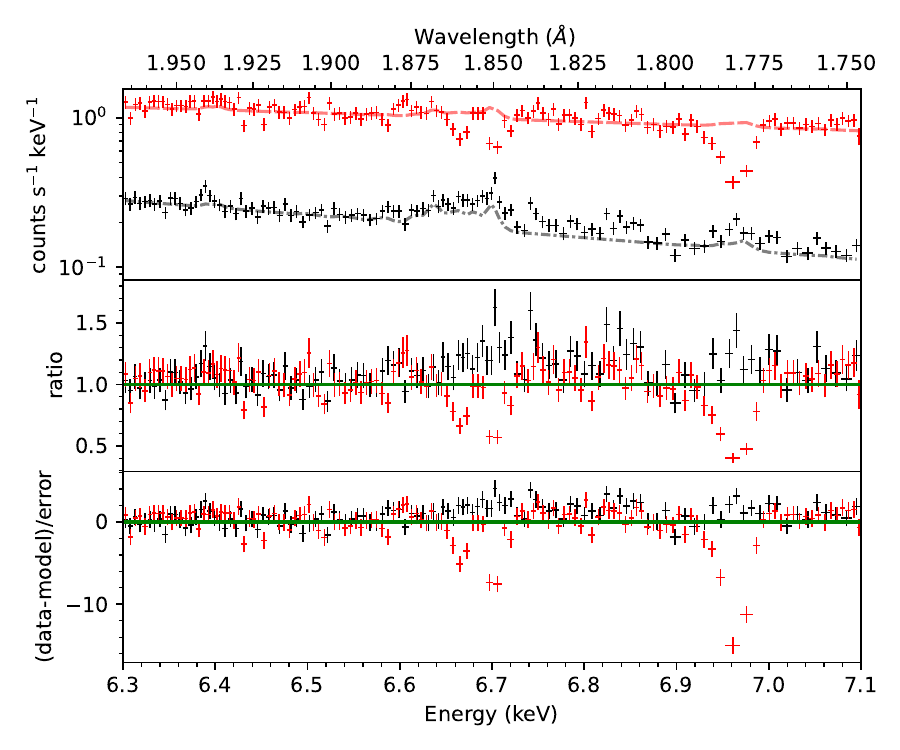}
    \includegraphics[clip,trim=2.cm 0cm 0.3cm 0.3cm,width=0.241\textwidth]{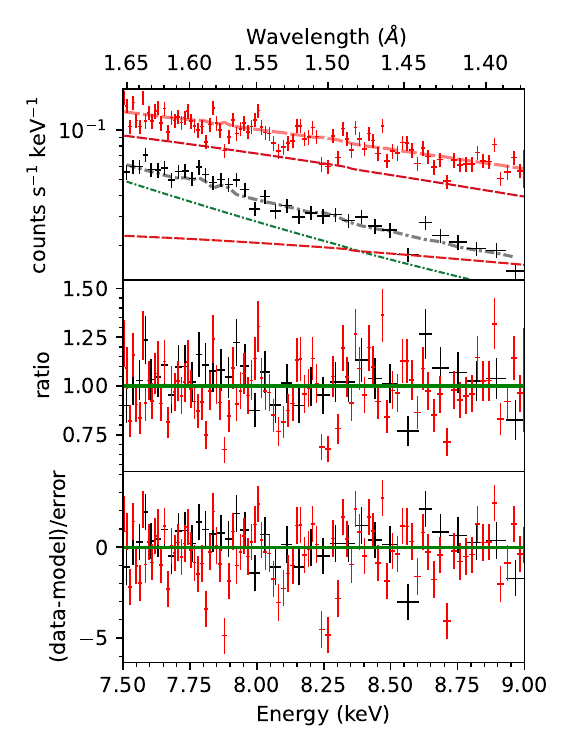}
\vspace{-1.em}
    \caption{Zooms of the Resolve spectra, ratio, and residuals for the "small" \src{} region and \axj{} region in the DDT observation, after the first step of their common continuum modeling, focusing on regions with narrow features. The spectra are visually rebinned independently in each panel to highlight the different lines, and model components are always shown at a 3$\sigma$ significance level. As the M7144 and AXJ count rates overlap around 6 keV, the count rate of AXJ is multiplied by 4 in the middle panel.}
    \label{fig:resolve_smallpix_empi_resid_conti_preline_zoom}
\end{figure*}

\begin{figure*}[t!]
\centering
    \includegraphics[clip,trim=0cm 0.3cm 0cm 0cm,width=0.49\textwidth]{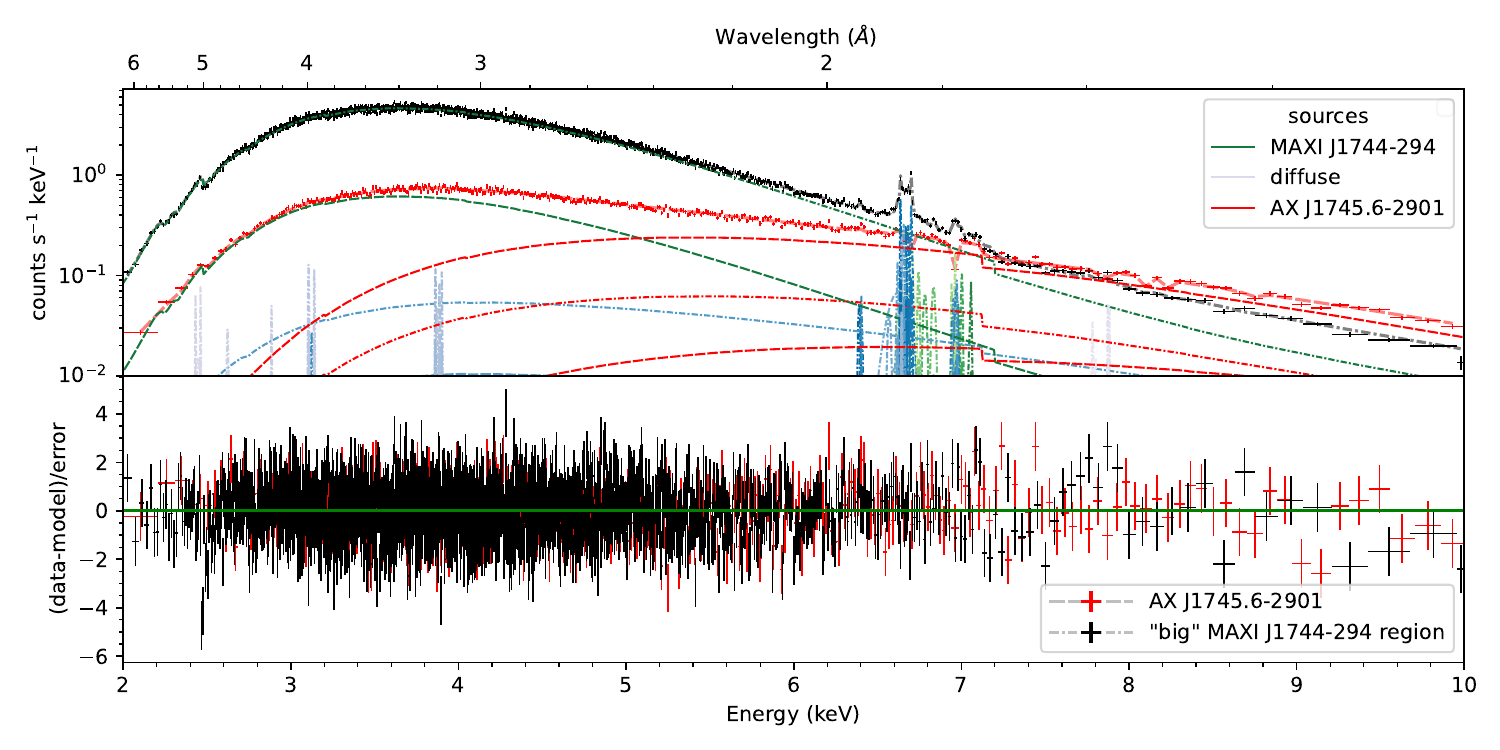}
    \includegraphics[clip,trim=0cm 0.3cm 0cm 0cm,width=0.49\textwidth]{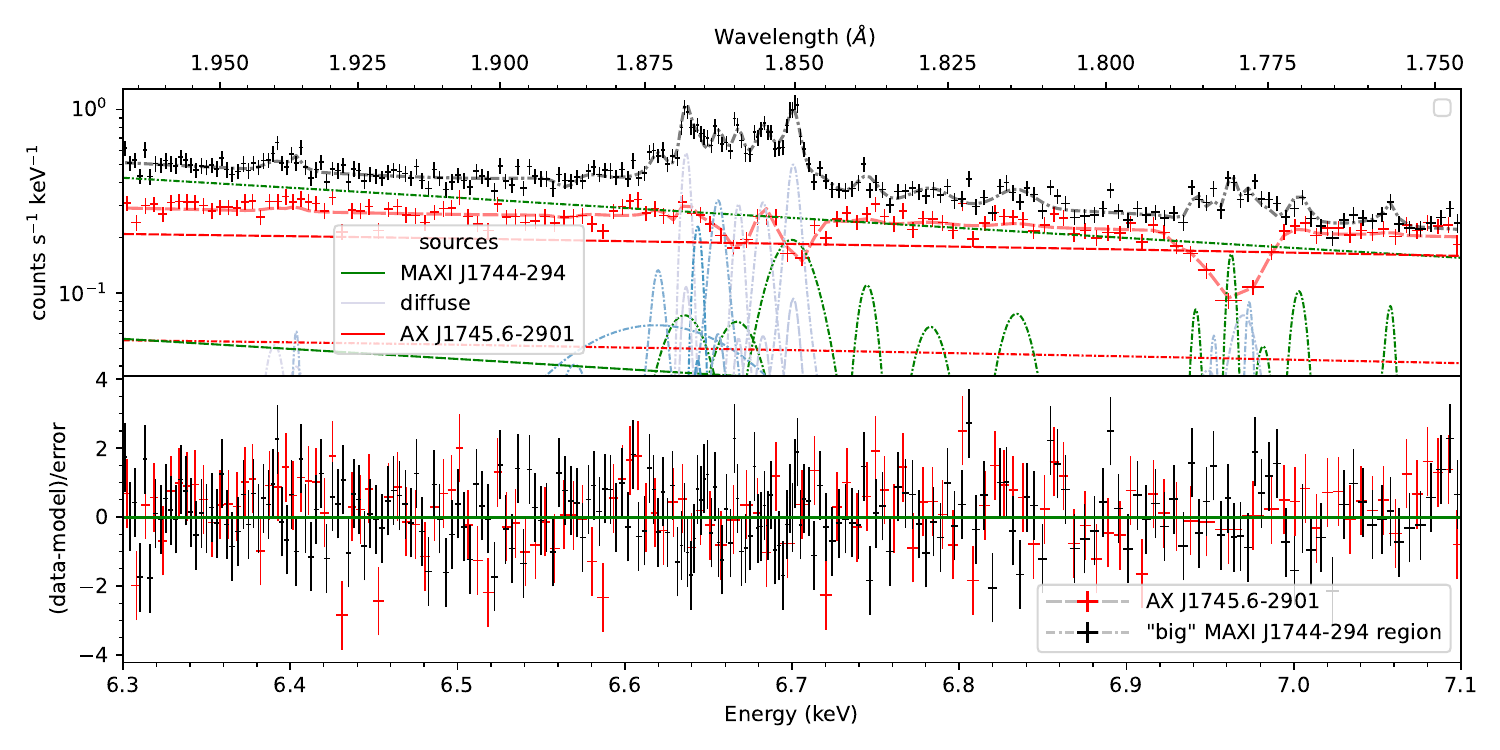}
\vspace{-1.em}
    \caption{Full Resolve residuals \textbf{(left)} and zoom in the 6.3-7.1 keV band \textbf{(right)} for the "big" \src{} region and \axj{} region in the DDT observation, after empirical modeling and having added all significant line features. For readability, the spectra are visually rebinned at a 20$\sigma$ and 10 $\sigma$ significance level in the left and right panel respectively, and model components are always shown at a 3$\sigma$ significance level.}
    \label{fig:resolve_bigpix_empi_resid_conti_postline}
\end{figure*}

\begin{figure*}[t!]
\centering
    \includegraphics[clip,trim=0cm 0.3cm 0cm 0cm,width=0.49\textwidth]{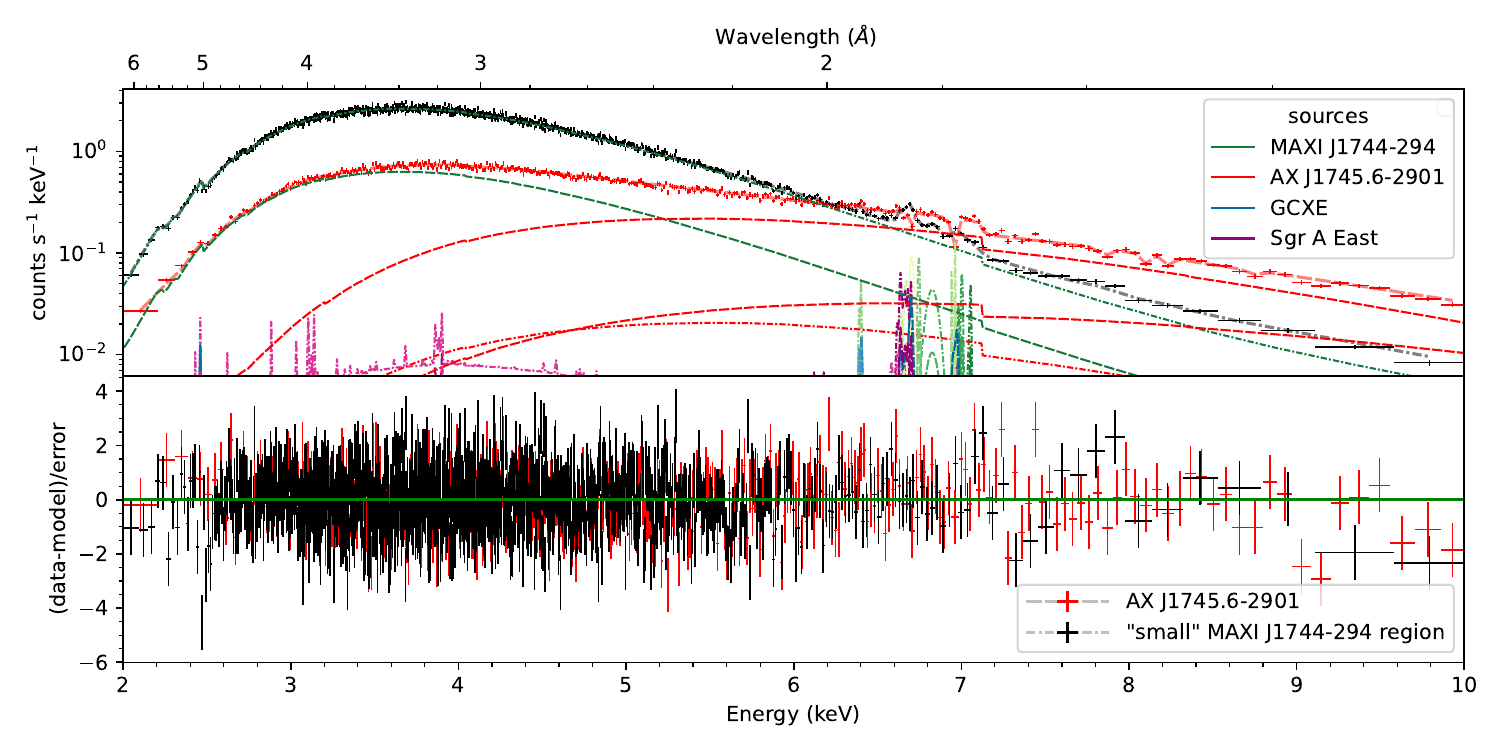}
    \includegraphics[clip,trim=0cm 0.3cm 0cm 0cm,width=0.49\textwidth]{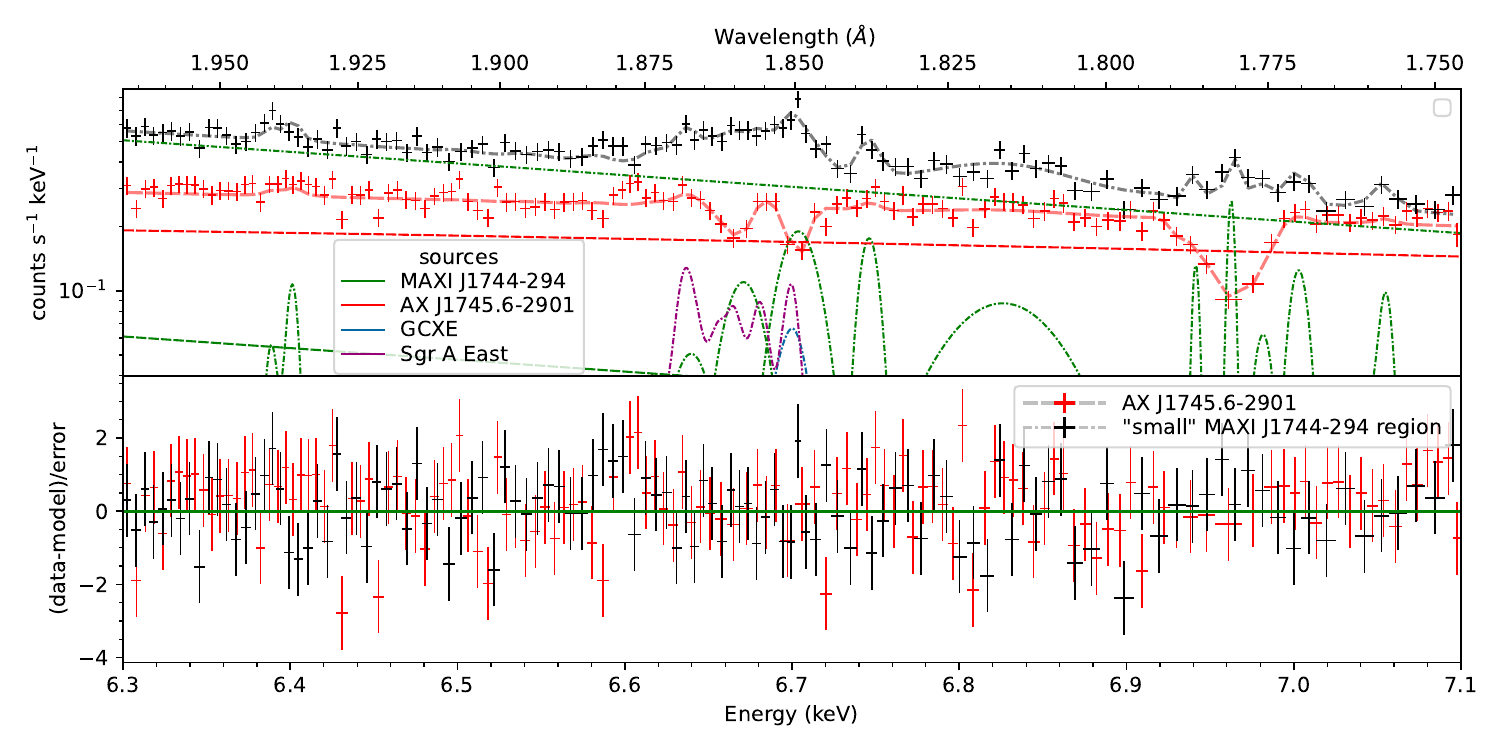}
\vspace{-1.em}
    \caption{Full residuals \textbf{(left)} and zoom in the 6.3-7.1 keV band \textbf{(right)} for the Resolve MAXI J1744-294 "small" M1744 region and NS region in the DDT observation, after empirical modeling and having added all significant line features. For readability, the spectra are visually rebinned at a 20$\sigma$ and 10 $\sigma$ significance level in the left and right panel respectively, and model components are always shown at a 3$\sigma$ significance level.}
    \label{fig:resolve_smallpix_empi_resid_conti_postline}
\end{figure*}

\begin{figure*}[h!]
\centering
    \includegraphics[clip,width=0.49\textwidth]{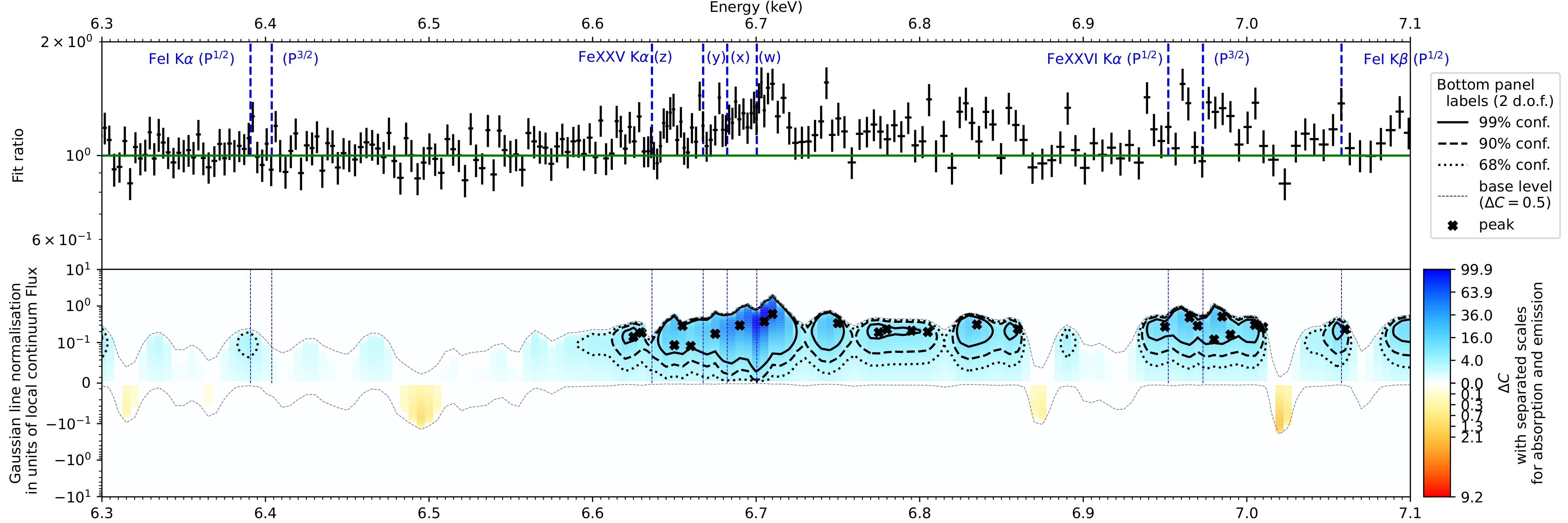}
    \includegraphics[clip,width=0.49\textwidth]{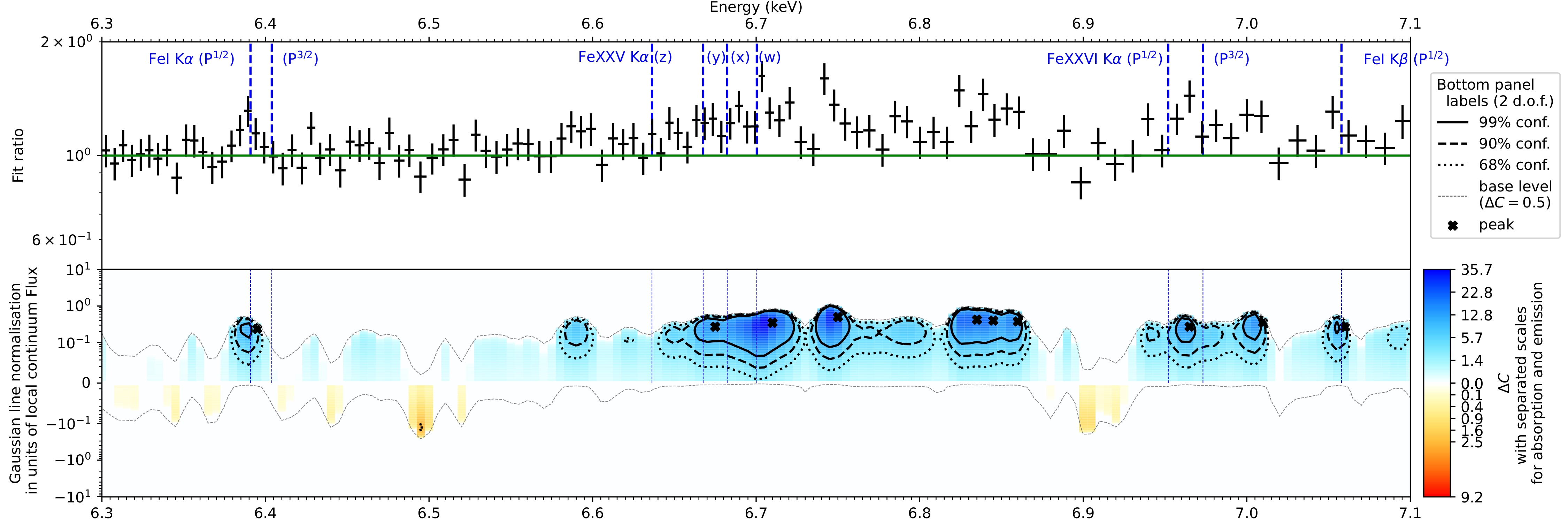}
    \includegraphics[clip,width=0.49\textwidth]{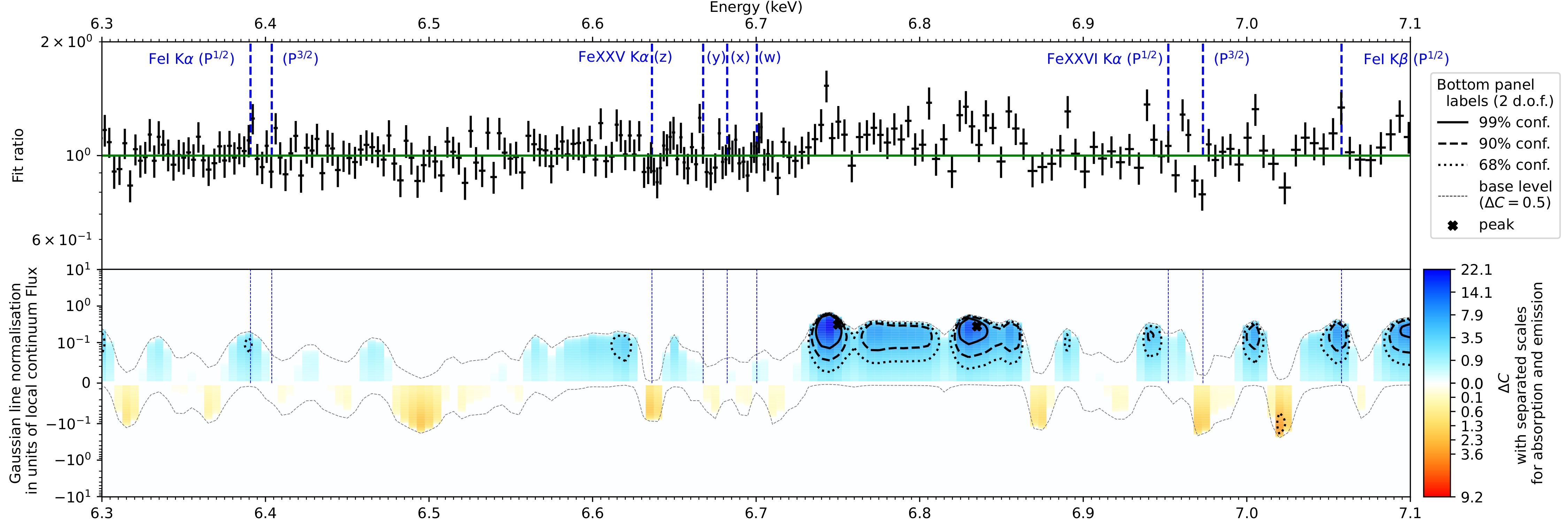}
    \includegraphics[clip,width=0.49\textwidth]{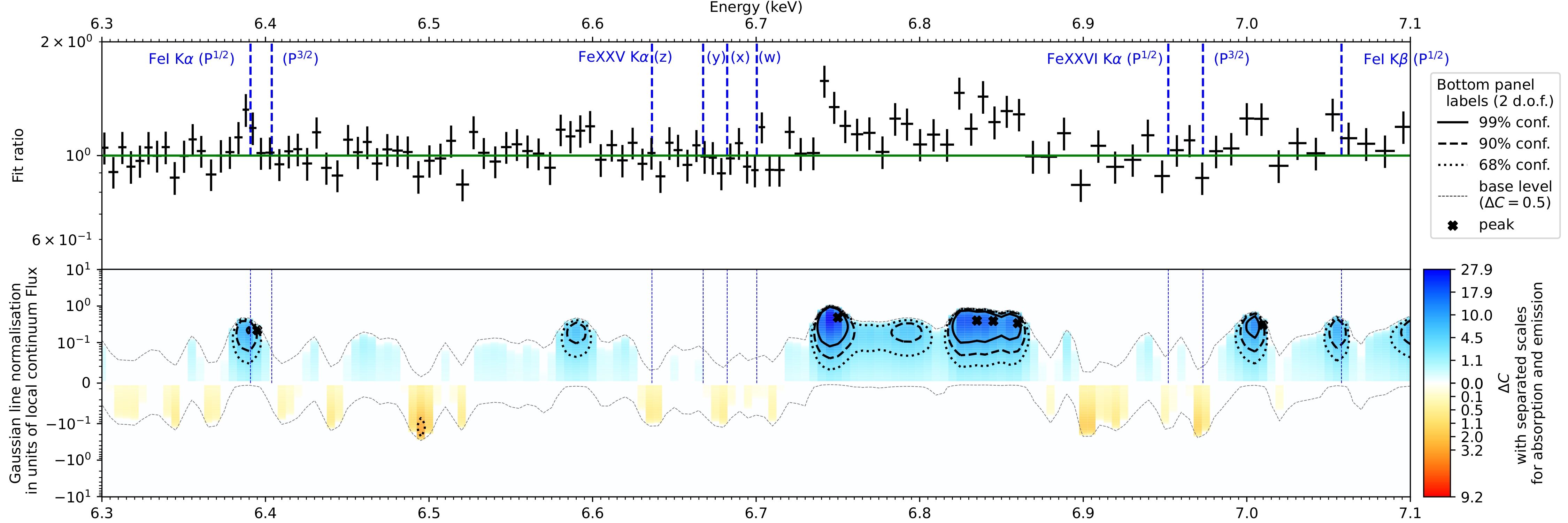}
    \includegraphics[clip,width=0.49\textwidth]{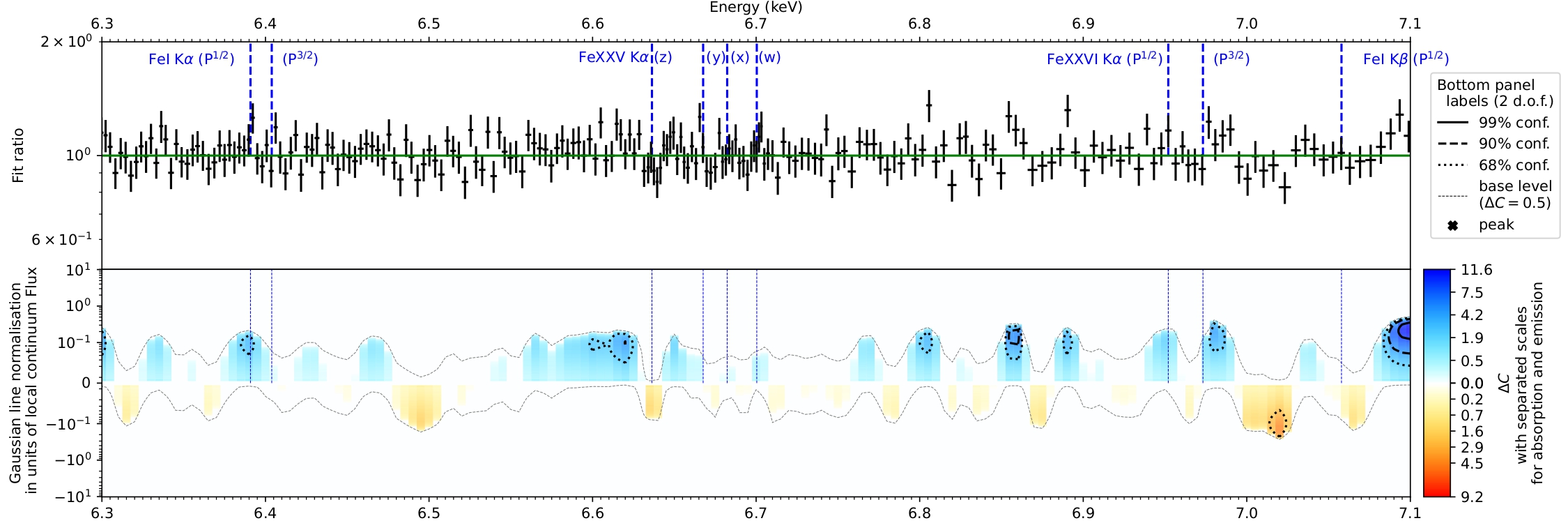}
    \includegraphics[clip,width=0.49\textwidth]{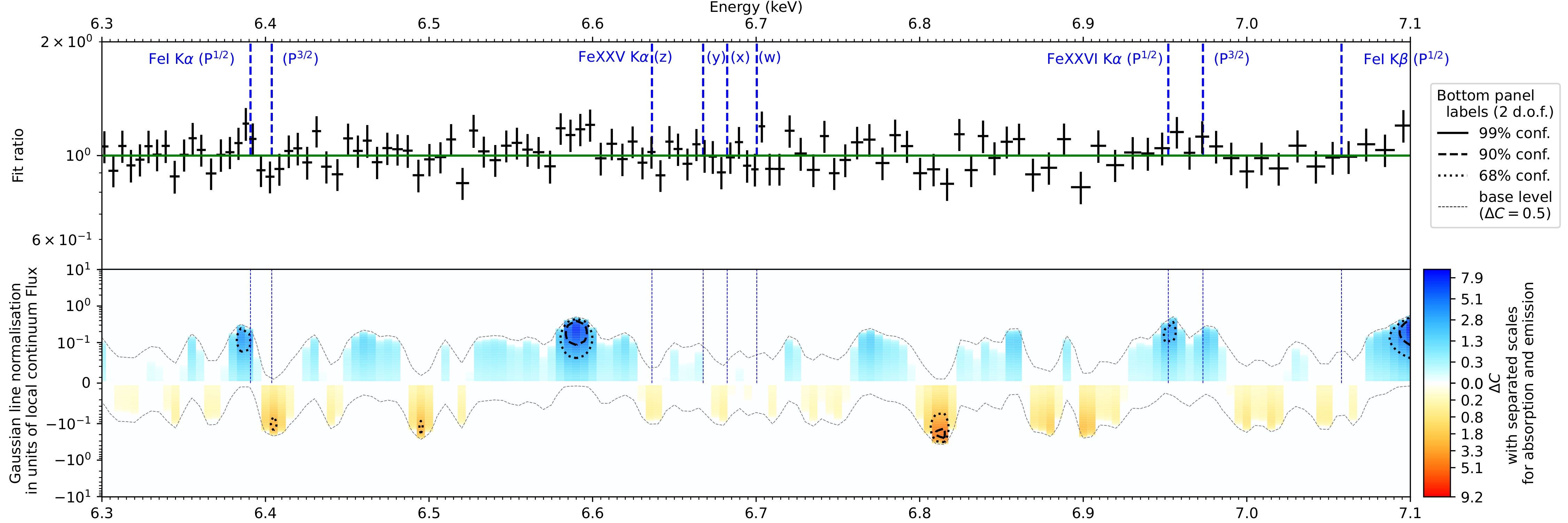}
    \caption{Blind searches for narrow line features in the 6.3-7.1 keV range along different steps of the empirical fits of the "big" \textbf{(left)} and "small" \textbf{(right)} MAXI J1744-294 region, in the DDT observation. The \textbf{(top)} panels are computed before adding any line in this band, the \textbf{(middle)} panels after adding the main static component, and the \textbf{(bottom)} panels after adding all significant secondary components. All spectra are rebinned at a 10 $\sigma$ significance level for visibility.}
    \label{fig:blind_search_empi_BH}
\end{figure*}

\begin{figure*}[h!]
\centering
    \includegraphics[clip,width=0.49\textwidth]{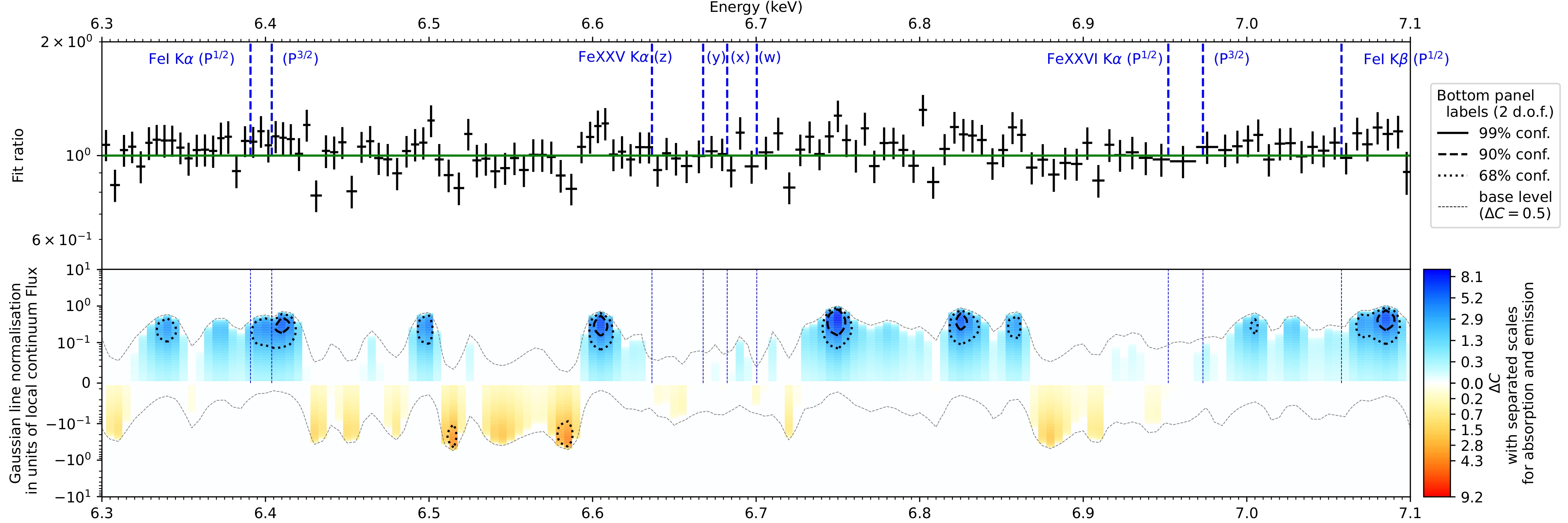}
    \includegraphics[clip,width=0.49\textwidth]{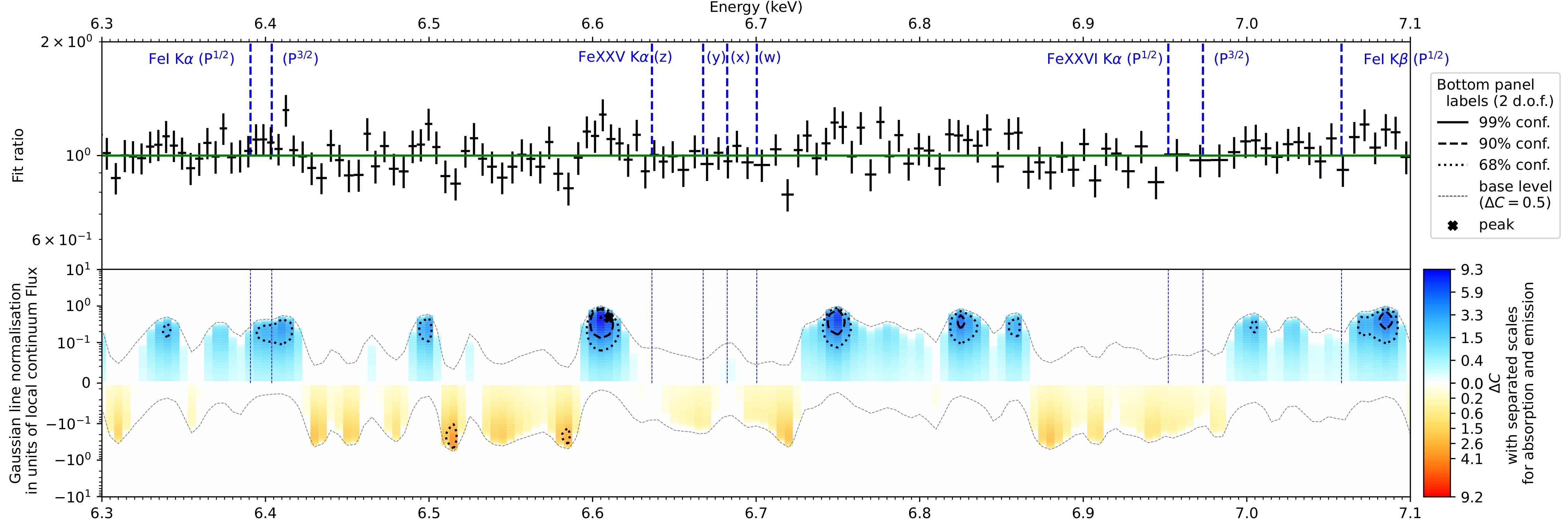}
    \includegraphics[clip,width=0.49\textwidth]{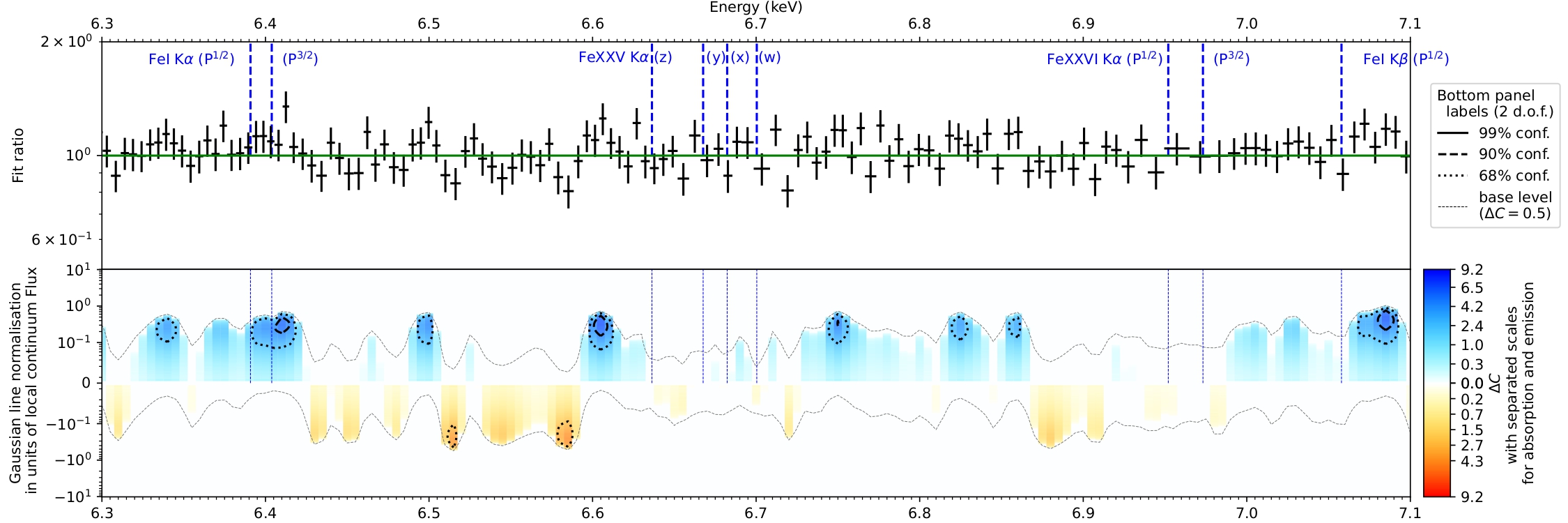}
    \includegraphics[clip,width=0.49\textwidth]{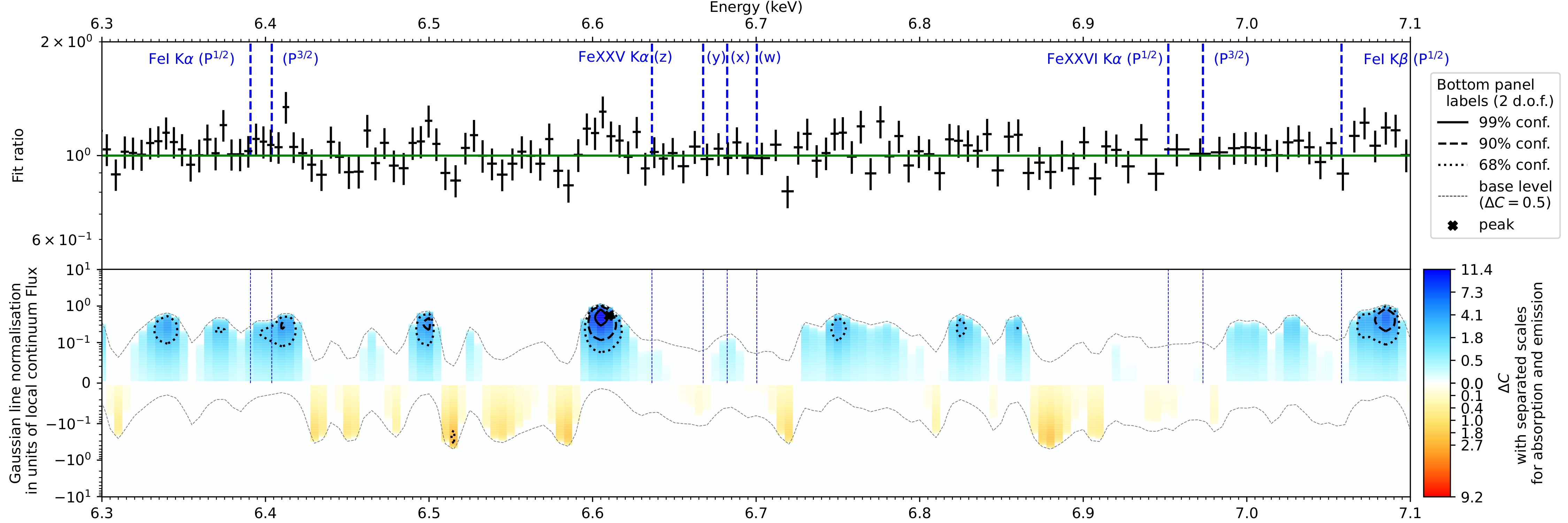}
    \caption{Blind searches for narrow line features in the 6.3-7.1 keV range along different steps of the empirical fits of the AX J1745.6-2901 region in the DDT observation, with simultaneous fitting to the "big" \textbf{(left)} and "small" \textbf{(right)} MAXI J1744-294 region. The \textbf{(top)} panels are computed after adding the main static component and all NS components, and the \textbf{(bottom)} panels after adding all significant secondary components. All spectra are rebinned at a 10 $\sigma$ significance level for visibility.}
    \label{fig:blind_search_empi_NS}
\end{figure*}

\clearpage

\begin{rotatepage}
\begin{table}
\movetabledown=7.5cm
\begin{rotatetable}
\begin{center}
\caption{Parameters of the empirical MAXI J1744-294 model, derived from fitting the "big" and "small" regions spectra in the XRISM DDT observation, simultaneously with the AX J1745.6-2901 spectrum.}
\label{tab:comp_param_empi_BH}
\begin{tabular}{lccccccccccccccccc}

\cline{4-10}
\cline{12-18}\vspace{-1.25em}\\
\cline{4-10}
\cline{12-18}
 & & & \multicolumn{7}{c}{"big" MAXI J1744-294 region} & & \multicolumn{7}{c}{"small" MAXI J1744-294 region}\T\B\\
\cline{4-10}
\cline{12-18}\vspace{-1.25em}\\
\cline{4-10}
\cline{12-18}

& & & \multicolumn{7}{c}{Empirical adjustments} & & \multicolumn{7}{c}{Empirical adjustments}\T\B\\
\cline{1-2}
\cline{4-10}
\cline{12-18}\vspace{-1.25em}\\
\cline{1-2}
\cline{4-10}
\cline{12-18}
\multicolumn{2}{c}{Parameter} 
& & C$_{M1744}$ & C$_{AXJ}$ &  A$_{Fe}$ & E$_{edge}$ & $\tau_{edge}$ & &
& & C$_{M1744}$ & C$_{AXJ}$ &  A$_{Fe}$ & E$_{edge}$ & $\tau_{edge}$ & &
\T\B\\
\multicolumn{2}{c}{Unit} 
& &  &  &  solar & eV &  & &
& &  &  &  solar & eV &  & &
\T\B\\
\multicolumn{2}{c}{Value} 
& & 1$^\dagger$ & $0.87_{-0.01}^{+0.01}$ &  0$^\dagger$ & $7198_{-29}^{+11}$ & $0.27_{-0.04}^{+0.04}$ & &
& & 1$^\dagger$ & $0.91_{-0.01}^{+0.01}$ & 1$^\dagger$ & / & / & &
\T\B\\
\cline{1-2}
\cline{4-10}
\cline{12-18}\vspace{-1.25em}\\
\cline{1-2}
\cline{4-10}
\cline{12-18}
& & & \multicolumn{7}{c}{Continuum} & & \multicolumn{7}{c}{Continuum}\T\B\\
\cline{1-2}
\cline{4-10}
\cline{12-18}\vspace{-1.25em}\\
\cline{1-2}
\cline{4-10}
\cline{12-18}
\multicolumn{2}{c}{Parameter} 
& & $N_H$ & kT$_{in}$ & $\Gamma$ & $f_{cov}$& kT$_e$ & F$_{2-10}^{abs}$$^\star$ & \multirow{2}{*}{$\dfrac{\mathrm{C-stat^*}}{\mathrm{d.o.f.}}$}
& & $N_H$ & kT$_{in}$ & $\Gamma$ & $f_{cov}$& kT$_e$ & F$_{2-10}^{abs}$$^\star$ & \multirow{2}{*}{$\dfrac{\mathrm{C-stat^*}}{\mathrm{d.o.f.}}$}
\T\B\\
\multicolumn{2}{c}{Unit} 
& & 10$^{22}$ cm$^{-2}$ & keV & & $10^{-3}$ & keV & 10$^{-10}$cgs & 
& & 10$^{22}$ cm$^{-2}$ & keV & & $10^{-3}$ & keV & 10$^{-10}$cgs & 
\T\B\\

\multicolumn{2}{c}{Value} 
& & $20.3_{-0.1}^{+0.1}$ & $0.62_{-0.01}^{+0.01}$ & 2.74$^\dagger$ & $7.7_{-0.5}^{+0.3}$ & 150$^\dagger$ & $9.5_{-0.1}^{+0.1}$ & $\dfrac{33930}{31920}$
& & $13.4_{-0.1}^{+0.2}$ & $0.62_{-0.01}^{+0.01}$ & 2.74$^\dagger$ & $7.3_{-0.5}^{+0.5}$ & 150$^\dagger$ & $9.4_{-0.1}^{+0.1}$ &  $\dfrac{33764}{31923}$
\vspace{0.3em}
\T\B\\

\cline{1-2}
\cline{4-10}
\cline{12-18}\vspace{-1.25em}\\
\cline{1-2}
\cline{4-10}
\cline{12-18}

& & & \multicolumn{7}{c}{Spectral lines} & & \multicolumn{7}{c}{Spectral lines}\T\B\\
\cline{1-2}
\cline{4-10}
\cline{12-18}\vspace{-1.25em}\\
\cline{1-2}
\cline{4-10}
\cline{12-18}
  
\multicolumn{2}{c}{Line}
& & $E_{\rm rest}$ & $v_{raw}\ddagger$ & $\sigma$  & norm  & EW  & $\Delta$C-stat & Sign. 
& & $E_{\rm rest}$ & $v_{raw}\ddagger$ & $\sigma$  & norm  & EW  & $\Delta$C-stat & Sign. \T\B \\
complex & ID
 & & eV & km s$^{-1}$ & eV & $10^{-5}$ & eV &  & MC$^\diamond$
 & & eV & km s$^{-1}$ & eV & $10^{-5}$ & eV &  & MC$^\diamond$ \T\B \\
 \cline{1-2}
  \cline{4-10}
   \cline{12-18}

\multirow{4}{*}{\Fexxv{}$_{0}$} & $z$ 
& & 6636.3 &  \multirow{4}{*}{$26_{-185}^{+258}$} & \multirow{4}{*}{$14_{-3}^{+6}$} & $2.8_{-1.2}^{+1.5}$ & $9_{-4}^{+4}$ & 17 & $>0.999$
& & 6636.3 & \multirow{4}{*}{$-152_{-206}^{+284}$} & \multirow{4}{*}{$12_{-3}^{+7}$} & $1.4_{-1.2}^{+1.2}$ & $4_{-4}^{+3}$ & 5 & $0.78$
\T\B \\
\multirow{4}{*}{He$\alpha$} & $y$
&  & 6667.6 &  &  & $2.5_{-2.4}^{+1.4}$ & $8_{-4}^{+6}$ & 15 & $>0.999$
&  & 6667.6 &  &  & $2.9_{-2.0}^{+1.1}$ & $9_{-4}^{+5}$ & 20 & $>0.999$
\T\B \\
& $x$ 
& & 6682.3 &  &  & $<1.3$ & $0^{+4}$ & 0 & /
& & 6682.3 &  &  & $<2.2$ & $0^{+4}$ & 0 & /
\T\B \\
& $w$ 
& & 6700.4 &  &  & $7.1_{-1.5}^{+2.3}$ & $25_{-6}^{+4}$ & 132 & $>0.999$
& & 6700.4 &  &  & $5.1_{-1.9}^{+1.9}$ & $17_{-5}^{+5}$ & 62 &  $>0.999$
\T\B \\
\cline{1-2}
\cline{4-10}
\cline{12-18}
\multirow{1}{*}{\Fexxvi{}$_{0}$} & 1/2  
& & 6952.0 &  \multirow{2}{*}{$446_{-53}^{+60}$} & \multirow{2}{*}{$1.7_{-1.7}^{+2.1}$} & $0.5_{-0.2}^{+0.6}$ & \multirow{2}{*}{$7_{-3}^{+3}$} & \multirow{2}{*}{39} & \multirow{2}{*}{$>0.999$}
& & 6952.0  & \multirow{2}{*}{$457_{-42}^{+36}$} & \multirow{2}{*}{$0.8_{-0.8}^{+2.2}$} & $0.6_{-0.2}^{+0.2}$ & \multirow{2}{*}{$9_{-3}^{+3}$} & \multirow{2}{*}{42} & \multirow{2}{*}{$>0.999$}
\T\B \\
\multirow{1}{*}{Ly$\alpha$} & 3/2 
&  & 6973.2 &  &  & $1.0_{-0.4}^{+1.2}$ &  &  & 
&  & 6973.2  &  &  & $1.2_{-0.4}^{+0.4}$ &  &  & 
\T\B \\
\cline{1-2}
\cline{4-10}
\cline{12-18}
\multirow{1}{*}{\Fexxv{}$_{1}$?} & $w$ 
& & $6741.9_{-2.9}^{+3.1}$ & $-1987_{-140}^{+130}$  & $5.6_{-2.2}^{+4.0}$ & $1.6_{-0.6}^{+0.8}$ & $6_{-2}^{+2}$ & 26 &  $>0.999$
& & $6746.1_{-2.7}^{+3.1}$ & $-2043_{-138}^{+120}$  & $5.6_{-2.0}^{+4.1}$ & $2.3_{-0.9}^{+1.1}$ & $8_{-3}^{+3}$ & 31 &  $>0.999$
\T\B \\
\cline{1-2}
\cline{4-10}
\cline{12-18}
\multirow{1}{*}{\Fexxv{}$_{2}$?} & $w$ 
& & $6783_{-10}^{+13}$ & $-3676_{-445}^{+574}$ & 10$^\dagger$ & $1.7_{-0.7}^{+0.7}$ & $6_{-2}^{+2}$ & 17 & $0.996$  
& &\multirow{2}{*}{$6826_{-20}^{+12}$} & \multirow{2}{*}{$-5621_{-556}^{+877}$} & \multirow{2}{*}{$36_{-13}^{+21}$} & \multirow{2}{*}{$7.1_{-2.1}^{+2.6}$} & \multirow{2}{*}{$27_{-7}^{+9}$} & \multirow{2}{*}{56} &  \multirow{2}{*}{$>0.999$}
\T\B \\\cline{1-2}
\cline{4-10}
\multirow{1}{*}{\Fexxv{}$_{3}$?} & $w$ 
& & $6827.7_{-5.2}^{+8.7}$ & $-5931_{-285}^{+337}$ & 10$^\dagger$  & $2.0_{-0.7}^{+0.7}$ & $8_{-3}^{+3}$ & 25 &  $>0.999$
\T\B \\
\cline{1-2}
\cline{4-10}
\cline{12-18}
\multirow{1}{*}{\Fexxvi{}$_{4}$?} & 1/2  
& & 6952.0 &  \multirow{2}{*}{$-1281_{-99}^{+115}$} & \multirow{2}{*}{$4.4_{-2.3}^{+2.6}$} & $0.6_{-0.2}^{+0.2}$ & \multirow{2}{*}{$9_{-3}^{+3}$} & \multirow{2}{*}{30} & \multirow{2}{*}{$>0.999$}
& & 6952.0  & \multirow{2}{*}{$-1277_{-131}^{+177}$} & \multirow{2}{*}{$5.4_{-5.4}^{+3.9}$} & $0.8_{-0.4}^{+0.3}$ & \multirow{2}{*}{$12_{-5}^{+3}$} & \multirow{2}{*}{27} & \multirow{2}{*}{$>0.999$}
\T\B \\
\multirow{1}{*}{Ly$\alpha$} & 3/2 
&  & 6973.2 &  &  & $1.2_{-0.4}^{+0.4}$ &  &  & 
&  & 6973.2  &  &  & $1.6_{-0.8}^{+0.6}$ &  &  & 
\T\B \\
\cline{1-2}
\cline{4-10}
\cline{12-18}
\multirow{1}{*}{\Fei{} K$\alpha$} & blend  
& \multicolumn{8}{c}{/}
& & blend  & \multirow{2}{*}{$101_{-707}^{+452}$} & \multirow{2}{*}{$2.0_{-2.0}^{+31}$} & $1.4_{-1.0}^{+1.6}$ & $3_{-2}^{+3}$ & 6 & $0.87$
\T\B \\
\multirow{1}{*}{\Fei{} K$\beta$} & blend 
&  & blend & $-6_{-83}^{+108}$ & $0_{-0}^{+4.3}$ & $1.1_{-0.5}^{+0.5}$ & $6_{-3}^{+3}$ & 14 & $0.998$
&  & blend  &  &  & $1.4_{-0.7}^{+1.9}$ & $7_{-4}^{+3}$ & 11 & $0.98$
\T\B \\

\cline{1-2}
\cline{4-10}
\cline{12-18}\vspace{-1.25em}\\
\cline{1-2}
\cline{4-10}
\cline{12-18}
& & & \multicolumn{7}{c}{With constrained FeI K$\beta$/FeI K$\alpha$ ratios} & & \multicolumn{7}{c}{With constrained FeI K$\beta$/FeI K$\alpha$ ratios }\T\B\\
\cline{1-2}
\cline{4-10}
\cline{12-18}\vspace{-1.25em}\\
\cline{1-2}
\cline{4-10}
\cline{12-18}
\multirow{1}{*}{\Fei{} K$\alpha$} & blend  
& & \multicolumn{7}{c}{/}
& & blend  & \multirow{2}{*}{$452_{-342}^{+333}$} & \multirow{2}{*}{$6.6_{-6.6}^{+12}$} & $2.4_{-1.1}^{+1.3}$ & $6_{-3}^{+3}$ &  \multirow{2}{*}{14} & \multirow{2}{*}{$0.997$}
\T\B \\
\multirow{1}{*}{\Fei{} K$\beta$} & blend 
&  & \multicolumn{7}{c}{/}
&  & blend  &  &  & $0.5_{-0.2}^{+0.3}$ & $2_{-1}^{+2}$ &  & 
\T\B \\
\cline{1-2}
\cline{4-10}
\cline{12-18}
\end{tabular}\\
\vspace{1em}
\raggedright
$\star$ computed from the full model, including the line components, and all sources applied to the two spectra of MAXI J1744-294 and AX J1745.6-2901.
$\dagger$ frozen or at the limit of the parameter space. 
$\ddagger$ does not include the correction of -28 km/s due to the relative motion of Earth in the Solar System on the date of the observation.
$^\diamond$ Significance computed from Monte-Carlo simulations to consider the look elsewhere effect. See the main text for details. 
We quote the rest energy of all lines fitted with Gaussians. 
The \Fei{} complexes are described with \texttt{feklor} and \texttt{fekblor}. Widths and velocities are linked within a single complex. The normalization of the He$\alpha$ complexes are tied to a 2-1 ratio, and the EW and significance are computed for both transitions. 

\end{center}
\end{rotatetable}
\end{table}
\end{rotatepage}

\subsubsection{Monte-Carlo simulations}\label{app:resolve_MC}

Owing to the differences in count rates and photon noise between our two methodologies, we computed Monte-Carlo simulations independently for the "big" and "small" M1744 regions, but with similar methodologies.  In each case, we start by refitting the continuum with a single empirical model applied to the M1744 set of spectral products, combining the contributions of M1744, AXJ, and the different sources of diffuse emission. We then simulated 1000 Resolve spectra using the \texttt{fakeit} command in xspec and allowing statistical fluctuations. For each simulated spectrum, we then refit the continuum to derive a new best fit and a base C-statistic, before scanning the parameter spaces of our detected lines for spurious C-statistic improvements when introducing emission lines. We use \texttt{gaussian} emission lines with widths fixed to those of the detected transitions, and scan a portion of the energy band around each line, which we detail below. The distribution of the maximum statistical improvement of these spurious lines ($\Delta$C-stat$_{fake}$) in each of the 1000 spectra can be compared to the statistical improvement of the "real" features detected in M1744 ($\Delta$C-stat$_{real}$). We then derive for each empirical feature a significance $P=1-N/1000$, with $N$ the number of simulated spectra where $\Delta$C-stat$_{real}<$$\Delta$C-stat$_{fake}$.

In the CCD era, it was common to scan the entirety of the high-energy band (e.g., 4-10 keV or 6-10 keV) when looking for lines. While this approach is fundamentally correct and conservative, here, we consider that in the micro-calorimeter era, the "look elsewhere" effect should be put into perspective of the proximity of empirical detections to expected strong lines. That is, while thousands of atomic transitions span the energy band of Resolve, detecting 3 individual transitions at less than 1 eV from the expected rest energies of the \Fexxv{} He$\alpha$ line is more "expected" than spurious transitions at very high and/or very low energies, with no other emission feature anywhere else in the spectrum. This can be justified physically, as the \Fexxv{} He$\alpha$ transitions are among the strongest lines seen in the parameter space of highly ionized plasma, due to the high abundances of iron and the high strength of the transitions themselves, and similar considerations can be applied to the other strong Fe transitions.

For this reason, we adopt two different significance tests for our different emission features. For the transitions up to $\sim\pm1000$ km/s of velocity shift, which include the static Fe components and the least blueshifted of the other transitions (namely \Fexxvi{}$_{4}$), we scan a $[-3000,3000]$ \kmps{} velocity shift band around the energies of the different lines, which we consider to be a conservative estimate of the band where any spurious detection would be \textit{unambiguously} associated to the relevant line. For the $3$ unidentified transitions with a much higher blueshift, we scan a $[-10^4,10^4]$ \kmps{} velocity shift band around the energies of \Fexxv{} K$\alpha$, as this transition is already unambiguously detected in the static layers, and this velocity band is a conservative estimate of the band where any transitions would be primarily associated to this transition. This significance estimate is thus tied to our proposed physical interpretation. For instance, in a scenario where these lines would be extremely blueshifted versions of SXV He$\alpha$ $w$ in a low-ionization plasma, the significance would have to be computed in light of the $\sim4.5$ keV separating the emission features and their supposed rest energy.

\clearpage
\section{Physical modeling}\label{app:phys_mod}

\begin{table}[h]
\movetabledown=7.5cm
\begin{center}
\caption{Parameters of the physical line models applied to the highly ionized lines of MAXI J1744-294, derived from fitting the "big" and "small" regions spectra in the XRISM DDT observation, simultaneously with the AX J1745.6-2901 spectrum. The marginally significant neutral \Fei{} K component is largely independent from the highly ionized lines and thus kept identical to the empirical fit.}
\label{tab:comp_param_phys_BH}
\begin{tabular}{l c ccccc c ccccc}

\cline{1-1}
\cline{3-13}\vspace{-1.25em}\\
\cline{1-1}
\cline{3-13}
Layer & & \multicolumn{5}{c}{"big" MAXI J1744-294 region} & & \multicolumn{5}{c}{"small" MAXI J1744-294 region}\T\B\\

\cline{1-2}
\cline{3-8}
\cline{9-13}\vspace{-1.25em}\\
\cline{1-2}
\cline{3-8}
\cline{9-13}
 & & \multicolumn{5}{c}{Photoionization layers} & & \multicolumn{5}{c}{Photoionization layers}\T\B\\
\cline{1-1}
\cline{3-7}
\cline{9-13}\vspace{-1.25em}\\
\cline{1-1}
\cline{3-7}
\cline{9-13}

Parameter 
& & log$\xi$ & $N_H$ & $v_{turb}^\ddagger$ & $v_{bulk}^\star$ & $\Omega^\diamond$
& & log$\xi$ & $N_H$ & $v_{turb}^\ddagger$ & $v_{bulk}^\star$ & $\Omega^\diamond$
\T\B \\

Unit
& &  & 10$^{22}$ cm$^{-2}$ & \kmps{} & \kmps{} & 
& &  & 10$^{22}$ cm$^{-2}$ & \kmps{} & \kmps{} & 
\T\B \\
\cline{1-1}
\cline{3-7}
\cline{9-13}

static 
& & $5.7_{-0.1}^{+0.2}$ & $8.4_{-1.7}^{+2.0}$ & $967_{-206}^{+33\dagger}$ & $201_{-188}^{+203}$ & 1$^\dagger$
& & $5.9_{-0.2}^{+0.2}$ & $7.4_{-1.9}^{+2.3}$ & $863_{-264}^{+137\dagger}$ & $191_{-236}^{+258}$ & 1$^\dagger$
\T\B \\
blue 1
& & $5.0_{-0.7}^{+0.8}$ & $0.48_{-0.29}^{+0.50}$ & $211_{-127}^{+175}$ & $-1967_{-193}^{+151}$ & 1$^\dagger$
& & $5.4_{-0.9}^{+0.6}$ & $1.4_{-0.8}^{+1.3}$ & $301_{-122}^{+205}$ & $-2072_{-169}^{+139}$ & 1$^\dagger$
\T\B \\
blue 2
& & $6.2_{-0.5}^{+0.6}$ & $2.0_{-1.1}^{+3.1}$ & $65_{-52}^{+507}$ & $-5570_{-295}^{+52}$ & 1$^\dagger$
& & \multirow{2}{*}{$6.0_{-0.3}^{+0.2}$} & \multirow{2}{*}{$4.0_{-1.4}^{+1.8}$} & \multirow{2}{*}{$972_{-328}^{+476}$} & \multirow{2}{*}{$-5927_{-320}^{+408}$} & \multirow{2}{*}{1$^\dagger$}
\T\B \\
blue 3
& & $4.4_{-0.5}^{+1.0}$ & $0.32_{-0.18}^{+0.36}$ & $120_{-84}^{+880\dagger}$ & $-6890_{-102}^{+98}$ & 1$^\dagger$ &
\T\B \\
\cline{1-1}
\cline{3-7}
\cline{9-13}\vspace{-1.25em}\\
\cline{1-1}
\cline{3-7}
\cline{9-13}
 & & \multicolumn{5}{c}{Remaining empirical component} & & \multicolumn{5}{c}{Remaining empirical component}\T\B\\
\cline{1-1}
\cline{3-7}
\cline{9-13}\vspace{-1.25em}\\
\cline{1-1}
\cline{3-7}
\cline{9-13}

Parameter
& & $E_{\rm rest}$ & $v_{raw}\ddagger$ & $\sigma$ & EW  & $\Delta$C-stat  
& & $E_{\rm rest}$ & $v_{raw}\ddagger$ & $\sigma$ & EW  & $\Delta$C-stat  
\T\B \\
Value
& & eV & km s$^{-1}$ & eV & eV & 
& & eV & km s$^{-1}$ & eV & eV &  
\T\B \\
 \cline{1-1}
\cline{3-7}
\cline{9-13}
\multirow{1}{*}{\Fexxvi{}$_{4}$?}   
& & 6952.0 &  \multirow{2}{*}{$-1311_{-60}^{+63}$} & \multirow{2}{*}{$0^{+10^\dagger}$} & \multirow{2}{*}{$4_{-2}^{+2}$} & \multirow{2}{*}{12}  
& & 6952.0  & \multirow{2}{*}{$-1313_{-108}^{+166}$} & \multirow{2}{*}{$0^{+2.7}$} & \multirow{2}{*}{$4_{-2}^{+3}$} & \multirow{2}{*}{13} 
\T\B \\
\multirow{1}{*}{Ly$\alpha$}  
&  & 6973.2 &  &  &   & 
&  & 6973.2  &  &  &  &  
\T\B \\

\cline{1-2}
\cline{3-8}
\cline{9-13}\vspace{-1.25em}\\
\cline{1-2}
\cline{3-8}
\cline{9-13}

 & & \multicolumn{5}{c}{Collisional ionization layers} & & \multicolumn{5}{c}{Collisional ionization layers}\T\B\\
\cline{1-1}
\cline{3-7}
\cline{9-13}\vspace{-1.25em}\\
\cline{1-1}
\cline{3-7}
\cline{9-13}

Parameter
& & kT & $n_{e}n_{H}V^+$ & $v_{turb}^\ddagger$ & $v_{bulk}^\star$ & $Z_{base}$
& & kT & $n_{e}n_{H}V^+$ & $v_{turb}^\ddagger$ & $v_{bulk}^\star$ & $Z_{base}$
\T\B \\

Unit
& & keV & $10^{58}$ cm$^{-5}$ & \kmps{} & \kmps{} & solar
& & keV & $10^{58}$ cm$^{-5}$ & \kmps{} & \kmps{} & solar
\T\B \\
\cline{1-1}
\cline{3-7}
\cline{9-13}
   
static 
& & $6.1_{-0.7}^{+0.5}$ & $1.6_{-0.2}^{+0.1}$ & $578_{-102}^{+191}$ & $0_{-122}^{+145}$ & $1^\dagger$
& & $9.1_{-2.0}^{+3.7}$ & $1.1_{-0.4}^{+0.4}$ & $963_{-396}^{+37\dagger}$ & $414_{-989}^{+179}$ & $1^\dagger$
\T\B \\
blue 1
& & $3.0_{-1.4}^{+0.8}$ & $0.68_{-0.24}^{+2.3}$ & $218_{-87}^{+141}$ & $-1972_{-158}^{+140}$ & $1^\dagger$
& & $1.5_{-0.2}^{+1.0}$ & $5.5_{-4.5}^{+5.5}$ & $112_{-112}^{+214}$ & $-2109_{-116}^{+81}$ & $1^\dagger$
\T\B \\
blue 2
& & $8.4_{-2.1}^{+1.7}$ & $0.34_{-0.08}^{+0.12}$ & $0^{+126}$ & $-5597_{-63}^{+62}$ & $1^\dagger$
& & \multirow{2}{*}{$6.7_{-1.4}^{+1.4}$} & \multirow{2}{*}{$1.0_{-0.2}^{+0.3}$} & \multirow{2}{*}{$829_{-261}^{+465}$} & \multirow{2}{*}{$-6053_{-288}^{+244}$} & \multirow{2}{*}{$1^\dagger$}
\T\B \\
blue 3
& & $2.7_{-1.0}^{+0.4}$ & $0.53_{-0.15}^{+1.6}$ & $40_{-40}^{+135}$ & $-6873_{-66}^{+71}$ & $1^\dagger$ &
\T\B \\

\cline{1-1}
\cline{3-7}
\cline{9-13}\vspace{-1.25em}\\
\cline{1-1}
\cline{3-7}
\cline{9-13}
 & & \multicolumn{5}{c}{Remaining empirical component} & & \multicolumn{5}{c}{Remaining empirical component}\T\B\\
\cline{1-1}
\cline{3-7}
\cline{9-13}\vspace{-1.25em}\\
\cline{1-1}
\cline{3-7}
\cline{9-13}

Parameter
& & $E_{\rm rest}$ & $v_{raw}\ddagger$ & $\sigma$ & EW  & $\Delta$C-stat  
& & $E_{\rm rest}$ & $v_{raw}\ddagger$ & $\sigma$ & EW  & $\Delta$C-stat  
\T\B \\
Value
& & eV & km s$^{-1}$ & eV & eV & 
& & eV & km s$^{-1}$ & eV & eV &  
\T\B \\
 \cline{1-1}
\cline{3-7}
\cline{9-13}
\multirow{1}{*}{\Fexxvi{}$_{4}$?}   
& & 6952.0 &  \multirow{2}{*}{$-1310_{-105}^{+94}$} & \multirow{2}{*}{$3.4_{-1.3}^{+2.6}$} & \multirow{2}{*}{$6_{-3}^{+3}$} & \multirow{2}{*}{17}  
& & 6952.0  & \multirow{2}{*}{$-1312_{-48}^{+46}$} & \multirow{2}{*}{$0^{+6.2}$} & \multirow{2}{*}{$6_{-3}^{+2}$} & \multirow{2}{*}{19} 
\T\B \\
\multirow{1}{*}{Ly$\alpha$}  
&  & 6973.2 &  &  &   & 
&  & 6973.2  &  &  &  &  
\T\B \\

\cline{1-1}
\cline{3-7}
\cline{9-13}
\end{tabular}\\
\vspace{1em}
\raggedright
$\dagger$ frozen or at the limit of the parameter space. 
$\ddagger$ does not consider the thermal broadening, included by default in the model, and of $\sim100$ \kmps{} at these temperatures. 
$\star$ does not include the correction of -28 km/s due to the relative motion of Earth in the Solar System on the date of the observation.
$^\diamond$ Fixed to 1 .w.r.t. a 4$\pi$ angular distribution at the distance of the galactic center.
$^+$ emission measure (normalization) normalized to the distance to the galactic center.

\end{center}
\end{table}

\begin{figure*}[t!]
\centering
    \includegraphics[clip,trim=0cm 0.3cm 0cm 0cm,width=0.49\textwidth]{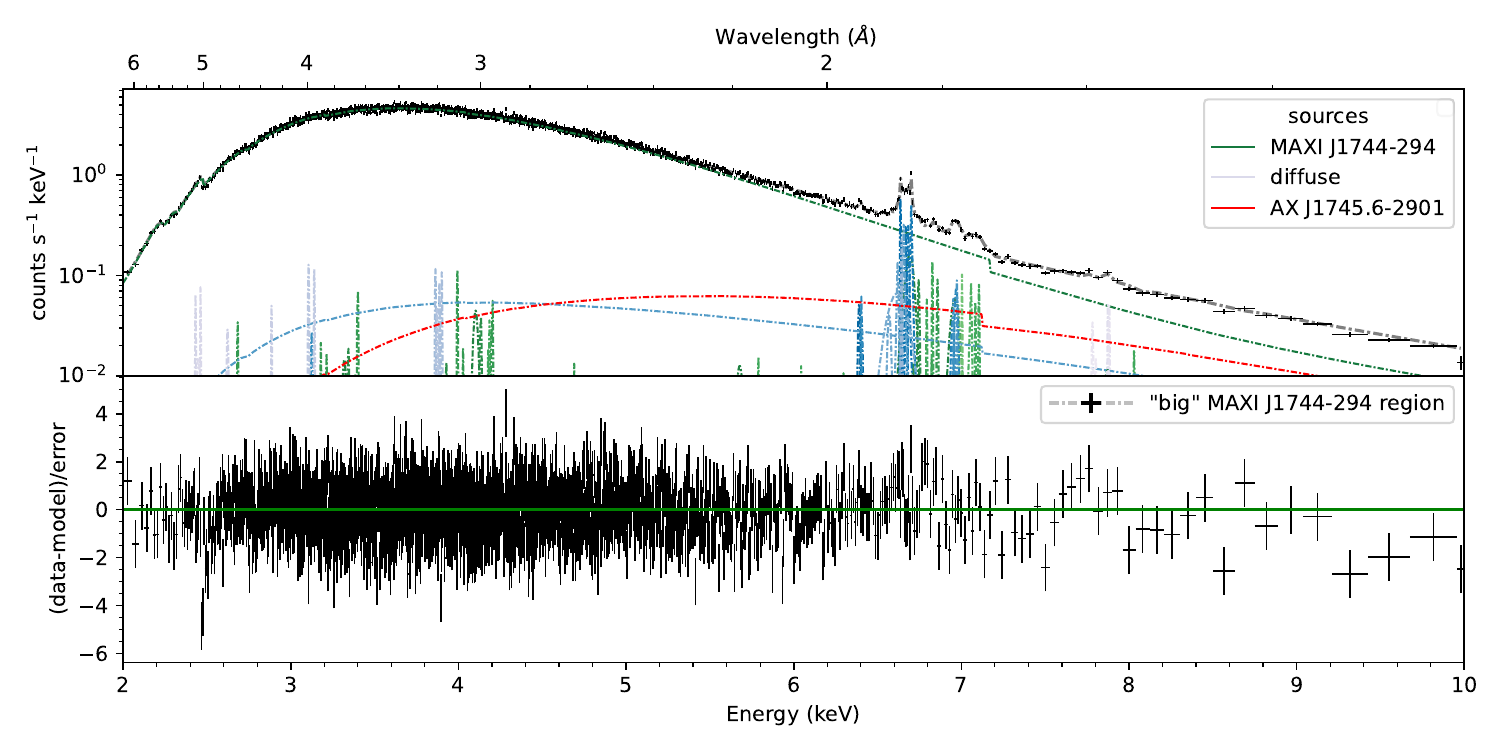}
    \includegraphics[clip,trim=0cm 0.3cm 0cm 0cm,width=0.49\textwidth]{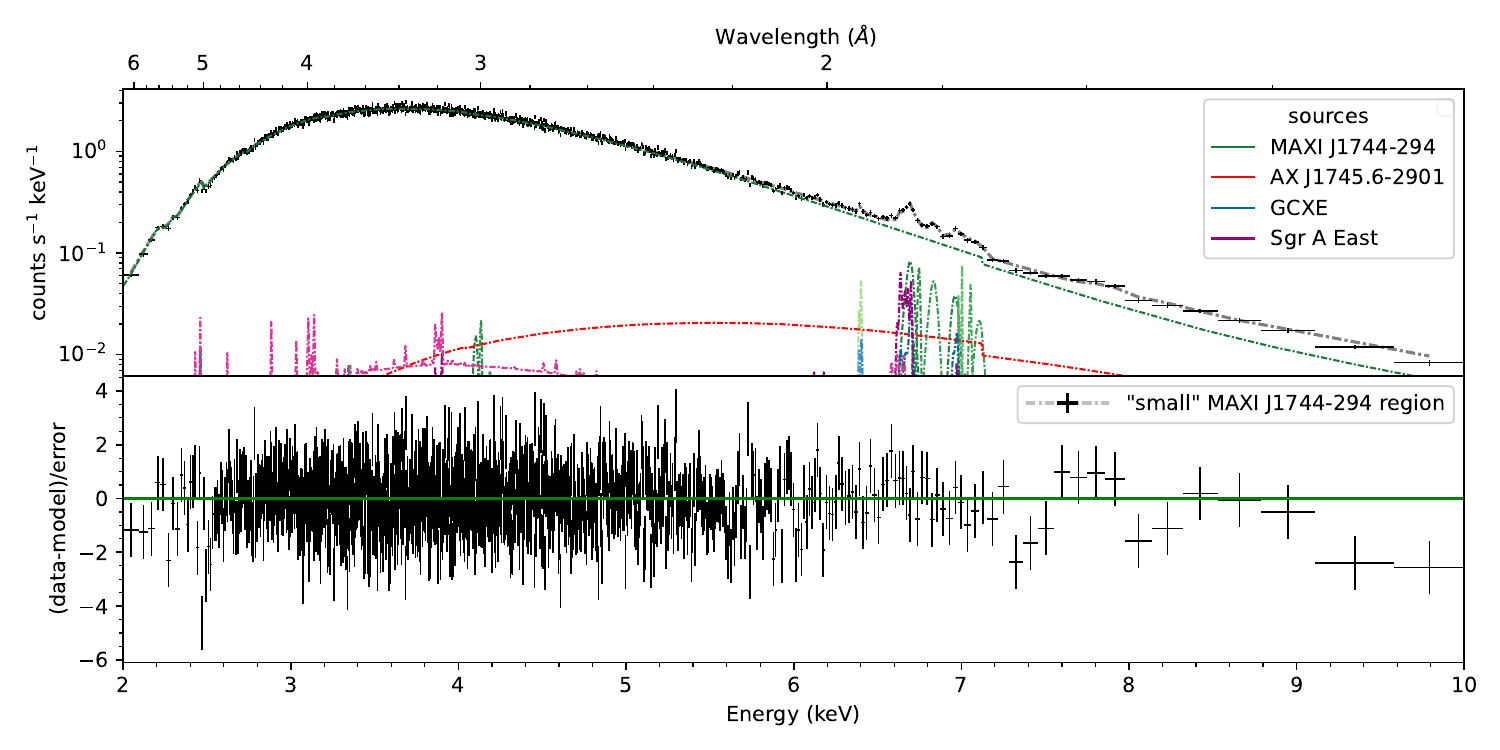}
\vspace{-1.em}
    \caption{Full residuals of the photoionization modeling of the line features in the "big" \textbf{(left)} and "small"  \textbf{(right)} MAXI J1744-294 region in the DDT observation. The residuals for the AX J1745.6-2901 spectrum, which are largely unaffected by the switch to physical models, are omitted for legibility. The spectra are visually rebinned at a 20$\sigma$ significance level, and model components at 3$\sigma$.}
    \label{fig:resolve_photo_resid_full}
\end{figure*}

\begin{figure*}[h!]
\centering
\vspace{-2em}
    \includegraphics[clip,width=0.49\textwidth]{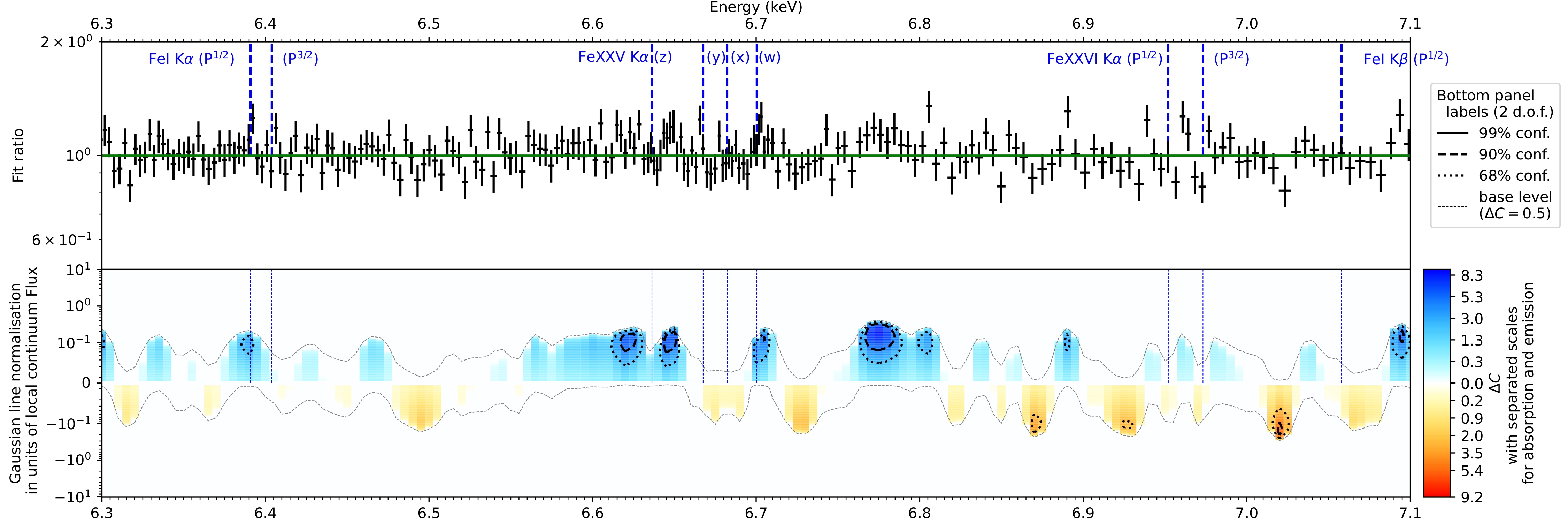}
    \includegraphics[clip,width=0.49\textwidth]{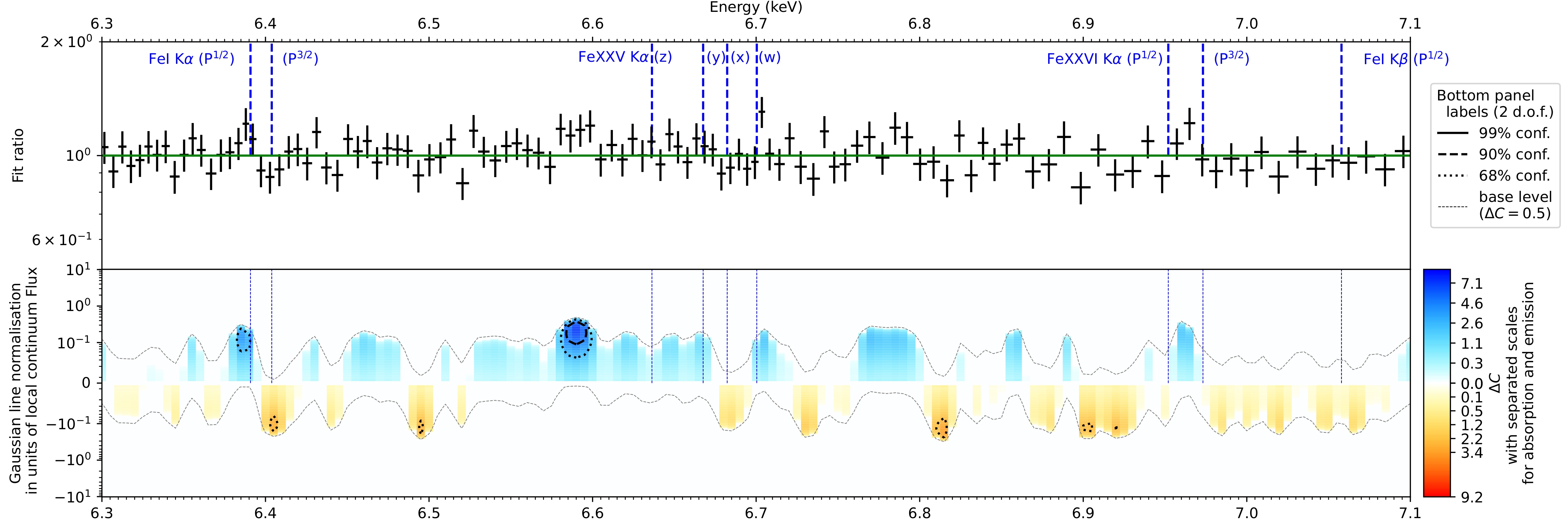}
    \caption{Blind searches for narrow line features in the 6.3-7.1 keV range after the photoionization modeling of the "big" \textbf{(left)} and "small" \textbf{(right)} MAXI J1744-294 region in the DDT observation. All spectra are rebinned at a 10$\sigma$ significance level for visibility.}
    \label{fig:blind_search_photo_BH}
\end{figure*}

\begin{figure*}[t!]
\centering
    \includegraphics[clip,trim=0cm 0.3cm 0cm 0cm,width=0.49\textwidth]{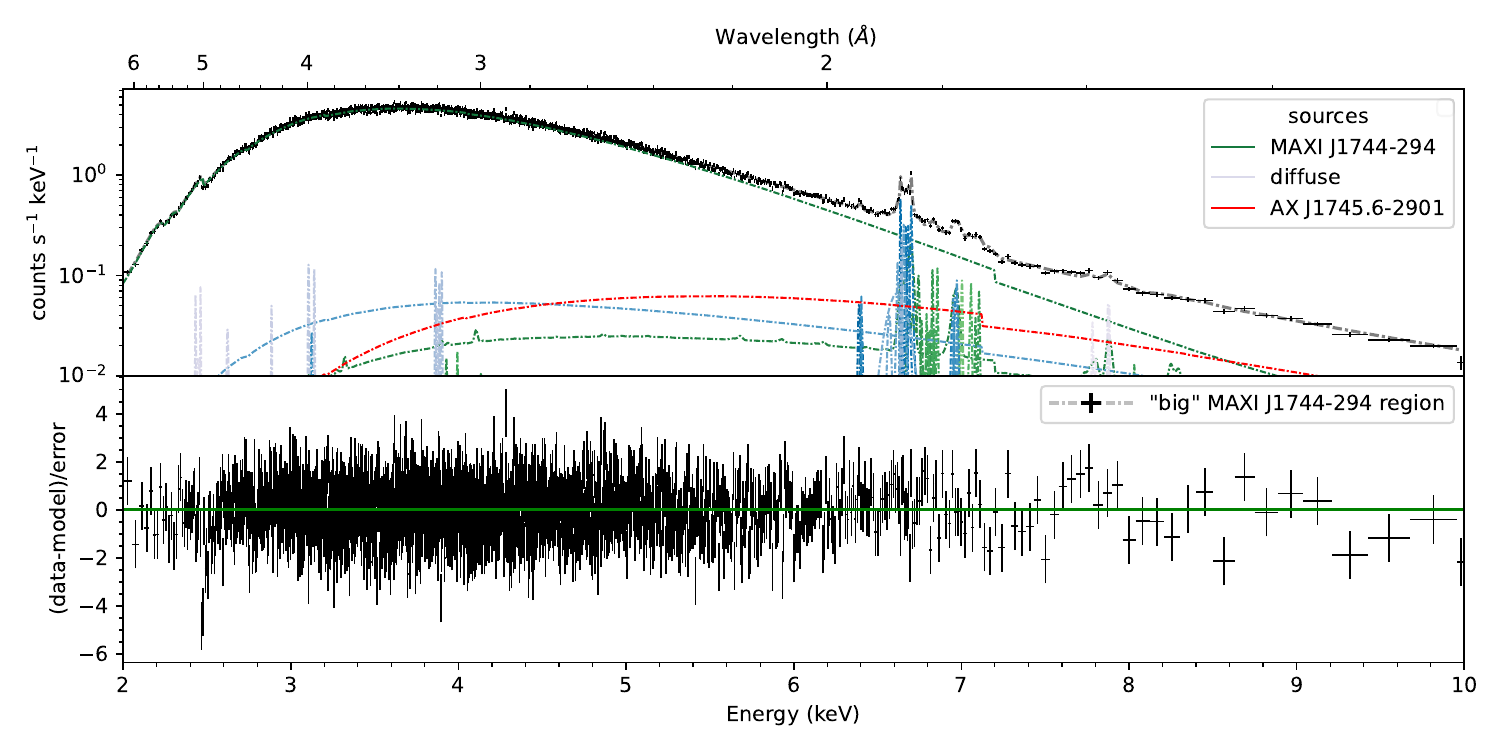}
    \includegraphics[clip,trim=0cm 0.3cm 0cm 0cm,width=0.49\textwidth]{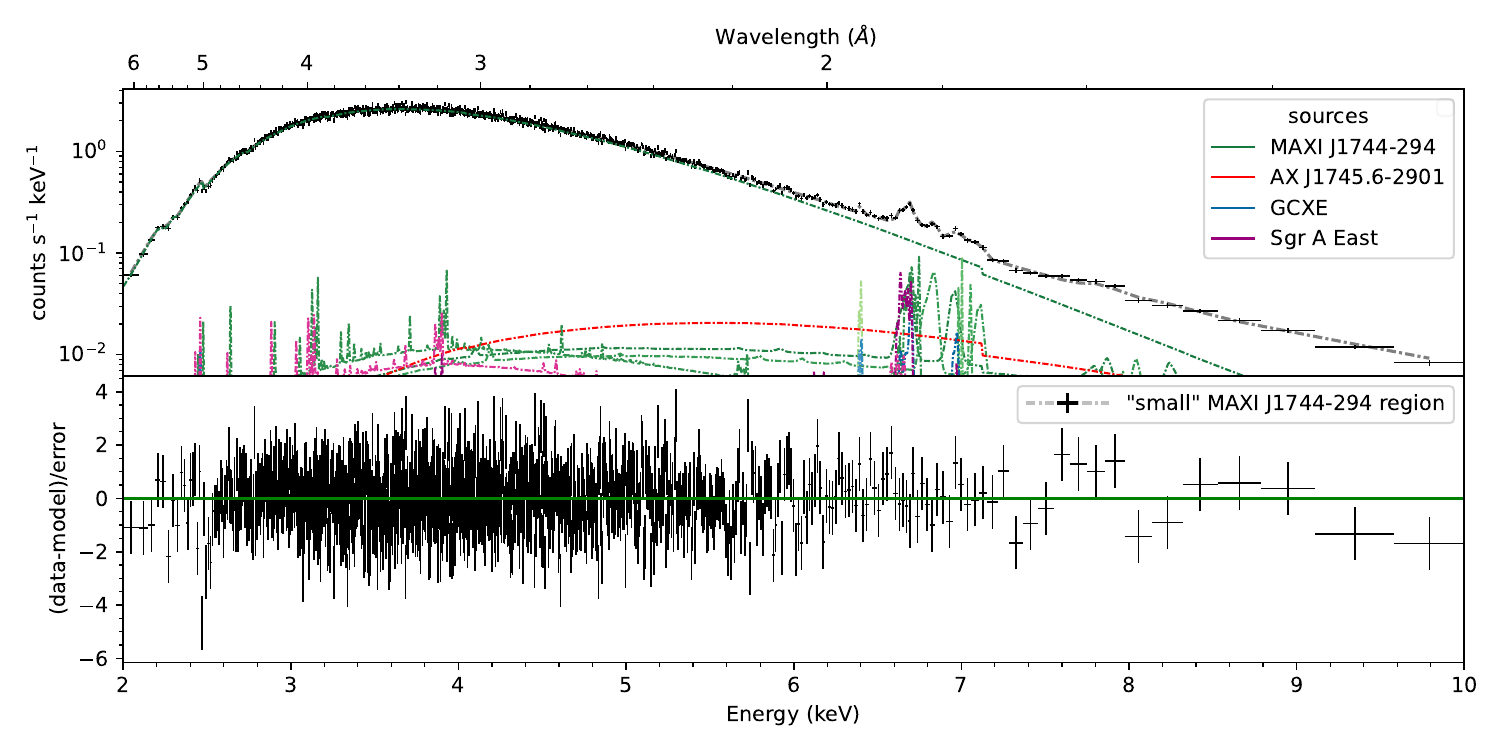}
\vspace{-1.em}
    \caption{Full residuals of the collisional ionization modeling of the line features in the "big" \textbf{(left)} and "small" \textbf{(right)} MAXI J1744-294 region in the DDT observation. The residuals for the AX J1745.6-2901 spectrum, which are largely unaffected by the switch to physical models, are omitted for legibility. The spectra are visually rebinned at a 20$\sigma$ significance level, and model components at 3$\sigma$.}
    \label{fig:resolve_CIE_resid_full}
\end{figure*}
 
\begin{figure*}[h!]
\centering
\vspace{-0em}
    \includegraphics[clip,width=0.49\textwidth]{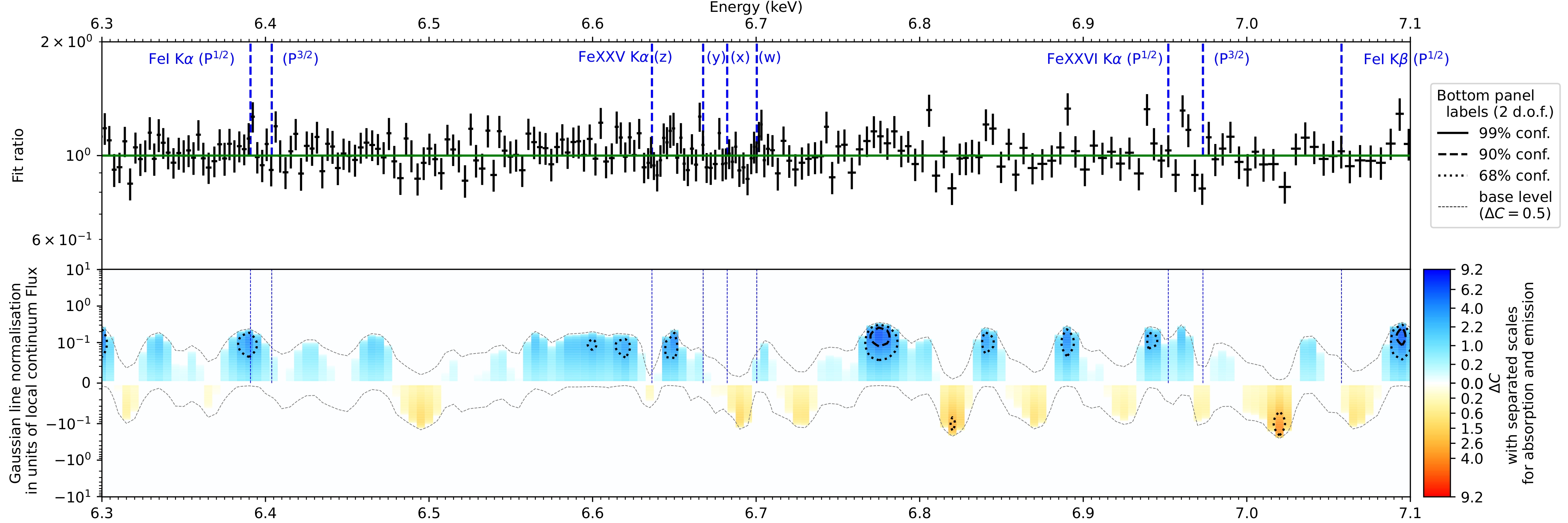}
    \includegraphics[clip,width=0.49\textwidth]{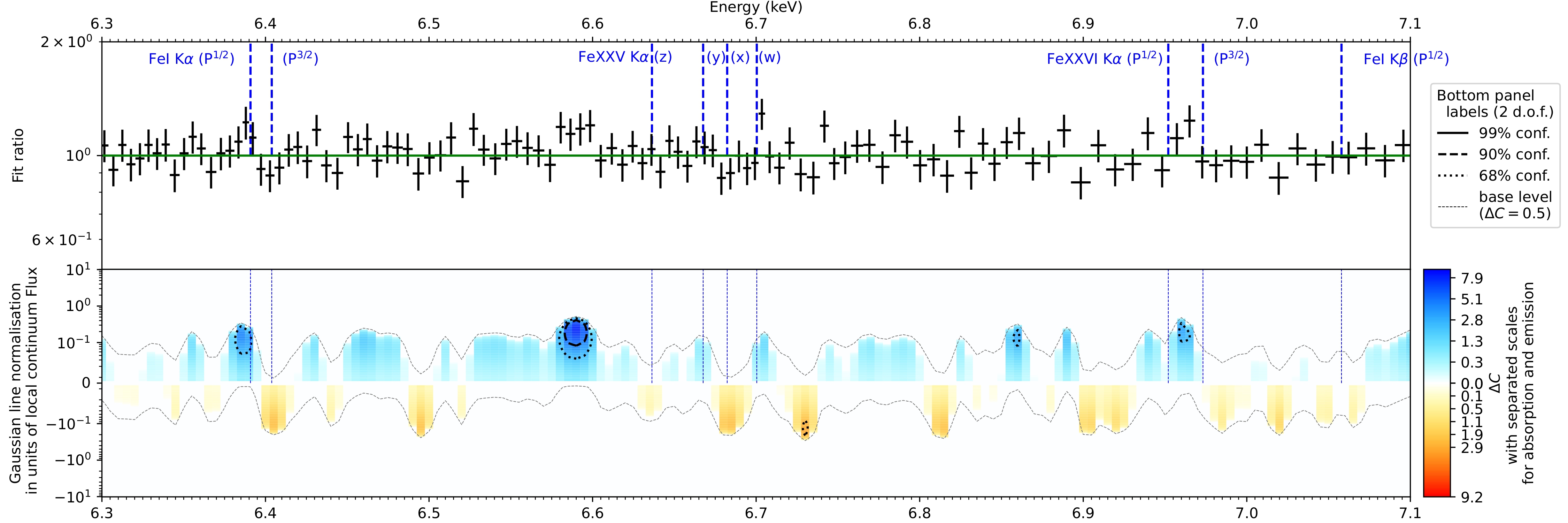}
    \caption{Blind searches for narrow line features in the 6.3-7.1 keV range after the collisional ionization modeling of the "big" \textbf{(left)} and "small"  \textbf{(right)} MAXI J1744-294 region in the DDT observation. All spectra are rebinned at a 10$\sigma$ significance level for visibility.}
    \label{fig:blind_search_CIE_BH}
\end{figure*}

\clearpage

\begin{figure*}[h!]
\centering
    \includegraphics[clip,width=0.49\textwidth]{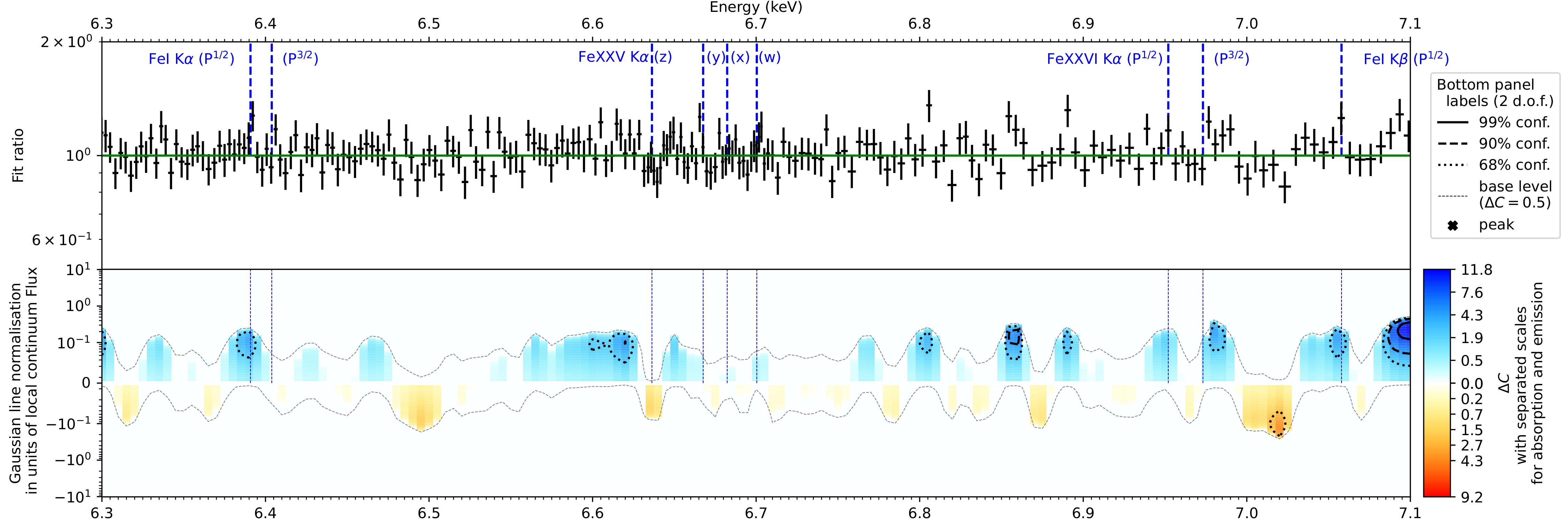}
    \includegraphics[clip,width=0.49\textwidth]{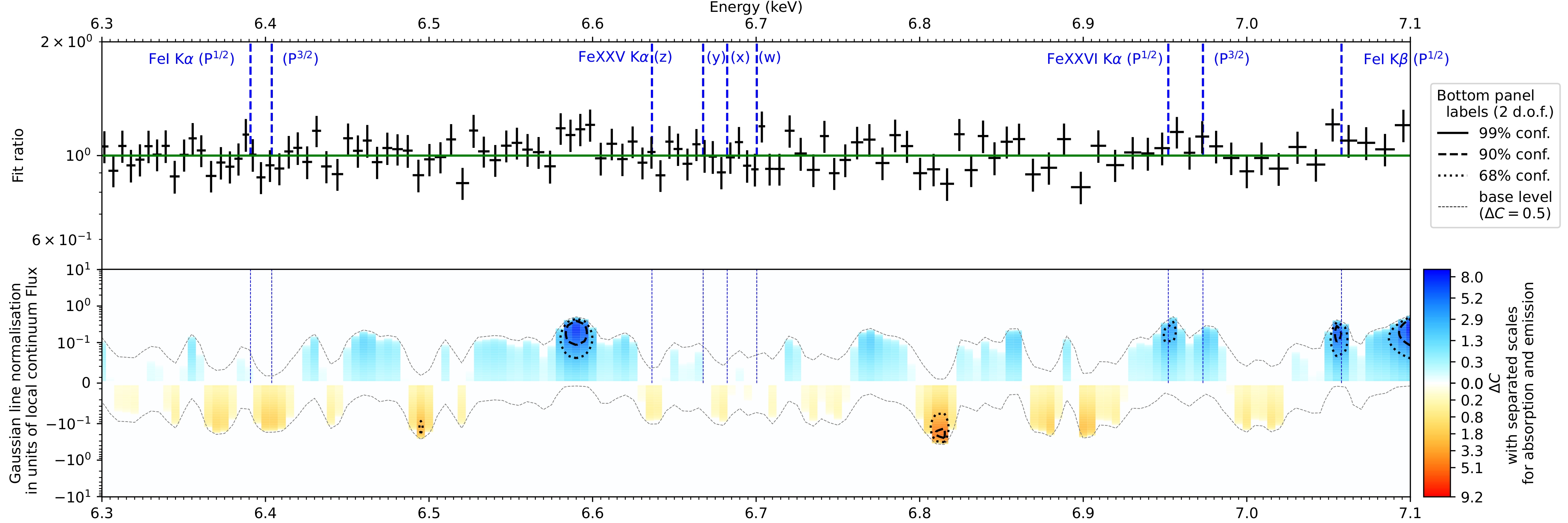}
    \caption{Blind searches for narrow line features in the 6.3-7.1 keV range after the adjusted empirical modeling of Section~\ref{sub:FeK_ratios}, for the "big" \textbf{(left)} and "small" \textbf{(right)} MAXI J1744-294 region in the DDT observation, enforcing conservative upper limits of 0.205 on the Fe I K$\beta$/K$\alpha$ flux ratios. All spectra are rebinned at a 10 $\sigma$ significance level for visibility.}
    \label{fig:blind_search_phys_fekratio}
\end{figure*}

\section{Background Rescaling}\label{app:bkg_rescale}

\begin{figure*}[h!]
\centering
    \includegraphics[clip,width=0.49\textwidth]{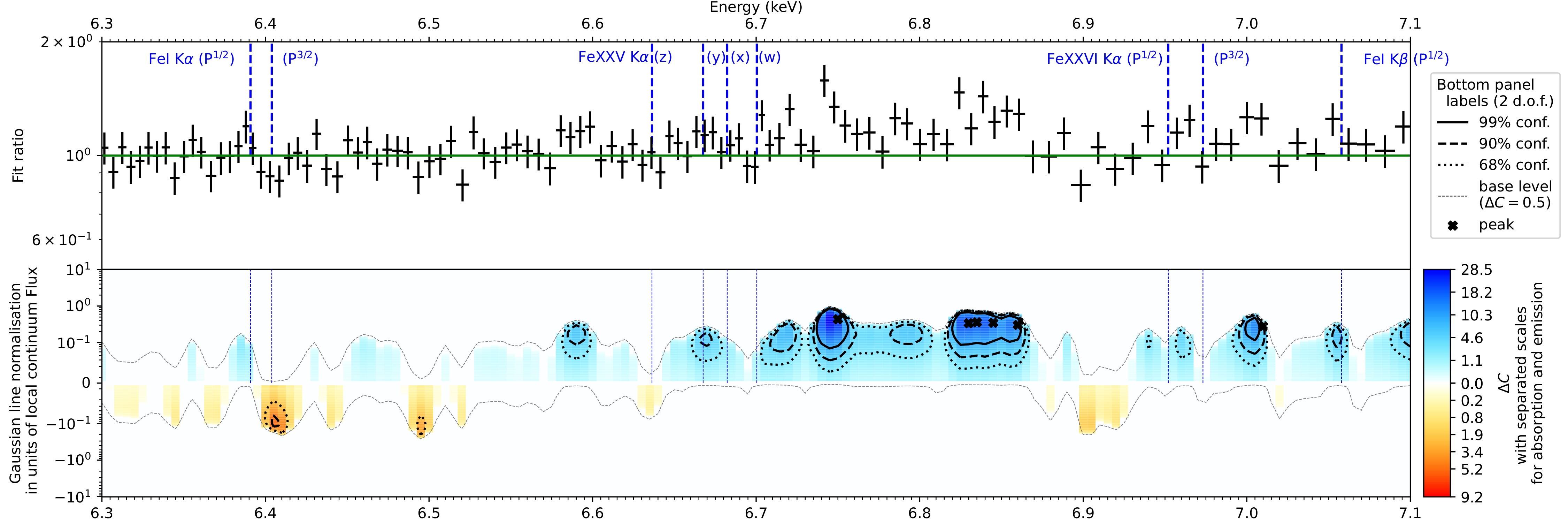}\\
    \includegraphics[clip,width=0.49\textwidth]{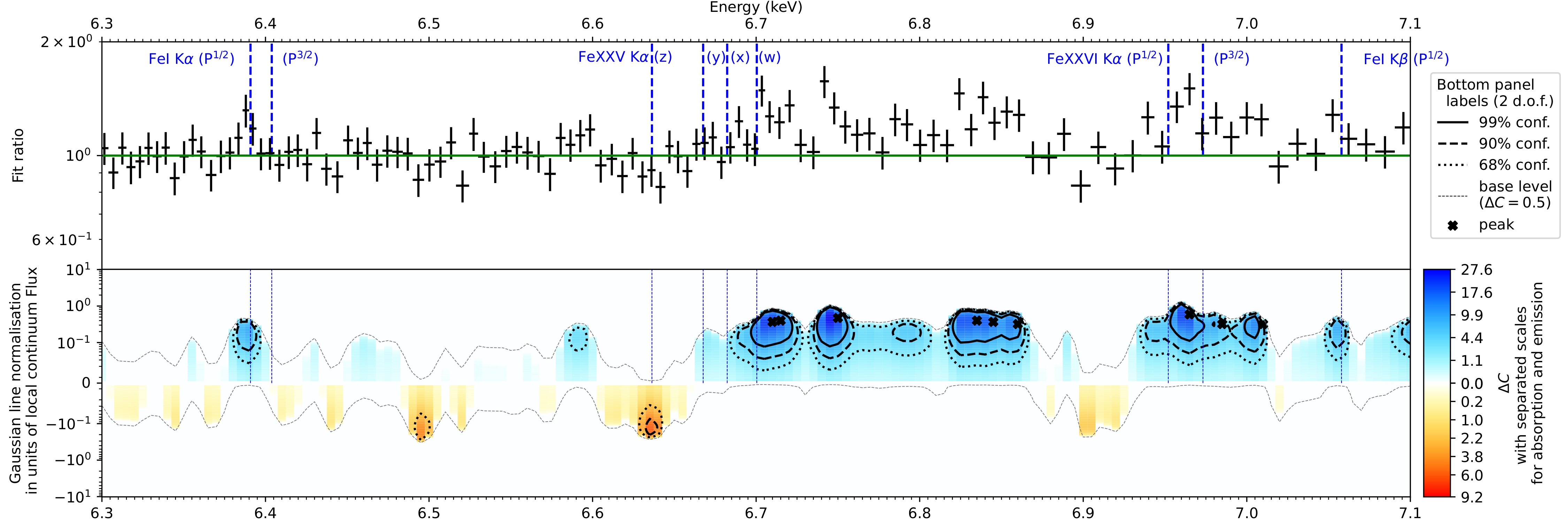}\\
        \includegraphics[clip,width=0.49\textwidth]{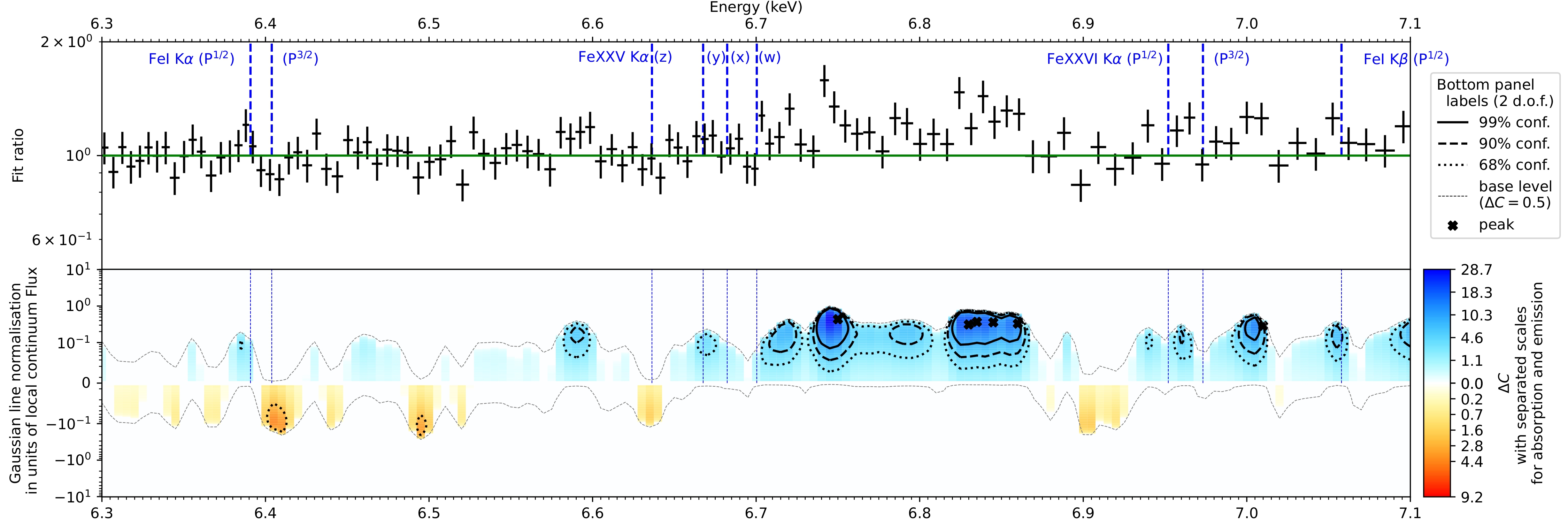}
    \caption{Blind searches for narrow line features in the 6.3-7.1 keV range after freely rescaling the GCXE (top), Sgr A East (middle), and both models (bottom), for the "small" MAXI J1744-294 region in the DDT observation. All spectra are rebinned at a 10 $\sigma$ significance level for visibility.}
    \label{fig:resid+blind_rescale_BH}
\end{figure*}




\end{document}